\documentclass[acmsmall, screen]{acmart}
\AtBeginDocument{%
  }

\setcopyright{acmcopyright}
\copyrightyear{2025}
\acmYear{2025}
\acmDOI{XXXXXXX.XXXXXXX}
\usepackage{amsthm}

\theoremstyle{definition}

\newtheorem{definition}{Definition}

\usepackage{listings}    
\usepackage{graphicx}    
\usepackage{subcaption}  

\usepackage{booktabs}    
\usepackage{xurl}        
\usepackage[htt]{hyphenat} 
\usepackage{microtype} 
\DisableLigatures{}    

\usepackage{caption}
\usepackage{setspace}
\usepackage{multirow}
\usepackage{enumerate}
\usepackage{pdflscape}
\usepackage{pgfplots} 
\usepackage{qcircuit}

\usepackage{xcolor}
\definecolor{codegreen}{rgb}{0.25,0.5,0.35}
\definecolor{codegray}{rgb}{0.5,0.5,0.5}
\definecolor{codepurple}{rgb}{0.6,0,0}
\definecolor{backcolour}{rgb}{0.95,0.95,0.92}
\definecolor{colorstring}{rgb}{0.5,0,0.35}
\definecolor{rltred}{rgb}{0.5,0,0}
\definecolor{rltgreen}{rgb}{0,0.5,0}
\definecolor{rltblue}{rgb}{0,0,0.5}
\definecolor{DarkGreen}{rgb}{0.00,0.60,0.00}
\definecolor{ScarletRed}{rgb}{0.80,0.00,0.00}
\definecolor{blizzardblue}{rgb}{0.67,0.9,0.93}
\definecolor{green-yellow}{rgb}{0.68,1.0,0.18}
\definecolor{dkgreen}{rgb}{0,0.6,0}
\definecolor{gray}{rgb}{0.5,0.5,0.5}
\definecolor{mauve}{rgb}{0.58,0,0.82}
\definecolor{lightgrey}{rgb}{0.90,0.90,0.90}
\definecolor{grey}{gray}{0.75}
\definecolor{light-gray}{gray}{0.80}
\usepackage{colortbl}

\lstdefinestyle{mystyle}{
    escapechar=©,
    language=Python,
	backgroundcolor=\color{backcolour},
    basicstyle=\footnotesize\ttfamily,
   	identifierstyle=\footnotesize\ttfamily,
	commentstyle=\color{codegreen},
	keywordstyle=\color{colorstring}\bfseries,
	morekeywords={OR, AND},
	numberstyle=\ttfamily\color{codegray},
	stringstyle=\ttfamily\color{DarkGreen},
	breaklines=true,
	captionpos=b,
	keepspaces=true,
	numbers=left,
	numbersep=2pt,
	showspaces=false,
	showstringspaces=false,
	showtabs=false,
	tabsize=2
}
\usepackage{braket}
\usepackage{amsfonts}  
\usepackage{mathtools}
\usepackage{pifont}

\usepackage{xspace}

\usepackage{tcolorbox}

\usepackage{ifthen}
\newboolean{showcomments}
\setboolean{showcomments}{false} 

\ifthenelse{\boolean{showcomments}}{
	\newcommand{\nbc}[3]{%
		{\colorbox{#3}{\bfseries\sffamily\scriptsize\textcolor{white}{#1}}}%
		{\textcolor{#3}{\sf\small$\langle$\textit{#2}$\rangle$}}%
	}
	
}{
	\newcommand{\nbc}[3]{}

}

\usepackage[ruled,linesnumbered]{algorithm2e}
\usepackage{algpseudocode}

\usepackage{enumitem}
\usepackage{ragged2e}

\newcommand{\Subtitle}[1]{\textbf{\textit{#1}.}\xspace}

\usepackage{makecell}

\usepackage{wrapfig}

\usepackage{titlecaps}  
\usepackage{textcase}   

\newcommand{\RQContent}[1]{%
    \ifnum#1=1
        What kinds of quantum programs were employed?
    \else\ifnum#1=2 
        What buggy variants were used for fault detection?
    \else\ifnum#1=3 
        How was the scalability of quantum programs under test handled?
    \else\ifnum#1=4 
        What test inputs were designed for the testing process?
    \else\ifnum#1=5 
        How was the test oracle problem handled for analyzing test outputs?
    \else\ifnum#1=6 
        Which metrics could be employed to quantify the performance of the proposed approach?
    \else\ifnum#1=7 
        What methods were adopted to compare the proposed approach with baselines?
    \else\ifnum#1=8 
        How were the statistical repetitions configured?
    \else\ifnum#1=9 
        What backends were used to execute the tested quantum circuits?
    \else\ifnum#1=10 
        What publicly available tools have been revealed in the primary studies?
    \fi\fi\fi\fi\fi\fi\fi\fi\fi\fi%
}

\newcommand{\RQShort}[1]{%
    \ifnum#1=1
        Quantum Programs
    \else\ifnum#1=2 
        Buggy Variants
    \else\ifnum#1=3 
        Scalability Issue
    \else\ifnum#1=4 
        Test Inputs
    \else\ifnum#1=5 
        Output Analysis
    \else\ifnum#1=6 
        Evaluation Metrics
    \else\ifnum#1=7 
        Contrastive Analysis
    \else\ifnum#1=8 
        Statistical Repetitions
    \else\ifnum#1=9 
        Execution Backends
    \else\ifnum#1=10 
        Available Tooling
    \fi\fi\fi\fi\fi\fi\fi\fi\fi\fi%
}

\newcommand{\RQSection}[1]{%
  \texorpdfstring{%
    RQ{#1}: \RQShort{#1}%
  }{%
    RQ{#1}: \RQShort{#1}%
  }%
}

\newcommand{\GroupObj}{Programs under test}
\newcommand{\GroupTest}{Testing process setups}
\newcommand{\GroupEval}{Approach evaluation}
\newcommand{\GroupExp}{Experimental configurations and resources}

\newcommand{\SearchRoundOne}{$\mathfrak{R}_{1}$\xspace}
\newcommand{\SearchRoundTwo}{$\mathfrak{R}_{2}$\xspace}

\newcommand{\KeywordEndTime}{2025-10-11\xspace}
\newcommand{\SnowballingEndTime}{2025-11-05\xspace}

\newcommand{\BibVenueOthers}{21\xspace}  
\newcommand{\BibTSE}{\PaperSeventyThree, \PaperEightyFour} 
\newcommand{\BibTOSEM}{\PaperThirtySix, \PaperFortyThree, \PaperFortySix, \PaperSeventyOne, \PaperEightyThree} 
\newcommand{\BibASE}{\PaperOne, \PaperThree, \PaperFive, \PaperSix, \PaperThirtyNine} 

\newcommand{\PaperNumKeywordInitial}{76\xspace}
\newcommand{\PaperNumKeywordSecond}{38\xspace}
\newcommand{\PaperNumKeywordTotal}{343\xspace}
\newcommand{\PaperNumSnowballingInitial}{39\xspace}
\newcommand{\PaperNumSnowballingSecond}{21\xspace}
\newcommand{\PaperNumSnowballingTotal}{41\xspace}

\newcommand{\SizeOfLiteraturePool}{384\xspace}
\newcommand{\NumberOfPrimaryStudies}{59\xspace}

\newcommand{\PaperOne}{mendiluze2022muskit}
\newcommand{\PaperThree}{wang2022quito}
\newcommand{\PaperFive}{wang2024qucat}
\newcommand{\PaperSix}{ye2024quratest}
\newcommand{\PaperTwelve}{huang2019statistical}
\newcommand{\PaperThirteen}{honarvar2020propertybased}
\newcommand{\PaperFourteen}{li2020projectionbased}
\newcommand{\PaperFifteen}{acharya2021test}
\newcommand{\PaperSeventeen}{wang2022qusbt}
\newcommand{\PaperTwenty}{wang2022origin}
\newcommand{\PaperTwentyOne}{wang2022generating}
\newcommand{\PaperTwentyFour}{abreu2023metamorphic}
\newcommand{\PaperTwentySeven}{costa2022asserting}
\newcommand{\PaperThirtyOne}{sato2024locating}
\newcommand{\PaperThirtyThree}{pontolillo2024delta}
\newcommand{\PaperThirtyFive}{fortunato2024gate}
\newcommand{\PaperThirtySix}{shi2025quantest}
\newcommand{\PaperThirtySeven}{chen2024automatic}
\newcommand{\PaperThirtyEight}{kang2024statistical}
\newcommand{\PaperThirtyNine}{muqeet2024quantum}
\newcommand{\PaperFortyOne}{yamaguchi2025practical}
\newcommand{\PaperFortyThree}{oldfield2025faster}
\newcommand{\PaperFortyFour}{tan2025hornbro}
\newcommand{\PaperFortyFive}{xia2025quantum}
\newcommand{\PaperFortySix}{li2025preparation}
\newcommand{\PaperFiftyOne}{guo2024on}
\newcommand{\PaperFiftyTwo}{huang2024a}
\newcommand{\PaperFiftyFour}{rovara2025a}

\newcommand{\PaperFiftyNine}{ali2021assessing}
\newcommand{\PaperSixty}{wang2021application}
\newcommand{\PaperSixtySeven}{pan2023understanding}
\newcommand{\PaperSeventy}{long2024equivalence}
\newcommand{\PaperSeventyOne}{long2024testing}
\newcommand{\PaperSeventyThree}{muqeet2024mitigating}
\newcommand{\PaperSeventyFour}{fortunato2022mutation}
\newcommand{\PaperSeventyEight}{usandizaga2025quantum}
\newcommand{\PaperEightyOne}{park2025squad}
\newcommand{\PaperEightyThree}{li2024automatic}
\newcommand{\PaperEightyFour}{sato2025buglocating}
\newcommand{\PaperEightySix}{wang2021poster}
\newcommand{\PaperNinetyFour}{pontolillo2025from}
\newcommand{\PaperOneHundredAndThree}{wang2018quanfuzz}
\newcommand{\PaperOneHundredAndFive}{jin2025novaq}
\newcommand{\PaperOneHundredAndSix}{long2025a}
\newcommand{\PaperOneHundredAndSeven}{shao2024a}
\newcommand{\PaperOneHundredAndEight}{long2022testing}
\newcommand{\PaperOneHundredAndNine}{ishimoto2025evaluating}
\newcommand{\PaperOneHundredAndTwelve}{miranskyy2025on}
\newcommand{\PaperOneHundredAndThirteen}{gil2024qcrmut}
\newcommand{\PaperOneHundredAndFourteen}{klymenko2025qut}
\newcommand{\PaperOneHundredAndFifteen}{oldfield2025bloch}
\newcommand{\PaperOneHundredAndSixteen}{virani2025distinguishing}
\newcommand{\PaperOneHundredAndSeventeen}{klymenko2025contextaware}
\newcommand{\PaperOneHundredAndNineteen}{pontolillo2025qucheck}
\newcommand{\PaperOneHundredAndTwentyOne}{li2019proq}
\newcommand{\PaperOneHundredAndTwentyTwo}{liu2020quantum}
\newcommand{\PaperOneHundredAndTwentyThree}{park2024aqua}

\usepackage{environ}

\newif\ifshowrevision
\showrevisionfalse        
\newcommand{\revise}[1]{%
  \ifshowrevision
    \textcolor{blue!70!black}{#1}%
  \else
    #1%
  \fi
}
\newcommand{\highlight}[1]{%
  \ifshowrevision
    \textcolor{magenta}{#1}%
  \else
    #1%
  \fi
}
\NewEnviron{EnvRevise}{
  \ifshowrevision
    {\color{blue!70!black}\BODY}
  \else
    \BODY
  \fi
}

\usepackage{natbib} 

\begin{document}

\title{\texorpdfstring{
A Methodological Analysis of Empirical Studies in Quantum Software Testing}
    {A Methodological Analysis of Empirical Studies in Quantum Software Testing}
}

\author{Yuechen Li}
\orcid{0009-0006-5109-3288}
\email{liyuechen@buaa.edu.cn}

\affiliation{%
  \institution{Beihang University}
  \city{Beijing}
  \country{China}
}

\affiliation{%
  \institution{Kyushu University}
  \city{Fukuoka}
  \country{Japan}
}

\author{Minqi Shao}
\orcid{0009-0001-4832-5256}
\email{shao.minqi.229@s.kyushu-u.ac.jp}
\affiliation{%
  \institution{Kyushu University}
  \city{Fukuoka}
  \country{Japan}
}

\author{Jianjun Zhao}
\authornote{Corresponding author}
\orcid{0000-0001-8083-4352}
\email{zhao@ait.kyushu-u.ac.jp}
\affiliation{%
  \institution{Kyushu University}
  \city{Fukuoka}
  \country{Japan}
}

\author{Qichen Wang}
\email{wang.qichen.256@s.kyushu-u.ac.jp}
\affiliation{%
  \institution{Kyushu University}
  \city{Fukuoka}
  \country{Japan}
}


\begin{abstract}
    In quantum software engineering (QSE), quantum software testing (QST) has attracted increasing attention as quantum software systems grow in scale and complexity. Since QST evaluates quantum programs through execution under designed test inputs, empirical studies are widely used to assess the effectiveness of testing approaches. However, the design and reporting of empirical studies in QST remain highly diverse, and a shared methodological understanding has yet to emerge, making it difficult to interpret results and compare findings across studies.
This paper presents a methodological analysis of empirical studies in QST through a systematic examination of \NumberOfPrimaryStudies primary studies identified from a literature pool of size \SizeOfLiteraturePool. We organize our analysis around ten research questions that cover key methodological dimensions of QST empirical studies, including objects under test, baseline comparison, testing setup, experimental configuration, and tool and artifact support. Through cross-study analysis along these dimensions, we characterize current empirical practices in QST, identify recurring limitations and inconsistencies, and highlight open methodological challenges. Based on our findings, we derive insights and recommendations to inform the design, execution, and reporting of future empirical studies in QST.

\end{abstract}





\maketitle















\section{Introduction}
\textbf{Quantum Computing (QC)}~\cite{feynman2018simulating} has emerged as a promising computational paradigm with the potential to outperform classical computing for certain classes of problems. In particular, gate-based quantum computing has attracted considerable attention due to its suitability for universal quantum computation and its support for programmable quantum software. With the rapid development of quantum \textbf{Software Development Kits (SDKs)}, such as Qiskit~\cite{aleksandrowicz2019qiskit}, Q\#~\cite{svore2018q}, Cirq~\cite{google2018cirq}, and PennyLane~\cite{bergholm2018pennylane}, quantum programs can now be implemented and executed on simulators or remote quantum hardware, making quantum software development increasingly accessible.

As quantum software systems grow in scale and complexity, ensuring their correctness and reliability has become a critical concern. \textbf{Quantum Software Testing (QST)} assesses whether the runtime behavior of quantum programs conforms to intended specifications or properties, and has therefore received increasing attention in recent years. Similar to classical software testing, QST relies on executing programs under designed test inputs and observing their outcomes; consequently, empirical studies are widely used to evaluate testing approaches and their practical effectiveness.
However, conducting empirical studies for QST poses challenges that go beyond those encountered in classical software testing. The probabilistic nature of quantum measurement, the destructive effect of observation on quantum states, and the dependence on underlying backend characteristics require repeated executions and carefully controlled experimental setups. Moreover, empirical evaluations in QST are often conducted using a combination of classical simulation and execution on \textbf{Noisy Intermediate-Scale Quantum (NISQ)} devices, further complicating the design, interpretation, and comparison of experimental results.

Despite the growing body of research on QST, we observe substantial diversity in empirical study design, experimental configurations, evaluation criteria, and reporting practices. Due to the interdisciplinary nature of QST, which lies at the intersection of software engineering and quantum computing, a shared methodological understanding of how empirical studies should be designed, conducted, and reported has yet to emerge. This lack of methodological consistency makes it difficult to interpret results across studies and assess the maturity and applicability of proposed testing approaches.
In this paper, we present a methodological analysis of empirical studies in QST. Our study is based on a systematic examination of \NumberOfPrimaryStudies primary studies published between 2018 and 2025, identified from a literature pool of size \SizeOfLiteraturePool. We organize our analysis around ten research questions that capture five key aspects of empirical study design in QST, including objects under test, baseline comparison, testing setup, experimental configuration, and tool and artifact support. Through cross-study analysis along these dimensions, we characterize current empirical practices in QST, identify recurring limitations and inconsistencies, and highlight open methodological challenges. Based on our findings, we derive insights and recommendations to inform the design, execution, and reporting of future empirical studies in QST. We make the data, code, and documentation used in our study publicly available at~\url{https://github.com/NahidaNahida/QST-empirical-study}, and also provide a long-term archive of the repository~\cite{li_2026_18159893}.

This paper makes the following contributions:
\begin{itemize}
    \item We conduct a systematic, methodology-oriented analysis of empirical studies in QST, focusing on how studies are designed, conducted, and reported.
    \item We propose ten research questions and a consistent extraction and categorization scheme that cover key methodological aspects of empirical QST studies, including objects under test, baseline comparison, testing setup (e.g., inputs and oracle handling), experimental configuration (e.g., repetitions and backends), and tool and artifact support.
    \item We synthesize evidence across studies to summarize current empirical practices, identify recurring limitations and inconsistencies, and provide recommendations and future research directions for improving empirical QST studies.
    \item We release the data, code, and documentation used in this study as a public artifact with a long-term archive to support transparency, reuse, and follow-up research.
\end{itemize}

The rest of this paper is organized as follows. Section~\ref{sec: pre} introduces background concepts in quantum computing and quantum information. Section~\ref{sec: method} describes our research methodology, including the research questions and the literature analysis process. Section~\ref{sec: bib} presents a bibliometric analysis of the selected studies. Sections~\ref{sec: objects} to~\ref{sec: settings} report and analyze the results, addressing each research question. Section~\ref{sec: threats} discusses threats to validity. Section~\ref{sec: discussion} reflects on the findings and outlines implications for future empirical studies in QST. Section~\ref{sec: related} discusses related work, and Section~\ref{sec: conclusion} concludes the paper.


\section{Preliminaries}
\label{sec: pre}
This section introduces the quantum computing and quantum information background needed to follow the rest of the paper, focusing on key mathematical and physical concepts relevant to empirical studies in quantum software testing.
\revise{Besides, we summarize several key concepts involved in software testing.}

\subsection{Basics of Quantum Information Theory}
This subsection provides a brief introduction to quantum information theory.
More details can be found in the book~\cite{nielsen2010quantum}.
\revise{The involved concepts are not concerned in testing classical programs, but as is shown in Section~\ref{sec: process}, they could help design and optimize the testing process for quantum programs.}

\Subtitle{Qubits and quantum states}
A qubit is a basic unit of quantum information, and quantum states characterize the behavior of a quantum system composed of qubits. A \textit{pure state} can be formally depicted by either a state vector $\ket{\psi}$ defined in the \textit{Hilbert space} $\mathcal{H}$ or a density operator $\rho$ defined in the space of linear operators on the Hilbert space $\mathcal{L}(\mathcal{H})$. Taking an example of a single-qubit system, the state vector $\ket{\psi}$ can be written in the span of the \textit{computational basis states}, i.e., $\ket{\psi}= \alpha\ket{0}+\beta\ket{1} (\alpha, \beta \in \mathbb{C}, |\alpha|^2+|\beta|^2=1)$, while the corresponding density operator turns out to be $\rho=\ket{\psi}\bra{\psi} =\alpha\alpha^*\ket{0}\bra{0}+\alpha^*\beta\ket{1}\bra{0} + \alpha\beta^*\ket{0}\bra{1}+\beta\beta^*\ket{1}\bra{1}$, where both ket $\ket{\cdot}$ and bra $\bra{\cdot}$ (i.e., the conjugate transpose of ket) are \textit{Dirac notations} used in quantum physics. 
By the way, from the perspective of a geometrical representation, a single-qubit pure state can be depicted by a \textit{Bloch sphere} with two real-number angles, i.e., $\ket{\psi}=\cos(\theta/2)\ket{0}+e^{i\phi}\sin(\theta/2)\ket{1}$.
In comparison, a \textit{mixed state} $\rho$ cannot be represented by a state vector and corresponds to a statistical ensemble $\{p_j, \ket{\psi_j}\} (p_j \ge 0, \sum_{j} p_j = 1)$, i.e. $\rho = \sum_j p_j \ket{\psi_j}\bra{\psi_j}$, which refers to the occurrence of the state $\ket{\psi_j}\bra{\psi_j}$ with probability $p_j$.

\Subtitle{Superposition and entanglement}
\textit{Quantum superposition} refers to a quantum state that, with respect to a chosen basis, is expressed as a linear combination of more than one basis state with non-zero amplitudes, rather than a single basis state. For example, the 2-qubit cat state $\ket{\text{cat}}_2\equiv(\ket{00}+\ket{11})/{\sqrt{2}}$ can be regarded to exist in superposition on a computational basis $\{\ket{00},\ket{01},\ket{10},\ket{11}\}$. Note that the interpretation of quantum superposition is not absolute, but relative to the selected basis states. For an $n$-qubit system, its Hilbert space can be represented by the tensor product of that for each component qubit, $\mathcal{H}=\mathcal{H}_{n-1}\otimes\cdots\otimes\mathcal{H}_{1}\otimes\mathcal{H}_0$. 
A multi-qubit quantum state is said to be \textit{genuinely multipartite entangled} if it cannot be written as a tensor product across any partition of the system. Otherwise, the state is \textit{partially separable} if it is separable with respect to at least one bipartition of the system (e.g., $(\ket{000}+\ket{011})/{\sqrt{2}}$ for $\mathcal{H}'\equiv \mathcal{H}_2\otimes\mathcal{H}_1\otimes\mathcal{H}_0$), and \textit{fully separable} if it can be expressed as a tensor product of states corresponding to individual subsystems (e.g., $\ket{011}$ for $\mathcal{H}'$). \revise{There are some exemplary genuinely multipartite entangled states, such as the typical Greenberger-Horne-Zeilinger (GHZ) state $\ket{\mathrm{GHZ}}\equiv(\ket{000}+\ket{111})/{\sqrt{2}}$, which is a 3-qubit generalization of one Einstein-Podolsky-Rosen (EPR) state $(\ket{00}+\ket{11})/{\sqrt{2}}$}.

\Subtitle{Quantum circuits}
A quantum circuit is an abstract model that is used to visualize the computational process in gate-based quantum computers. It consists of lines representing qubits, a temporal sequence of quantum gates, and quantum measurement operations. 
In a quantum circuit, qubit operations are displayed from left to right, which determines the order of their application.
Key features of a quantum circuit include the \textit{width}, meaning the number of qubits; the \textit{size}, defined as the total number of gates; and the \textit{depth}, which corresponds to the number of sequential time steps or layers in the longest sequence of dependent gate operations. These quantifiable features can be used to analyze the complexity and resource requirements of quantum algorithms. 

\Subtitle{Quantum gates and quantum channels}
Following the discrete-time Schr\"{o}dinger equation, a \textit{quantum gate} governs the time evolution of a quantum state over a single discrete time step. Formally, a quantum gate can be denoted as a unitary operator $U$ that satisfies $UU^{\dagger}=I$. 
The unitarity guarantees the reversibility of the evolution and the conservation of total probability. Specifically, a quantum gate transforms an initial pure state $\ket{\psi}$ into $U\ket{\psi}$, while in the density matrix formalism it transforms $\rho$ into $U\rho U^\dagger$.
Within the quantum circuit, a quantum gate acting on $n$ qubits is mathematically instantiated by a $2^n\times 2^n$ unitary matrix, where this matrix is defined in the special unitary group of degree $2^n$ (i.e., $U\in\mathrm{SU}(2^n)$), such as the following basic quantum gates,
$$
H \equiv \frac{1}{\sqrt{2}}
\begin{bmatrix}
1 & 1 \\
1 & -1
\end{bmatrix},
X \equiv \begin{bmatrix}
  0 & 1 \\ 1 & 0
\end{bmatrix},
R_x(\theta) \equiv 
\begin{bmatrix}
\cos(\theta/2) & -i\sin(\theta/2) \\
-i\sin(\theta/2) & \cos(\theta/2)
\end{bmatrix},
CNOT \equiv  \begin{bmatrix}
    1&0&0&0\\
    0&1&0&0\\
    0&0&0&1\\
    0&0&1&0
\end{bmatrix},
$$
where the operations of some gates, such as the X-axis rotation gate $R_x(\theta)$, rely on angle parameters.
Compared to the quantum gate, a \textit{quantum channel} $\mathcal{E}\in\mathcal{L}(\mathcal{L}(\mathcal{H})$, defined in the space of linear operators on $\mathcal{L}(\mathcal{H})$, is a universal representation of quantum operations, which can support a formal description of a non-unitary operation caused by the interaction between the outer environment and the closed quantum system. 
More particularly, a quantum channel is formalized as a completely positive trace-preserving map from an input density operator $\rho$ to an output $\mathcal{E}(\rho)$. 
When the evolution is unitary, the corresponding quantum channel is implemented by a unitary operator.
More generally, the action of a quantum channel $\mathcal{E}(\rho)$ can be denoted using the operator-sum representation or equivalently by the Choi matrix~\cite{wood2011tensor}.

\Subtitle{Quantum measurements}
\textit{Quantum measurement} is the only means by which observers in the classical world can access information about a quantum system. 
Although the qubit evolution within a quantum circuit can be described in a deterministic manner, the quantum measurement brings uncertainty to the outcomes of a quantum system.
In detail, quantum measurement can be represented by a collection of measurement operators $\{M_m\}_m\subset\mathcal{L}(\mathcal{H})$, which satisfy the completeness relation (i.e., $\sum_m M_m^{\dagger}M_m =I$).
There are specific quantum measurements widely adopted, such as the projective measurement, which specifies the orthogonality of different operators, and the more general positive operator-valued measurement, which requires that all operators are positive semi-definite.
The probability of obtaining the outcome $m$ after measuring a pure state $\ket{\psi}$ turns out to be $p_M(m)=\braket{\psi|M_m^{\dagger}M_m|\psi}$, while the density-operator representation of measuring the state $\rho$ is $p_M(m)=\mathrm{Tr}(M^{\dagger}_m M_m \rho)$, where $\mathrm{Tr}(\cdot)$ refers to the trace operation in linear algebra.
For example, given that the Pauli-Z measurement with $M_m=\ket{m}\bra{m}(m\in\mathbb{B})$\footnote{$\mathbb{B}$ indicates the Boolean set throughout our paper, i.e., $\mathbb{B}\equiv\{0,1\}$.}, the probabilities of measurement outcomes for $\alpha\ket{0}+\beta\ket{1}$ are $p_M(0)=|\alpha|^2$ and $p_M(1)=|\beta|^2$.
Once a measurement outcome $m$ is produced, the quantum state will update to $\rho'_m = M_m\rho M_m^{\dagger}/p_M(m)$. 
In QC, the term \textit{shots} refers to the number of independent repetitions per quantum circuit run, each of which yields a single outcome from every measurement action in the circuit.
Owing to the probabilistic nature of quantum measurements, we usually rely on multiple shots to extract statistical quantities about quantum states, such as probability distributions and expectation values. 

\Subtitle{Quantum tomography}
\textit{Quantum state tomography} is a technique for reconstructing an unknown quantum state $\rho$ by performing measurements on multiple identical copies of the state. More specifically, it involves performing projective measurements in different measurement bases and using the resulting statistics to reconstruct the density operator. For an $n$-dimensional quantum system, the density operator is fully characterized by $n^2-1$ independent real parameters~\cite{nielsen2010quantum}.
 In addition to that, \textit{quantum process tomography} aims to reconstruct an unknown quantum channel $\mathcal{E}$. This is achieved by (1) preparing a tomographically complete set of input states; (2) processing these input states through the quantum channel $\mathcal{E}$; (3) performing quantum state tomography on the corresponding output states; and (4) using the resulting data to solve a system of linear equations that fully characterizes the underlying linear map between inputs and outputs, thus obtaining a complete description of the quantum process~\cite{o2004quantum}.

\subsection{Quantum Programs and Their Implementation}
\revise{This subsection introduces the basics of quantum programs along with their runtime execution. The involved basic knowledge is associated with several aspects (e.g., selection of programs under test and configuration of execution backends) that should be particularly considered in empirical studies on QST.}

\Subtitle{Quantum algorithms}
\textit{Quantum algorithms} are algorithms that run on any realistic model of quantum computation, and the most commonly used model is the quantum circuit~\cite{mosca2008quantumalgorithms}.
\textit{Quantum programs} are concrete implementations of quantum algorithms written in quantum programming languages, such as OpenQASM, an assembly-based language tailored to quantum computers.
For gate-based quantum computers, a quantum algorithm consists of three basic steps: (1) encoding the classical or quantum data into the state of the input qubits, (2) implementing a sequence of quantum gates applied to these input qubits, and (3) performing quantum measurements on more than one qubit at the end to obtain a classically interpretable result~\cite{abhijith2022quantum}. 
As a prominent class of quantum algorithms, \textit{variational quantum algorithms} adopt a hybrid classical-quantum paradigm, combining a \textit{parameterized quantum circuit}, also referred to as an \textit{ansatz}, with a classical optimizer to iteratively update the circuit parameters to minimize or maximize a predefined objective function. 
An ansatz typically consists of parameterized single-qubit rotation gates and entangling operations.
\textit{Quantum neural networks}~\cite{kak1995quantum} are a class of quantum machine learning models often trained using variational quantum algorithms, in which the parameters are optimized from training data using gradients computed via the parameter shift rule~\cite{liu2018differentiable}. 

\Subtitle{Backends}
The implementation of quantum programs depends on the specific \textit{backends} that determine whether the computation is performed via logical simulation on classical hardware (e.g., CPU or GPU) or via physical implementation on quantum hardware (i.e., QPU).
The feasibility of classical simulators in academic research mainly stems from the limited qubit count and connectivity of current devices, as well as the presence of non-negligible quantum noise in current NISQ devices. 
\revise{The native gate sets vary with NISQ devices, implying that quantum circuits derived from quantum programs must be carefully transpiled and optimized to match the hardware-specific gate primitives.}
Meanwhile, practical considerations, such as the high overhead of accessing real quantum hardware, further motivate the use of classical simulators.
Among the classical simulation backends, ideal simulators and noisy simulators are the two main approaches. Ideal simulators model the qubit evolution of a theoretically closed quantum system, whereas noisy simulators aim to approximate the behavior of current NISQ devices in the presence of quantum noise.

\begin{EnvRevise}
\subsection{Several Concepts of Software Testing}
In this section, we briefly introduce several software testing concepts widely recognized in the \textbf{Software Engineering (SE)} community. This not only facilitates understanding of the remainder of the paper but also inspires adapting SE practices to QST.

\Subtitle{Mutation testing}
Mutation testing is a fault-based testing technique to evaluate the fault-detection capability of a test suite by introducing small syntactic changes into a program \cite{jia2010analysis}. Each modified version of the original program is referred to as a \textit{mutant}, which simulates faults that may occur during software development using predefined \textit{mutation operators}.  
In particular, a mutant is considered killed if a test case produces different outputs for the mutant and the original program, and the mutation score is typically used as the evaluation metric, computing the ratio of killed mutants. 
Moreover, mutation testing has been widely applied to various SE problems, such as fault localization~\cite{papadakis2015metallaxis}. More recently, it has been applied to intelligent software, such as deep neural networks \cite{ma2018deepmutation}, with domain-specific mutation operators.

\Subtitle{Program specification and test oracle}
A \textit{program specification} refers to a statement of requirements for a program, an expression of a design for a program, or a (formal or informal) statement of conditions against which the program can be verified~\cite{10.5555/1074100, barr2014oracle}. Guided by program specifications, \textit{test oracles}, including both automated implementations and human evaluations, are constructed to determine whether observed program outputs conform to expected behavior. Regarding the effectiveness of a test oracle, it is considered \textit{sound} if it does not incorrectly classify failing tests as passing (i.e., producing no false negatives). It is considered \textit{complete} if it does not incorrectly classify passing tests as failing (i.e., producing no false positives).
However, test oracle design is widely recognized as a challenging problem in SE, and it is tough to ensure both soundness and completeness simultaneously~\cite{barr2014oracle}. 

\Subtitle{Complexity measure}
Program complexity is often characterized by size-based and structural metrics, which provide a practical basis for assessing how test approaches scale as programs become more complex. Size-based metrics, such as lines of code, the number of functions, and the number of statements~\cite{shen1983software, halstead1977elements, weyuker1988evaluating}, are widely used to approximate software size and testing effort. Structural complexity is measured using cyclomatic complexity~\cite{mccabe1976complexity,zhu1997software}, which quantifies the number of independent execution paths in a program based on its control-flow graph. It estimates the testing difficulty and the minimum number of test cases required to adequately test a program. Additionally, more control-flow metrics, including the number of branches, loops, and nesting depth, provide further insights into program structure and testing complexity. Related coverage criteria~\cite{zhu1997software} measure the extent to which execution paths are exercised during testing.

\Subtitle{Cost-effectiveness analysis}
Test cost is typically measured using metrics such as execution time and memory or resource usage. The number of program executions also affects execution costs, especially in mutation testing with a large number of mutated programs~\cite{jia2010analysis}. 
Test effectiveness is commonly assessed using metrics including the percentage of failed tests, mutation score, and structural coverage criteria. The percentage of failed tests reflects the ability of a test suite to reveal embedded faults, while the mutation score, tailored to mutation testing, indicates test adequacy. Coverage criteria, such as statement, branch, and path coverage, further evaluate the extent to which program structures have been explored during testing, as higher coverage might expose more potential faults.
In practice, cost-effectiveness analysis is a broadly discussed issue in CST~\cite{basili2006comparing, zhou2018cost}, such as the prospect of executing a small number of test cases (i.e., test cost) to detect a large number of failures (i.e., test effectiveness), using typical evaluation metrics including average percentage of faults detected (i.e., abbreviated as APFD). Various techniques, such as test prioritization~\cite{rothermel2001prioritizing} and mutant quality evaluation~\cite{jia2010analysis}, have been proposed to balance this trade-off by reducing test cost while maintaining fault detection capability.

\end{EnvRevise}

\section{Research Methodology}
\label{sec: method}
This section describes the research methodology used to support a systematic, reproducible analysis of empirical studies in QST. We first define the \textbf{Research Questions (RQs)} that guide our analysis of how empirical QST studies are designed, conducted, and reported. We then introduce the process for collecting and analyzing primary studies.

\subsection{Research Questions}

\begin{table}[!t]
    \small
    \centering
    \caption{Research questions proposed in this study}
    \label{tab: rq_list}
    \resizebox{.98\columnwidth}{!}{
        \begin{tabular}{p{0.05\columnwidth}  p{0.9\columnwidth}}
    \toprule[1pt]
    \textbf{\S{}\ref{sec: objects}} & \textbf{\GroupObj}\\
    \textit{RQ1} & \textit{\RQContent{1}} \\
    \textit{RQ2} & \textit{\RQContent{2}} \\
    \textit{RQ3} & \textit{\RQContent{3}} \\
    
    \cmidrule(lr){1-2}
    \textbf{\S{}\ref{sec: process}} & \textbf{\GroupTest}      \\ 
    \textit{RQ4} & \textit{\RQContent{4}} \\
    \textit{RQ5} & \textit{\RQContent{5}} \\

    \cmidrule(lr){1-2}
    \textbf{\S{}\ref{sec: evaluation}} &  \textbf{\GroupEval}      \\ 
    \textit{RQ6} & \textit{\RQContent{6}} \\
    \textit{RQ7} & \textit{\RQContent{7}} \\
    
    \cmidrule(lr){1-2}
    \textbf{\S{}\ref{sec: settings}} &  \textbf{\GroupExp}      \\ 
    \textit{RQ8} & \textit{\RQContent{8}} \\
    \textit{RQ9} & \textit{\RQContent{9}} \\
    \textit{RQ10} & \textit{\RQContent{10}} \\
    \bottomrule[1pt]
\end{tabular}

    }
\end{table}

In our paper, we focus on empirical research on QST and examine primary studies from a methodological perspective. The RQs are designed to analyze how empirical studies in QST are structured and reported, rather than to evaluate the effectiveness of specific testing techniques. 
From the perspective of systematic study design and evidential rigor, we propose ten RQs, as summarized in Table~\ref{tab: rq_list}.
\revise{Some RQs are inspired by common issues in experiments on \textbf{Quantum Software Engineering (QSE)} according to a recent survey~\cite{zhang2025empirical}, such as available tools promoted by SE and execution backends largely specific to QC.}
\revise{Also, we conducted an initial literature analysis to identify QST-specific RQs, such as RQ4 and RQ5, that may require exploring concepts from quantum information theory and QC techniques for the design and analysis of test inputs and outputs.}

\revise{More specifically, }RQ1--RQ3 investigate how tested quantum programs are selected and characterized in empirical QST studies, including the types of quantum programs considered, the presence of faults or bugs in fault-detection scenarios, and program scale as reflected by circuit complexity. RQ4 and RQ5 examine how testing processes are designed in empirical evaluations, with particular attention to the construction of test inputs and the handling of the test oracle problem. RQ6 and RQ7 focus on how proposed testing approaches are evaluated and compared in empirical studies, including the choice of evaluation metrics and baseline techniques. Finally, RQ8--RQ10 address experimental resources and infrastructure, covering experimental configurations, execution backends, and tool support commonly adopted in the \revise{QSE} community.

\subsection{Paper Collection}

\begin{figure}
    \centering
    \includegraphics[width=0.9\textwidth]{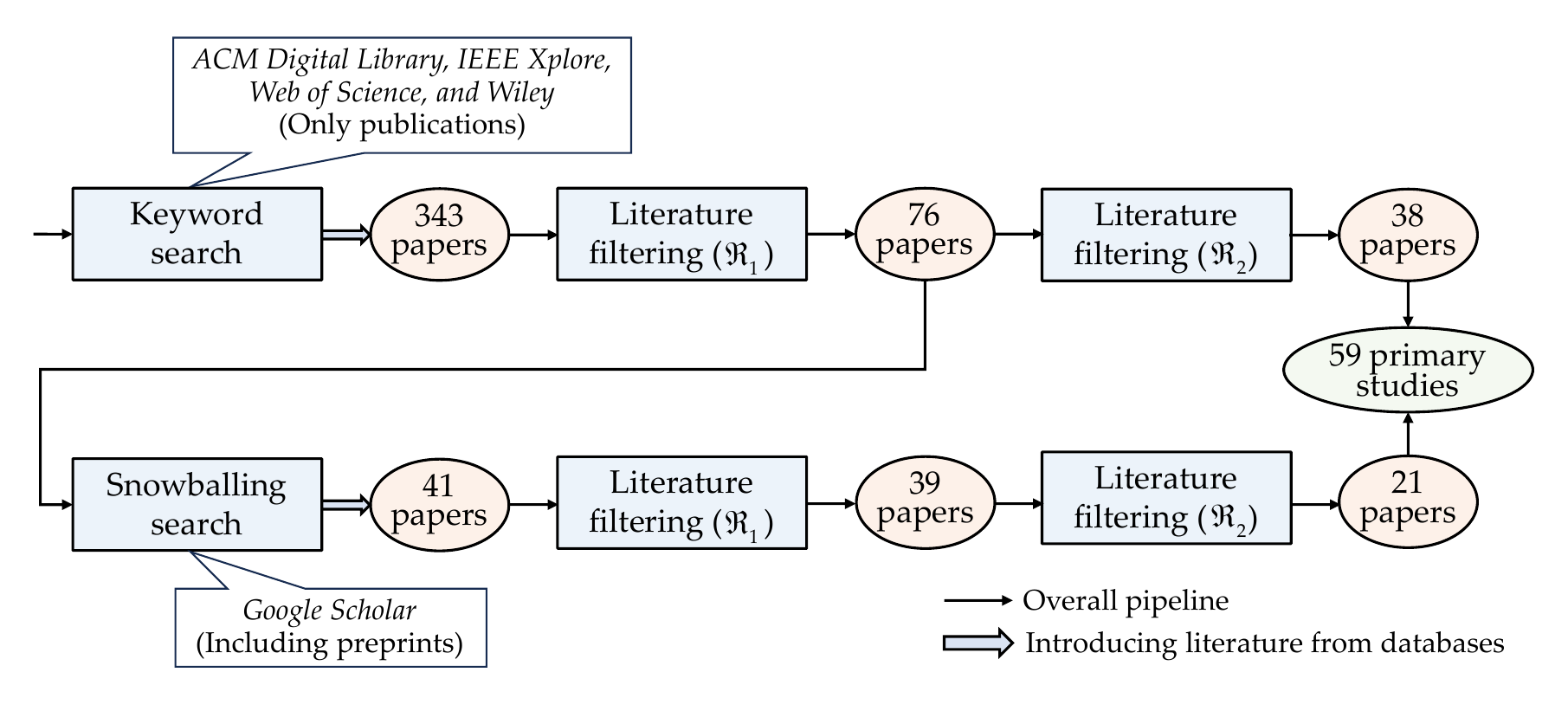}
    \caption{\revise{Pipeline of paper collection}}
    \Description{A figure displaying the pipeline of collecting primary studies.}
    \label{fig: pipeline}
\end{figure}

To support a comprehensive and reproducible methodological analysis, we adopt a two-stage literature collection process consisting of a keyword-based search followed by snowballing. \revise{Figure~\ref{fig: pipeline} visualizes the pipeline for collecting primary studies, illustrating the number of papers collected and retained in each phase.}

\subsubsection{Inclusion Criteria}

\begin{table}[!t]
    \small
    \centering
    \caption{Passing criteria for two rounds of literature filtering}
    \label{tab: criteria}
    \resizebox{.98\columnwidth}{!}{
        \begin{tabular}{c  p{0.85\columnwidth}}
    \toprule[1pt]
    \multicolumn{1}{c}{\textbf{Rounds}} & \multicolumn{1}{c}{\textbf{Criteria}} \\
    \cmidrule(lr){1-1} \cmidrule(lr){2-2}
    \SearchRoundOne, \SearchRoundTwo &  The papers must be written in English with complete and accessible texts.    \\
    \SearchRoundOne, \SearchRoundTwo & The papers must be technically research studies other than book chapters, surveys, talks, tutorials, and future ideas.   \\
    \SearchRoundOne, \SearchRoundTwo & The papers must be preprints, early-access publications, or formal publications available between 2015 and 2025. \\
    \SearchRoundOne & The papers must demonstrate a focus relevant to software testing for quantum systems. \\
    \SearchRoundTwo & The papers must concentrate on issues strongly associated with QST, i.e., dynamically testing quantum software, programs, or circuits. \\
    \SearchRoundTwo & The papers must incorporate empirical studies for involved testing approaches, where small-scale case studies are still acceptable.  \\
    \bottomrule[1pt]
\end{tabular}
 
    }
\end{table}

Based on the defined research scope, we apply two rounds of literature filtering, denoted as \SearchRoundOne and \SearchRoundTwo, corresponding to the keyword search and snowballing stages, respectively. Papers retained after \SearchRoundTwo constitute the primary studies analyzed in this work. Table~\ref{tab: criteria} summarizes the criteria used for each round.

To capture recent developments in QST, both rounds retain papers published within the past decade up to 2025. \SearchRoundOne adopts a broader scope and includes studies related to software testing for quantum systems, providing a sufficiently large literature pool for subsequent snowballing. \SearchRoundTwo applies stricter criteria consistent with the focus of our study, retaining only empirical studies on QST. In particular, studies focusing on testing quantum hardware or quantum software stacks (e.g., compilers and simulators) are excluded. Moreover, studies without dynamic execution of quantum programs, such as those based on bug reports, static analysis, or formal verification and validation, are filtered out in \SearchRoundTwo.

\subsubsection{Keyword Search}

We conducted the keyword search over four widely used digital libraries: \textit{ACM Digital Library}, \textit{IEEE Xplore}, \textit{Web of Science}, and \textit{Wiley}. The search was performed on~\KeywordEndTime, covering publications from 2015-01-01 \revise{(one year just before the launch of IBM Quantum Experience service\footnote{https://web.archive.org/web/20160504214945/http://www-03.ibm.com/press/us/en/pressrelease/49661.wss})} to that date, in accordance with the temporal constraints in Table~\ref{tab: criteria}.
To ensure both efficiency and relevance, the keyword search targets abstracts using the following query:
\textit{("Abstract":"quantum") AND ("Abstract":"software" OR "Abstract":circuit* OR "Abstract":program*) AND ("Abstract":test*)}.
This query captures the core concepts of QST while accounting for semantically close terminologies such as software, programs, and circuits. For \textit{IEEE Xplore}, which provides high-level index terms, the same query is additionally applied to the \textit{"Index Terms"} field.

The identification of papers passing \SearchRoundOne and \SearchRoundTwo was conducted by multiple authors, with final decisions reached through discussion and consensus. Among the~\PaperNumKeywordTotal papers retrieved via keyword search,~\PaperNumKeywordInitial papers passed \SearchRoundOne, and~\PaperNumKeywordSecond papers remained after \SearchRoundTwo.

\subsubsection{Snowballing Search}
Following established guidelines for systematic literature studies~\cite{wohlin2014guidelines} and recent systematic literature reviews in \revise{SE}~\cite{marcen2024systematic, he2025llm}, we employed both backward and forward snowballing to identify additional relevant studies, including high-quality or high-impact preprints not indexed in the selected digital libraries. Forward snowballing identifies papers citing those retained after \SearchRoundOne using \textit{Google Scholar}, while backward snowballing examines the reference lists of these papers. \revise{Especially, the snowballing search, based on papers after~\SearchRoundOne rather than~\SearchRoundTwo, facilitates a more comprehensive collection of potentially relevant literature. For example, a primary study included in our final literature pool may be cited by another related work that, however, did not contain empirical studies and was therefore excluded after~\SearchRoundTwo.}

After completing the snowballing process up to~\SnowballingEndTime, we identified~\PaperNumSnowballingTotal additional candidate papers. Only the most recent preprint versions were considered. In total, the literature pool comprises~\SizeOfLiteraturePool(\PaperNumKeywordTotal+\PaperNumSnowballingTotal) papers. Among these,~\PaperNumSnowballingInitial and~\PaperNumSnowballingSecond papers passed \SearchRoundOne and \SearchRoundTwo, respectively. Ultimately, this study analyzes~\NumberOfPrimaryStudies(\PaperNumKeywordSecond+\PaperNumSnowballingSecond) primary studies that satisfy all inclusion criteria, \revise{i.e., passing both rounds}.

\subsection{Data Extraction}

All evidence used to answer the research questions is derived from systematically defined data extraction items. We first specified the information required for each RQ, such as the number and types of quantum programs under test for RQ1. Detailed definitions of extraction items are provided in our artifact repository~\cite{li_2026_18159893}. Multiple authors independently annotated the primary studies, and discrepancies were resolved through discussion and consensus.

The extracted metadata includes both original and coded data. Original data were retained when primary studies explicitly reported information that directly matched our extraction items, such as qubit counts for RQ3. When consolidation or categorization was required, we applied data coding to unify terminology and remove redundancy, for example, when grouping evaluation metrics for RQ6 or classifying test oracle types for RQ5. The full texts of primary studies served as the primary data source for all RQs except RQ10, which required additional inspection of publicly available repositories beyond published articles.





\section{Bibliometric Analysis}
\label{sec: bib}
\begin{figure}[!t]
    \centering
    \begin{subfigure}[b]{0.485\textwidth}
        \centering
        \includegraphics[width=0.8\textwidth]{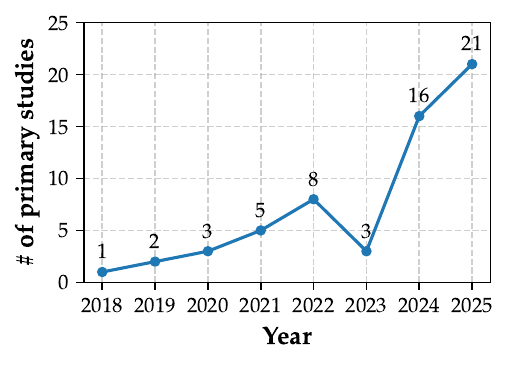}
        \caption{Temporal trend}  
        \label{fig: bib_year}
    \end{subfigure}
    \hfill
    \begin{subfigure}[b]{0.485\textwidth}
        \centering
        \includegraphics[width=0.6\textwidth]{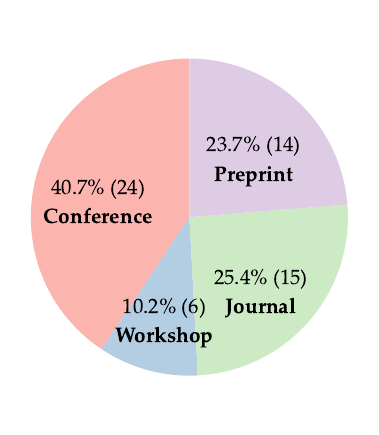}
        \caption{Venue types}  
        \label{fig: bib_venue_type}
    \end{subfigure}

    \begin{subfigure}[b]{0.485\textwidth}
        \centering
        \includegraphics[width=0.98\textwidth]{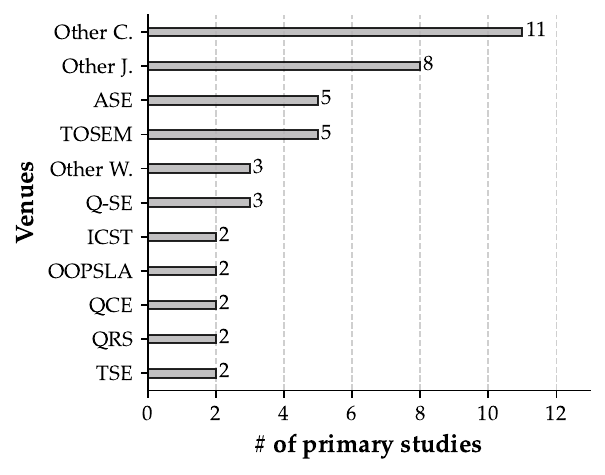}
        \caption{Publication venues}  
        \label{fig: bib_venue_name}
    \end{subfigure}
    \hfill
    \begin{subfigure}[b]{0.485\textwidth}
        \centering
        \includegraphics[width=0.62\textwidth]{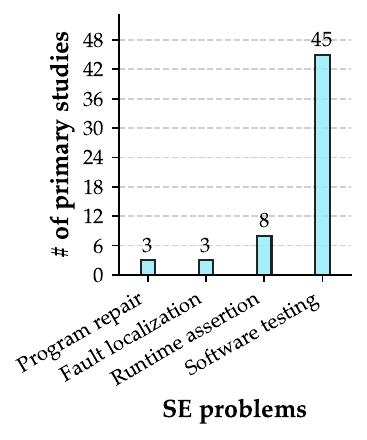}
        \caption{Fined-grained problems}  
        \label{fig: bib_se_problems}
    \end{subfigure}
    {\justify
        {\small\selectfont
            Full names of the publication venues, which are displayed in alphabetical order with the same frequency:
            \begin{itemize}[leftmargin=*] 
                \item Other C./J./W.: Corresponding conferences/journals/workshops with only one occurrence among the primary studies,
                \item ASE: \textit{International Conference on Automated Software Engineering},
                \item TOSEM: \textit{ACM Transactions on Software Engineering and Methodology},
                \item Q-SE: \textit{International Workshop on Quantum Software Engineering},
                \item ASPLOS: \textit{International Conference on Architectural Support for Programming Languages and Operating Systems},
                \item ICST: \textit{International Conference on Software Testing, Verification, and Validation},
                \item OOPSLA: \textit{International Conference on Object-Oriented Programming, Systems, Languages, and Applications},
                \item QCE: \textit{International Conference on Quantum Computing and Engineering},
                \item QRS: \textit{International Conference on Software Quality, Reliability and Security},
                \item TSE: \textit{IEEE Transactions on Software Engineering}.
            \end{itemize}
        }
    \par}
    \caption{Bibliometric analysis of the~\NumberOfPrimaryStudies primary studies}
    \Description{This composite figure provides a bibliometric overview of the \NumberOfPrimaryStudies primary studies, including their temporal distribution, venue types, and publication venues, as well as the fine-grained software engineering problems they address.}
    \label{fig:overall}
\end{figure}

Before addressing the proposed RQs, we conducted a bibliometric analysis of the collected literature. 
In this subsection, we aim to identify research trends of primary studies regarding the temporal trend, the published venues, and the detailed SE problems.
\revise{These three aspects help us understand the significance of empirical studies in QST, thereby motivating us to establish comprehensive guidelines and further foster community consensus through the formulation of ten RQs.}

\subsection{Temporal Trend}

Figure~\ref{fig: bib_year} visualizes the research trend over the years. The earliest study can be traced to the preprint version~\cite{\PaperOneHundredAndThree} available in 2018. 
Despite the decline observed in 2023, in general, empirical studies on QST have steadily gained increasing attention since 2018. 
Apart from the importance of ensuring software reliability and quality inherent in SE, the yearly uptrend of primary studies can also be attributed to the breakthrough of quantum computing hardware, the advancement of quantum software ecosystems, and the increase in engaged researchers from diverse communities.
\revise{At the same time, the relatively recent emergence of this trend suggests that empirical research in QST is still in an early stage of development. As a result, there is a need for a systematic investigation to better understand existing research practices and identify potential gaps, which motivates us to present this study.}

\subsection{Publication Venues}
 According to Figure~\ref{fig: bib_venue_type}, we observed that the conference was the most prevailing venue for publication owing to its fast impact. In addition, workshop publications account for around ten percent of the primary studies, which can be interpreted by the fact that several workshops are specifically dedicated to QSE, like~\cite{\PaperTwentyFour, \PaperThirtyThree, \PaperFiftyOne} \textit{International Workshop on Quantum Software Engineering} (Q-SE) and~\cite{\PaperTwentySeven} \textit{International Workshop on Quantum Programming for Software Engineering} (QP4SE).
 
 Regarding the concrete venues shown in Figure~\ref{fig: bib_venue_name}, the publication venues manifest the diversity, as there are~\BibVenueOthers (10+8+3) primary studies published at completely different conferences, journals, or workshops. \highlight{The venues across communities of software engineering, quantum computing, programming language, etc., can further support the interdisciplinary nature of QST. Besides, we found that a moderate number of studies were published at commonly acknowledged top-tier journals and conferences, such as~\cite{\BibTSE} at TSE,~\cite{\BibTOSEM} at TOSEM, and~\cite{\BibASE} at ASE for the SE community, indicating that the topic of our paper has begun to gain recognition within highly influential venues and gradually establish its academic significance.}
 
\revise{Despite influential and top-tier venues, their diversity suggests that research practices may vary across communities, potentially leading to inconsistencies in empirical methodologies and peer-review standards. This fragmentation highlights the need for a structured analysis to consolidate existing knowledge and systematically examine empirical practices for QST research, which we address through the proposed RQs and subsequent meta-analysis.}

\subsection{\revise{Fine-grained Testing-related SE Tasks}}
\revise{As a key activity in the quantum software development life cycle, QST provides execution-based evidence that can support subsequent quality assurance tasks, such as diagnosing failures and fixing defects.}
\revise{Considering a broader scope of testing-related tasks in SE and their shared empirical study practices, we further identify several fine-grained tasks that are strongly connected to QST and commonly evaluated through runtime execution.}

\begin{EnvRevise}
Based on an initial literature analysis and author discussion, we consider the following four fine-grained tasks that are closely related to software testing:
\begin{itemize}[leftmargin=*]
    \item \textit{Software testing}: Execute a quantum program under designed test cases to assess whether the program satisfies expected behaviors or specified properties. It provides primary evidence for correctness and defect detection.
    \item \textit{Runtime assertion}: Specify and check properties or conditions during program execution, producing explicit violation signals or diagnostic information. Such checks can serve as lightweight oracles and help interpret failures.
    \item \textit{Fault localization}: Identify likely root causes of observed failures by analyzing execution evidence (e.g., failing tests, assertion violations, and related execution context) to pinpoint program elements responsible for defects.
    \item \textit{Program repair}: Automatically or semi-automatically generate program modifications to fix defects, typically validated against a test suite (and related checks) to avoid regressions.
\end{itemize}

These tasks are closely connected and often studied together in execution-based empirical settings. Their empirical evaluations rely on running quantum programs on test cases and checking expected behaviors or properties, which makes it reasonable to discuss them within the scope of our analysis.
\end{EnvRevise}

\revise{Illustrated in Figure~\ref{fig: bib_se_problems}, beyond primary studies that focus on \revise{testing quantum programs (i.e., software testing)}, we also identified 14 out of the~\NumberOfPrimaryStudies primary studies that address program repair (3), fault localization (3), or runtime assertion (8). Including these studies is reasonable because the empirical evaluation of these tasks still relies on execution-based testing practices, such as running quantum programs with test cases and checking expected behaviors or properties. In addition, key empirical study practices used in software testing (e.g., test design, oracle handling, and experimental configuration) are largely shared by these three tasks. For example, primary studies on program repair~\cite{\PaperFortyFour, \PaperFiftyOne, \PaperEightyThree} not only generate patches to fix defects but also validate patched programs through testing. Overall, analyzing these studies can help inform and encourage future empirical work on these testing-related tasks, which have received limited attention in the current QST literature.}

\section{\texorpdfstring{\titlecap{\GroupObj}}{Programs Under Test}}
\label{sec: objects}

\subsection{\texorpdfstring{\titlecap{programs under test versus circuits under test}}{PUT vs. CUT}}

For the preliminary design of empirical studies on QST, we need to specify which quantum program is selected as \textbf{Programs Under Test (PUTs)}. The PUTs should incorporate quantum subroutines, and the quantum subroutines can be modeled by quantum circuits. Herein, we can call the quantum circuits involved in PUTs as \textbf{Circuits Under Test (CUTs)}.

In detail, a CUT for \textbf{universal} QST can be formally represented by a quantum channel $\mathcal{E}_{c}\in \mathcal{L}(\mathcal{L}(\mathcal{H}))$, which is determined by a pair of the PUT and the possible classical input $c$ specified by a test case $t$. We prefer the representation of a quantum channel $\mathcal{E}_c$ instead of a unitary operator $U_c$, since special quantum programs like Quantum Teleportation involve intermediate quantum measurement during the qubit evolution, which mathematically breaks down the unitarity of the quantum operation.

\begin{figure}
    \centering
    \includegraphics[width=0.75\textwidth]{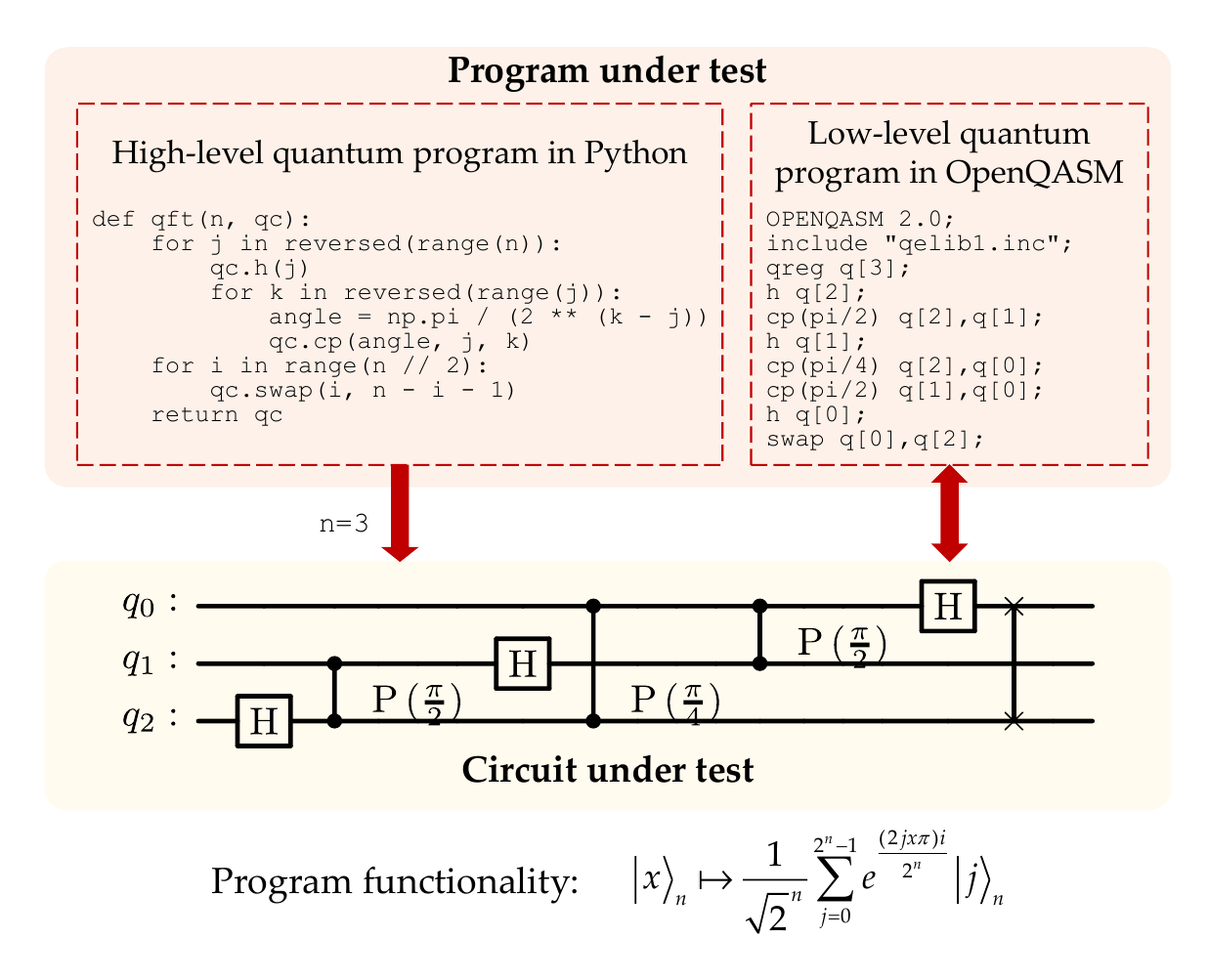}
    \caption{Example of PUTs and the corresponding CUT for a 3-qubit Quantum Fourier Transform}
    \Description{Diagram illustrating the relationship between PUTs and the corresponding CUT in a 3-qubit Quantum Fourier Transform circuit.}
    \label{fig: PUT_versus_CUT}
\end{figure}

\revise{The two concepts, PUT and CUT, help distinguish among quantum programs with different code abstractions, and outline the underlying gap between code structures and abstract models.}
Figure~\ref{fig: PUT_versus_CUT} presents an example of a 3-qubit \revise{Quantum Fourier Transform (QFT)}, where the displayed two PUTs are written in different programming languages but can be mapped to the same quantum circuits. 
More particularly, we classify the two PUTs into a \textit{low-level quantum program} and a \textit{high-level quantum program} according to their degree of code abstraction.
A high-level quantum program specifies quantum algorithms or subroutines using abstract constructs and may incorporate classical control logic, whereas a low-level quantum program explicitly describes quantum circuits at the gate level, exposing qubits and quantum operations.
Note that the above-mentioned level is not determined by the programming language, as code written in a high-level language may still exhibit a low level of abstraction. For example, the following two code snippets manifest the same level of code abstraction: ``\texttt{qc.h(2)}'' in Python and ``\texttt{h q[2]}'' in OpenQASM. Back to the figure, the left PUT is a high-level quantum program written in Python, which is scalable for execution and readable for developers. 
\highlight{The quantum subroutine of this PUT is associated with the argument \texttt{n} that indicates the number of input qubits. Valid arguments yield a family of potential quantum circuits (i.e., CUTs), and only the instance \texttt{n=3} results in the CUT shown at the bottom of this figure. By contrast, the PUT in the right is a low-level quantum program written in OpenQASM and commonly serves as an intermediate representation targeting quantum hardware. Mostly, there is a one-to-one correspondence between a low-level quantum program and its quantum circuit. This is because OpenQASM programs do not rely on external arguments, and their code models the behavior of a fixed circuit, which explains why we do not consider the classical inputs $c$ in test cases for OpenQASM PUTs.}

\subsection{\texorpdfstring{\RQSection{1}}{RQ1}}\label{sec: rq1}

Regarding the selection of PUTs, each of the employed quantum programs should correspond to a specific and unique functionality that manifests distinctions from the others to ensure the test diversity. The requirement of sufficiency also expects a reasonable number of quantum programs used to evaluate the generalization of the proposed test approach. In this section, we zoom in on the type and quantity of the quantum programs used in the primary studies. By this means, the currently preferred configuration of quantum programs can be captured as a guideline for future studies.

\subsubsection{Quantum Algorithms and Subroutines}
\label{sec: rq1_algorithms}

Testing quantum algorithms and subroutines plays a part in current studies on QST. \revise{A quantum algorithm may comprise multiple quantum subroutines, each of which can be encapsulated as an abstract model that implements a specific subtask, such as state preparation. We group the two together, because of their analogy to the \textit{traditional software} widely acknowledged in \textbf{Classical Software Engineering (CSE)}~\cite{pei2017deepxplore, du2019deepstellar}. The program logic of quantum algorithms and subroutines can be directly specified by developers instead of being driven by data.} The latent bugs to be disclosed in such PUTs, for instance, can be attributed to the incorrect statements in the view of code or the corresponding undesired quantum gates from the perspective of quantum circuits.

\begin{figure}
    \centering
    \includegraphics[width=0.97\textwidth]{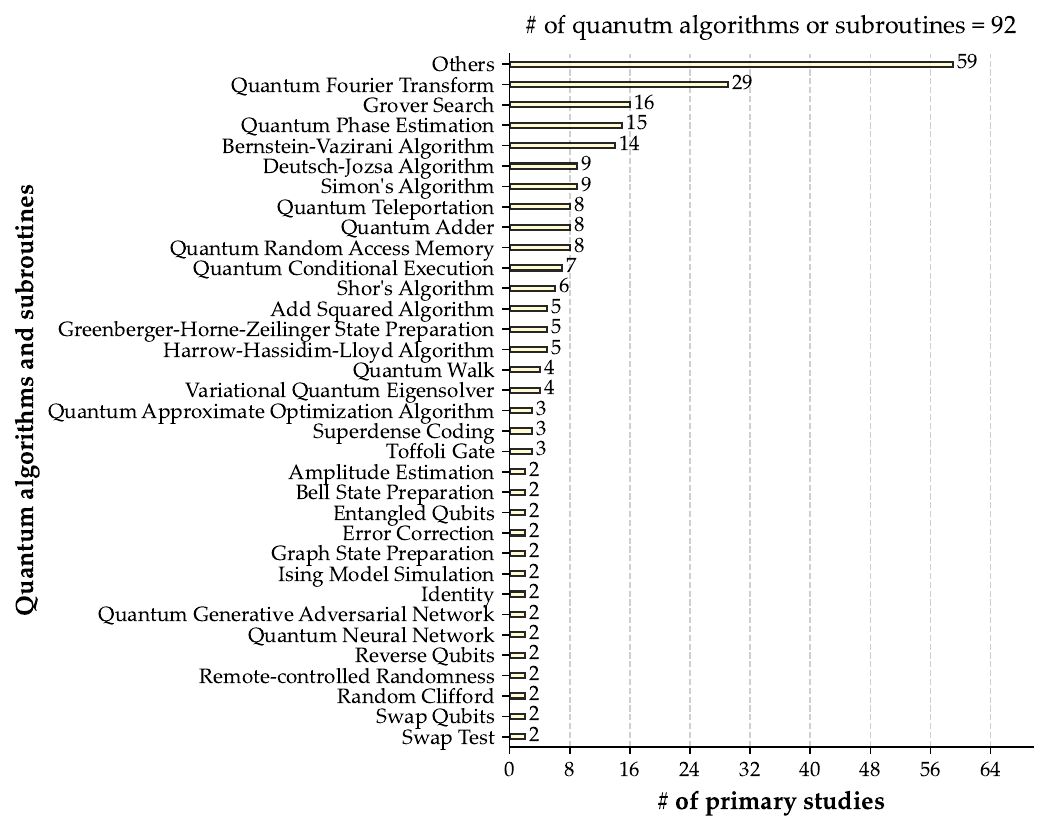}
    \caption{Quantum algorithms and subroutines used in the primary studies}
    \Description{A figure displaying the quantum algorithms and subroutines along with their frequencies in the primary studies.}
    \label{fig: rq1_program_names}
\end{figure}

\revise{Figure~\ref{fig: rq1_program_names} displays the quantum algorithm and subroutines employed by primary studies. Our counting was merely based on the algorithm or subroutine type per study, such as PUTs that differ in circuit scale but implement the same algorithmic functionality were treated as a single type. From this figure,} we observed that there were 92 different quantum algorithms or subroutines used in the primary studies, indicating rather diverse program sources for testing. To reduce the number of candidate program types, we reasonably merged closely related variants into their more general, strongly relevant prototypes, such as the Quantum Fourier Transform, subsuming its inverse implementation. The ``Others'' in this figure, incorporating 59 samples, refers to the collection of significantly different programs that appeared only once in the primary studies. 

In terms of the quantity, QFT was employed in 29 papers, showing its dominance over the other types of quantum programs. This is because QFT not only serves as a key subroutine in famous quantum algorithms, like Shor's Algorithm and Quantum Phase Estimation, but also manifests the quantum-specific characteristics in the output state, such as quantum superposition and relative phases. The three programs following QFT are based on typical quantum algorithms as well, i.e., Grover Search, Quantum Phase Estimation, and Bernstein-Vazirani Algorithm. \revise{The broad inclusion of the three algorithms is likely tied to their usage for solving particular computational problems and the potential to demonstrate quantum advantages.} For example, Grover Search offers a quadratic speedup for searching an unstructured database in comparison to its classical counterpart. \revise{Meanwhile, such algorithms are usually used in canonical textbooks for the pedagogical purpose of QC, but some struggle to reflect the complex real-world quantum software systems applicable in the NISQ era, like the Bernstein-Vazirani Algorithm, which is merely designed for an artificial mathematical problem.
In the context of benchmark selection, this implies an underlying bias between evaluation in academic research and application in industrial showcases. The resultant threat to the external validity of empirical studies cannot be effectively mitigated by blindly increasing the number of programs.}

Besides, we discovered that some primary studies accounted for quantum subroutines just for the preparation of exemplary quantum states, including the cat state, Bell state, and graph state. Apart from those, two ans\"atze for quantum machine learning, i.e., Quantum Generative Adversarial Network and Quantum Neural Network, were tested similarly to other quantum algorithms and subroutines. 
However, these studies examined whether the fixed ans\"atze obey the assumed expected states instead of testing the performance of models trained by the corresponding quantum algorithms. 

\subsubsection{Quantum Machine Learning Models}
\label{sec: quantum learning-based models}
As the quantum counterpart of \textit{intelligent software} proposed in CSE, quantum machine learning models are mainly built on specific quantum algorithms and commonly rely on training data to determine the parameters of ans\"atze. Like CSE, testing such learning-based models typically targets identifying the functional faults dependent on the training phase, rather than the implementation-level bugs in the underlying code. Differently, testing quantum machine learning models would be impacted by the quantum measurement for the decision on expectation values
and the quantum noise involved in the execution backend.

\begin{table}[!t]
    \small
    \centering
    \caption{Quantum machine learning models involved in the primary studies}
    \label{tab: rq1_learning_models}
    \resizebox{.98\textwidth}{!}{
        
\begin{tabular}{p{0.6\columnwidth}  p{0.3\columnwidth} c}
    \toprule[1pt]
    \multicolumn{1}{c}{\textbf{Quantum machine learning models}} & \multicolumn{1}{c}{\textbf{Primary studies}} & \multicolumn{1}{c}{\textbf{\#}} \\
    \cmidrule(lr){1-1} \cmidrule(lr){2-2} \cmidrule(lr){3-3}  
    Quantum Convolutional Neural Network & \cite{\PaperThirtySix ,\PaperOneHundredAndSeven ,\PaperOneHundredAndTwentyThree} & 3 \\ 
    Quantum Circuit Learning & \cite{\PaperThirtySix ,\PaperOneHundredAndSeven} & 2 \\ 
    Circuit-centric Quantum Classfiers & \cite{\PaperThirtySix} & 1 \\ 
    Quantum Distributed Learning Software & \cite{\PaperEightyOne} & 1 \\ 
    Circuit-centric Quantum Classfier & \cite{\PaperOneHundredAndSeven} & 1 \\ 
    Hierarchical Circuit Quantum Classifier & \cite{\PaperOneHundredAndSeven} & 1 \\ 
    Hybrid Quantum Neural Network & \cite{\PaperOneHundredAndSeven} & 1 \\
    \bottomrule[1pt]
\end{tabular}
        
    }
\end{table}

Due to the early stage of testing quantum machine learning models, there are only six models clarified in four primary studies~\cite{\PaperThirtySix, \PaperEightyOne, \PaperOneHundredAndSeven, \PaperOneHundredAndTwentyThree}. From Table~\ref{tab: rq1_learning_models}, we found that three out of the four studies took Quantum Convolutional Neural Networks~\cite{cong2019quantum} as PUTs, which are motivated by classical Convolutional Neural Networks and play a part in addressing the barren plateau problem in quantum machine learning~\cite{mcclean2018barren}. Besides, many of the tested quantum machine learning models, such as Circuit-centric Quantum Classifier and Hierarchical Circuit Quantum Classifier, are particularly suited for classification tasks. This finding for the current state of testing quantum machine learning models is still convincing, as a large number of studies on machine learning testing in CSE also focus on classification compared to other tasks like regression and clustering~\cite{zhang2020machine}. 

\revise{Generally speaking, observations from existing studies on testing quantum machine learning models highlight their scarcity, along with a research gap in intelligent software testing in CSE. Nevertheless, this line of research is promising and deserves further attention, as these models are typically built on hybrid classical–quantum approaches well suited to applications in the current NISQ era. This further underscores the underlying practical value of relevant QST research.}

\subsubsection{Quantity Analysis for Employed Quantum Programs}
\label{sec: rq1_learning}

In empirical SE research, \textit{external validity}, which reflects the generalizability of results beyond the studied context, has received considerable attention in the community~\cite{lago2024threats, wright2010validity}. Regarding QST, the quantity of quantum programs under test is crucial for the reliability of the evidence presented in empirical studies, which supports the claimed generalizability of the proposed test approaches. Motivated by this, we collected the program counts configured in the primary studies.

\begin{figure}[!t]
    \centering
    \begin{subfigure}[b]{0.485\textwidth}
        \centering
        \includegraphics[width=\textwidth]{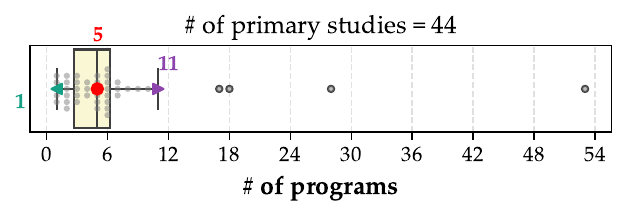}
        \caption{Quantum algorithms and subroutines}  
        \label{fig: rq1_object_number_algorithms}
    \end{subfigure}
    \hfill
    \begin{subfigure}[b]{0.485\textwidth}
        \centering
        \includegraphics[width=\textwidth]{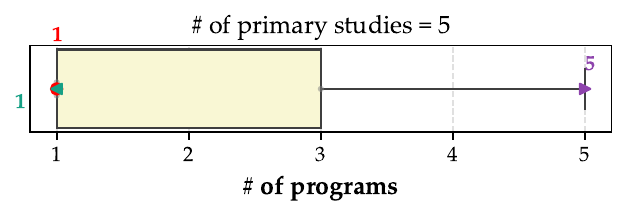}
        \caption{Quantum machine learning models}  
        \label{fig: rq1_object_number_learning}
    \end{subfigure}

    \begin{subfigure}[b]{0.49\textwidth}
        \centering
        \includegraphics[width=\textwidth]{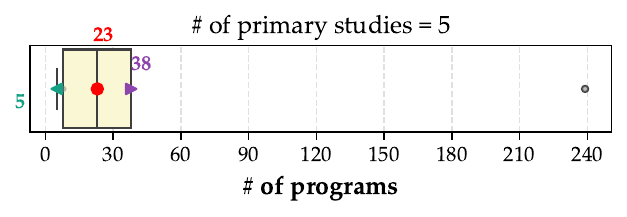}
        \caption{Real-world benchmarks}  
        \label{fig: rq1_object_number_real-world}
    \end{subfigure}
    \hfill
    \begin{subfigure}[b]{0.49\textwidth}
        \centering
        \includegraphics[width=\textwidth]{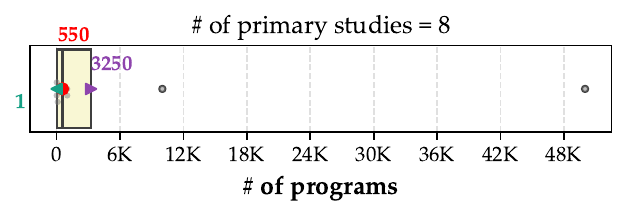}
        \caption{Artificial programs}  
        \label{fig: rq1_object_number_artificial}
    \end{subfigure}

    \vspace{0.3em}
    \begin{subfigure}[b]{0.8\textwidth}
        \centering
        \includegraphics[width=0.6\textwidth]{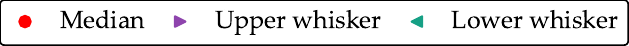}
    \end{subfigure}
    \caption{Boxplots for the numbers of adopted quantum programs}  
    \label{fig: rq1_object_number}
    \Description{Four boxplots illustrating the numbers of programs in the context of quantum algorithms and subroutines, quantum machine learning models, real-world benchmarks, and artificial programs.}
\end{figure}

Observed from Figure~\ref{fig: rq1_object_number}, we categorize the quantum programs into four types to draw an unbiased and accurate conclusion. In addition to quantum algorithms and subroutines as well as quantum machine learning models discussed in Sections~\ref{sec: rq1_algorithms} and~\ref{sec: rq1_learning}, \textit{real-world benchmarks} refer to the PUTs sourced from available benchmarks collecting real-world programs, while \textit{artificial programs} indicate the ones whose quantum circuits are generated by the primary studies but not guaranteed to yield practical functionalities. 
 
According to the boxplots, the median number of \revise{quantum algorithms and subroutines adopted in a single primary study} is 5, while the upper whisker excluding outliers is 11. By contrast, the median number and the upper whisker for quantum machine learning models turn out to be lower. 
The concern of cost is validated in the context of artificial programs, where these programs are easily synthesized through random combinations of basic quantum gates. Their median number and upper whisker increase to 550 and 3250, respectively, and the study~\cite{\PaperSixtySeven} even produced $5\times10^{4}$ random instances. 

Additionally, the limited pool of available programs would restrict the real-world benchmarks for QST. Only five studies~\cite{\PaperFortyFour, \PaperFiftyOne, \PaperEightyThree, \PaperEightyFour, \PaperOneHundredAndNine} on fault localization and program repair mentioned the number of instances selected from real-world benchmarks, and the employed programs in four of the five studies were not limited to the benchmarks. 
The scarcity of such primary studies, especially those with a concentration only on software testing, may result from the fact that current real-world benchmarks are not created particularly for testing quantum programs, where many of the included bugs are traced to simulators and compilers in the quantum software stacks. Bugs4Q, which was used in all the aforementioned five studies and whose repository remained available throughout our literature review, includes only 42 real-world bugs in its latest release~\cite{zhao2023bugs4q}. 
\revise{In this latest Bugs4Q and its publication, 17 bugs were identified as throwing exceptions and 16 as producing incorrect outputs, which are the two most common categories, while only 8 quantum programs were validated through empirical studies to be compatible with prior QST approaches.}
This limited scale could indirectly explain why the upper whisker (i.e., 38) lies just below 42, and the only outlier is due to one primary study~\cite{tan2025hornbro} utilizing more than one benchmark. 
\revise{Generally, the limited use of real-world benchmarks that align with the research problem may threaten the external validity of proposed approaches, as there is insufficient convincing evidence to support their applicability to complex, real-world scenarios. Artificial programs, however, can neither exactly represent the quantum programs developed in the real world nor effectively capture the bug types that developers are prone to introducing.}

\subsubsection{Summary of RQ1}
\text{}

\Subtitle{Takeaway 1-1} The employed quantum algorithms and subroutines manifest significant diversity up to 92 types. The top 4 frequently used were Quantum Fourier Transform, Grover Search, Quantum Phase Estimation, and Bernstein-Vazirani Algorithm, listed in descending order.

\Subtitle{Takeaway 1-2} There are scarce primary studies on testing quantum machine learning models, where only 6 models were considered, and the Quantum Convolutional Neural Network was the most employed.

\Subtitle{Takeaway 1-3} The number of opted quantum programs significantly varies among the four program types (i.e., quantum algorithms and subroutines, quantum machine learning models, real-world benchmarks, and artificial programs). We underline that the referable median numbers of programs for quantum algorithms and subroutines and artificial programs are 5 and 550, respectively. Moreover, only a small proportion of primary studies (i.e., 5) acquired PUTs from real-world benchmarks.


\subsection{\texorpdfstring{\RQSection{2}}{RQ2}}\label{sec: rq2}

\begin{figure}[!t]
    \centering
    \begin{subfigure}[b]{0.49\textwidth}
        \centering
        \includegraphics[width=0.72\textwidth]{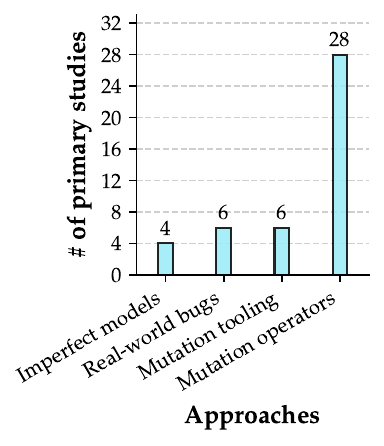}
        \caption{Approaches for generating buggy variants}  
        \label{rq2_distribution_of_generation_approaches}
    \end{subfigure}
    \hfill
    \begin{subfigure}[b]{0.49\textwidth}
        \centering
        \includegraphics[width=0.72\textwidth]{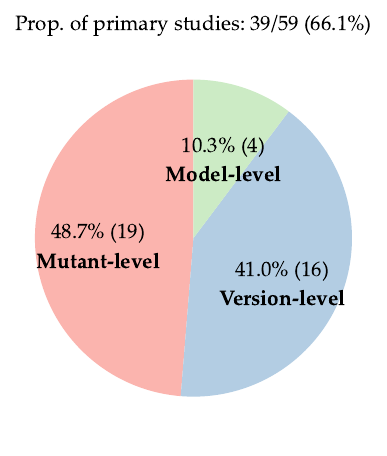}
        \caption{Different levels of buggy variants}  
        \label{rq2_variant_distribution}
    \end{subfigure}
    \caption{Statistics related to the buggy variants for fault detection}  
    \Description{A figure including two subfigures, where the left one is a bar chart of generation approaches for buggy variants, while the right one offers a pie chart for buggy variants with different levels in empirical studies.}
    \label{fig: rq2}
\end{figure}



Fault detection is one purpose of QST, which aims to trigger the program failure through executing the test suites. 
For controlled experiments on evaluating the fault-detection capability of test approaches, the programs actually being tested should contain controllable and practical bugs, trying to imitate real-world cases in quantum software development. To be clarified, we entitle such a program as one \textit{buggy variant} of the original PUT investigated in Section~\ref{sec: rq1}, where an original PUT and its buggy variant correspond to the same program functionality. 
For a comprehensive evaluation, the construction of buggy variants should consider diverse pre-defined bug types and various locations where these bugs are implanted.
Numerous techniques have been widely adopted in \textbf{\highlight{Classical Software Testing (CST)}}, like applying mutation operators and reusing real-world bugs, and they still have the potential for application to QST. 
Motivated by this, RQ2 revisits primary studies to investigate the approaches for generating buggy variants and varying levels of buggy variants for evaluation. In the following part, we will first overview the generation approaches and the variant levels. Then, specific techniques for each generation approach and the experiment scales in the context of the defined variant levels will be discussed in order.


\subsubsection{Overview of Generation Approaches and Variant Levels}

In Figure~\ref{rq2_distribution_of_generation_approaches}, we only investigated primary studies that focused on fault detection and clarified their approaches for introducing buggy variants. 
One primary study could adopt more than one generation approach, such as~\cite{\PaperEightyThree} that took both mutation operators and real-world bugs into account.
In detail, the four kinds of approaches for generating buggy variants are defined as follows:
\begin{itemize}[leftmargin=*]
    \item \textit{Mutation operators}: Denote the operators used in mutation testing and analysis, which are designed or introduced in primary studies to imitate the bugs in real-world scenarios as well as match the PUTs written in specific programming languages or exhibiting particular structures.
    \revise{For example, creating the mutation operator of replacing a quantum gate can mimic the real-world bug pattern ``incorrect gate'' that is a subclass of math-related bugs~\cite{luo2022comprehensive}.}
    \item \textit{Mutation tooling}: Indicates accessible and integrated software tooling that provides a pool of candidate mutation operators and even automatically generates buggy variants based on these defined operators. Compared to \textit{mutation operators}, the operators provided by \textit{mutation tooling} do not guarantee evolutionary extension of diversity over time, nor are they guaranteed to be applicable to generic PUTs.
    \item \textit{Real-world bugs}: Refer to real-world bugs or faulty programs collected from available repositories, programming platforms, or industrial software projects.
    \item \textit{Imperfect models}: Particularly mean the quantum machine learning models introduced in Section~\ref{sec: quantum learning-based models}, which exhibit imperfect performance after training, typically reflected as wrong predictions. 
\end{itemize}

Among these four approaches, we found that most studies (34 out of 42, where 34=28+6) adopted mutation-based generation of buggy variants, where the majority (i.e., 28 studies) considered particularly designed mutation operators. This is mainly because mutation testing and analysis are mature and popular techniques in SE to approximate real-world faults and evaluate the quality of test suites. The mechanism of implanting bugs applies subtle and syntax-preserving transformations to the original bug-free program while maintaining its overall structure. 
Currently, the lack of program benchmarks further amplifies the reliance on mutation-based generation and calls for future work on designing more diverse and realistic mutation operators.
Regarding the other two approaches that gained less attention than mutation-based ones, imperfect models were considered in 4 studies, whose scarcity is strongly related to their specificity to only quantum machine learning models.
Meanwhile, the consideration of real-world bugs depends on open-source resources. 
The corresponding studies with a limited number may reflect a shortage of publicly available resources suitable for general testing scenarios.


In view of the design of controlled experiments, we classified the buggy variants into three different levels, where they are defined as follows:
\begin{itemize}[leftmargin=*]
    \item \textit{Mutant-level variants}: Denote the \textit{mutants} generated and evaluated through mutation testing and analysis. In methodology, one mutant commonly corresponds to one application of a mutation operator \revise{mostly in an automatic way}. The evaluation is frequently based on the mutation-specific metrics (e.g., mutation score) associated with the number of mutants being killed by a given test suite, i.e., the existence of faults being triggered.
    \item \textit{Version-level variants}: 
    \revise{Point to program versions including one or more bugs that have been preliminarily confirmed to alter program functionality\footnote{Compared to mutant-level variants, one may argue that automatic implementation of a mutation operator cannot be ensured to implant a bug in the mutant, given the possibility of equivalent mutants that demonstrate an identical functionality to the original program.}.  
    \textit{Version-level variants} can be obtained via manual seeding with mutation operators or extracted from real version histories.}
    Besides, the evaluation can be more finely grained \revise{than mutant-level variants}, such as measuring how many test cases can trigger the faults.  
    \item  \textit{Model-level variants}: Incorporate the \textit{imperfect models} discussed above. In comparison to the other two levels, variants at this level are analyzed at a much coarser granularity and beyond the program code, with an emphasis on deficiencies in overall model behavior.
\end{itemize}

Figure~\ref{rq2_variant_distribution} shows the distribution of buggy variants in terms of our defined three levels over 39 primary studies. Nearly half (i.e., 48.7\%) of the involved studies investigated mutant-level variants, indicating the popularity of mutation testing and analysis in current QST empirical studies. Meanwhile, there are still 41.0\% of the involved studies employing version-level variants. The comparable proportions of variants in the two levels suggest that both have become conventional choices in empirical studies on testing general quantum programs, except for the machine learning models, with neither exhibiting a clear dominance over the other.

\subsubsection{Composition of Mutation Operators}

\begin{figure}
    \centering
    \includegraphics[width=0.6\textwidth]{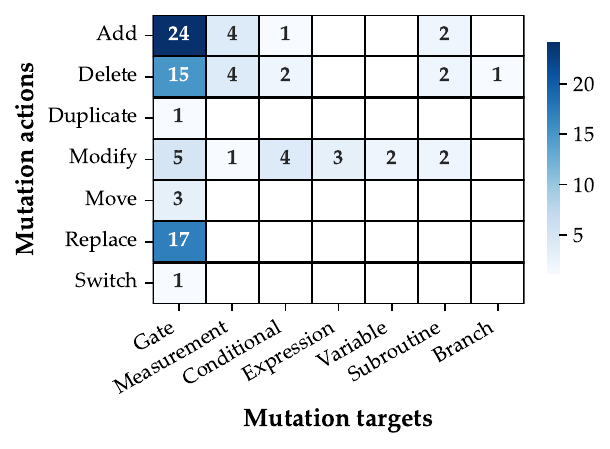}
    \caption{Combinations of mutation actions and mutation targets adopted in primary studies, where the numbers refer to the frequencies of corresponding mutation operators.}
    \label{fig: rq2_mutation_operators}
    \Description{A heatmap showing the frequency of combinations of mutation actions and mutation targets}
\end{figure}

In this part, we delve into specific techniques for mutation operators, owing to the popularity and importance of mutation testing and analysis. We decompose each mutation operator into a mutation action and its mutation target, such as ``delete'' for an action and ``gate'' for the target, given a mutation operator ``delete a quantum gate''. As detailed in Figure~\ref{fig: rq2_mutation_operators}, we identified seven mutation targets and seven mutation actions involved in the primary studies.

Regarding the diverse and fragmented distribution of potential mutation targets, we can classify them into three groups based on the involvement of elements in quantum and classical workflows, i.e., \textit{quantum mutation targets}, \textit{classical mutation targets}, and \textit{hybrid mutation targets}. For mutation targets involved in Figure~\ref{fig: rq2_mutation_operators}, (quantum) gates and measurements are categorized into quantum mutation targets, since they are rooted in the quantum circuit. Classical mutation targets include conditional, expression, and variable, as these are fundamental elements of classical workflows and do not exhibit strong relevance to quantum workflows. 
In primary studies, conditional refers to \texttt{if} statements whose predicates depend on classical variables, while expressions mainly indicate mathematical operators and constants.
Although variables in quantum programs should be quantum or classical in general, we categorize them as classical mutation targets, motivated by the sole primary study~\cite{\PaperSeventyOne} mentioning variables for mutation operation and specifying them as classical.
Subroutines and branches are grouped into hybrid mutation targets, and one reason could be that they represent highly abstracted structures in high-level quantum programs and thus have the potential to span both classical and quantum workflows. 


Regarding the mutation actions, the action names are reused from the primary studies. It is worth noting that the meaning of ``modify'' varies across studies, but this action consistently focuses on input arguments of interfaces\revise{.}
In comparison, the ``replace'' action substitutes a whole function with a syntactically equivalent and compatible one, but possibly makes semantic changes. For example, changing the code snippet ``\texttt{qc.rx(theta, q0)}'' into ``\texttt{qc.ry(theta, q0)}'' and into ``\texttt{qc.rx(theta, q1)}'' correspond to ``replace'' and ``modify'', respectively.

Figure~\ref{fig: rq2_mutation_operators} illustrates that quantum mutation targets demonstrate a heavy concentration, especially for quantum gates. 
This could be mainly because low-level quantum programs have been greatly employed in current QST empirical studies, but they almost exclude classical subroutines compared to high-level quantum programs, thereby making quantum mutation targets dominate the other two groups. 
Gates support a wider range of mutation actions than measurements while still ensuring that the resulting mutants do not crash, for example, adding an extra measurement operation that could result in a mismatch between the number of measured qubits and that of the allocated classical registers.
In turn, several combinations of targets and actions being currently absent may relate to the non-trivial effort required to protect the program logic without producing crashes. 
For example, compared to gates, the mutation operation, i.e., replacing a subroutine, depends on a careful design of detailed implementation and its interface (e.g., the number and type of arguments) to avoid incompatibilities and exceptions.
In another view, some combinations are still valid in theory, e.g., delete an expression as a potentially valid mutation operator but absent from primary studies, whereas they go beyond the low-level quantum programs that only focus on the quantum mutation targets.
The community would benefit from exploring high-level quantum programs with hybrid subroutines and proposing a more accurate and unified taxonomy of mutation actions and mutation targets.


\subsubsection{Specific Techniques with Outer Available Sources}
\begin{table}[!t]
    \small
    \centering
    \caption{Approaches and sources for generating variants, where we focus on the other three approaches except \textit{mutation operators}}
    \label{tab: rq2_sources_for_variants}
    \resizebox{.98\textwidth}{!}{
        
\begin{tabular}{>{\centering\arraybackslash}p{0.3\columnwidth}
>{\centering\arraybackslash}p{0.3\columnwidth}
p{0.32\columnwidth}
>{\centering\arraybackslash}p{0.05\columnwidth}}
    \toprule[1pt]
    \multicolumn{1}{c}{\textbf{Approaches}} & \multicolumn{1}{c}{\textbf{Sources}} & \multicolumn{1}{c}{\textbf{Primary studies}} & \multicolumn{1}{c}{\textbf{\#}} \\
    \cmidrule(lr){1-1} \cmidrule(lr){2-2} \cmidrule(lr){3-3} \cmidrule(lr){4-4}  
    Real-world bugs & Bugs4Q & \cite{\PaperFortyFour, \PaperFortyFive, \PaperFiftyOne, \PaperEightyThree, \PaperEightyFour, \PaperOneHundredAndNine} & 6 \\ 
     & Qbugs & \cite{\PaperFortyFour, \PaperEightyThree, \PaperEightyFour} & 3 \\ 
    \cmidrule(lr){1-1} \cmidrule(lr){2-2} \cmidrule(lr){3-3} \cmidrule(lr){4-4} 
    Imperfect models & MNIST & \cite{\PaperThirtySix, \PaperOneHundredAndSeven} & 2 \\ 
     & Fashion-MNIST & \cite{\PaperThirtySix, \PaperOneHundredAndSeven} & 2 \\ 
     & CIFAR10 & \cite{\PaperOneHundredAndSeven} & 1 \\ 
    \cmidrule(lr){1-1} \cmidrule(lr){2-2} \cmidrule(lr){3-3} \cmidrule(lr){4-4} 
    Mutation tooling & QMutPy & \cite{\PaperSix, \PaperNinetyFour, \PaperOneHundredAndNine, \PaperOneHundredAndNineteen} & 4 \\ 
     & Muskit & \cite{\PaperThirtyNine, \PaperOneHundredAndSixteen} & 2 \\
    \bottomrule[1pt]
\end{tabular}
        
    }
\end{table}


Compared to mutation operators, the techniques for all of real-world bugs, imperfect models, and mutation tooling share the dependency on the outer available sources, such as benchmarks and tools. 
Sources for generating variants of the mentioned three approaches are listed in Table \ref{tab: rq2_sources_for_variants}. The real-world bugs are sourced from two benchmarks, i.e., Bugs4Q~\cite{zhao2021bugs4q, zhao2023bugs4q} and Qbugs~\cite{campos2021qbugs}. 
For imperfect models, the three sources are datasets specific to image classification and used for model training and testing. These datasets have been widely adopted for testing deep neural networks in CSE. 
QMutpy~\cite{fortunato2022qmutpy} and Muskit~\cite{mendiluze2022muskit} are identified as automated and well-archived tools for mutation testing and analysis, where both are mainly gate-oriented and support common operators.
They also provide a general mutation testing workflow, covering operator definition, operator application (e.g., the number and locations of target gates), mutation score computation, etc.

However, compared to simply using manually defined mutation operators, the studies that explored the available outer sources remained relatively scarce. This may reflect the lack of publicly available and generally usable benchmarks and tools for real-world bugs and mutation tooling, respectively. In terms of imperfect models, although the candidate datasets for the classification tasks are abundant, the research on testing quantum machine learning models is still at an early stage, with a limited number of related primary studies.

\subsubsection{Number of Adopted Buggy Variants}
\begin{figure}[!t]
    \centering
    \begin{subfigure}[b]{0.485\textwidth}
        \centering
        \includegraphics[width=\textwidth]{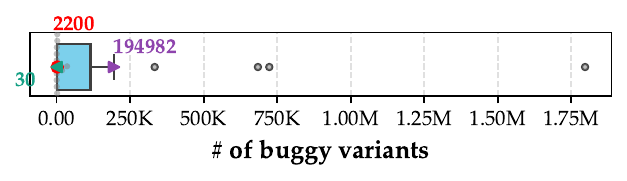}
        \caption{Mutant-level variants}  
        \label{fig: rq2_variant_number_Mutation-level}
    \end{subfigure}
    \hfill
    \begin{subfigure}[b]{0.485\textwidth}
        \centering
        \includegraphics[width=\textwidth]{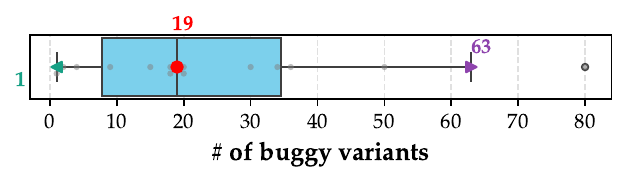}
        \caption{Version-level variants}  
        \label{fig: rq2_variant_number_Version-level}
    \end{subfigure}

    \vspace{0.3em}
    \begin{subfigure}[b]{0.8\textwidth}
        \centering
        \includegraphics[width=0.6\textwidth]{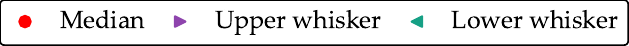}
    \end{subfigure}
    \caption{Numbers of mutant- and version-level buggy variants}  
    \label{fig: rq2_variant_number}
\end{figure}



Considering the potential distinctions between buggy variants at different levels, we collected the numbers of variants for mutant-level variants and version-level variants, respectively. As depicted in Figure~\ref{fig: rq3_scalability_metrics}, the medium numbers of mutant-level and \revise{version-level} variants are respectively 2,200 and 19. Especially, the upper whisker for mutant-level ones reaches 194,982, and one study~\cite{miranskyy2025on} even contributes to a tremendous maximum as 1,796,880. The above results explicitly indicate a substantial distinction in experimental scale between buggy variants of the two levels.
Reasonably, this gap in quantity mainly stems from their different evaluation granularity and objectives.
Empirical studies on mutant-level variants commonly assess how many mutants can be killed by the given test suite without detailing the execution results of individual test cases. 
The coarse-grained evaluation with the only focus on the number of killed mutants allows covering a wide range of mutation operators by generating mutants in large quantities.
In contrast, when employing version-level variants in empirical studies, we can zoom into the distinction among these variants through fine-grained evaluation metrics, such as the number of faults triggered by executing a given number of test cases.
However, this detailed evaluation hardly supports a large number of the version-level variants, which would otherwise introduce overwhelming fine-grained data and substantially hinder the effective presentation as well as analysis of the experimental results.


\subsubsection{Summary of RQ2}
\text{}


\Subtitle{Takeaway 2-1} Existing approaches for generating buggy variants for PUTs are dominated by custom mutation operators. The exploration of real-world bugs, mutation tooling, and imperfect models, which rely on available outer sources, remains comparably limited.  
With respect to the levels of buggy variants, we found that mutant-level and version-level exhibited close proportions that exceeded 40\%.


\Subtitle{Takeaway 2-2} We identified seven mutation actions and seven mutation targets, which constitute mutation operators. The mutation targets could be marked as classical, quantum, and hybrid types based on the involvement of classical and quantum workflows. Moreover, the quantum mutation targets, including gates and measurements, were more frequently used than the other two types, and gates were the most employed mutation targets.


\Subtitle{Takeaway 2-3} We found that Bugs4Q and Qbugs were the two benchmarks as the sources of real-world bugs. QMutPy and Muskit were considered in empirical studies as the mutation tooling to perform mutation testing and analysis. The training and testing of imperfect models only involved three datasets for image classification, i.e., MNIST, Fashion-MNIST, and CIFAR10.

\Subtitle{Takeaway 2-4} The medium numbers of mutant-level and version-level variants employed in one primary study were 2,200 and 19, respectively. This substantial difference in quantity suggests that the configuration of variant counts should take the evaluation granularity of buggy variants into account.
For example, a large quantity is suggested for mutant-level variants, as only the killability of mutants is evaluated, which corresponds to a coarse-grained evaluation.

\subsection{\texorpdfstring{\RQSection{3}}{RQ3}}\label{sec: rq3}
Outlined by the QSE community, the test scalability specifies the issue to address the challenge brought by the quantum programs with high-dimensional spaces, relevant to factors like the number of qubits, and the states in superposition or entanglement~\cite{murillo2025quantum}. Moreover, this challenge would be reflected in the cost and effectiveness of the test process. Although the complexity of quantum software can also be measured by metrics shared in general SE~\cite{zhao2021some}, like lines of code and number of statements~\cite{shen1983software, halstead1977elements, weyuker1988evaluating}, 
current works on QST prefer to investigate the complexity of corresponding quantum circuits, which is raised from the unique character of QC. To this end, this RQ investigates the scalability of CUTs involved in the primary studies. In detail, we will demonstrate whether the scalability issue was paid attention to in these studies and what scales of CUTs were employed for the empirical studies.

\subsubsection{Considerations for Scalability}
\begin{figure}[!t]
    \centering
    \begin{subfigure}[b]{0.49\textwidth}
        \centering
        \includegraphics[width=0.62\textwidth]{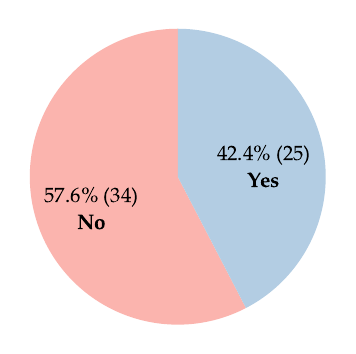}
        \caption{Pie chart about the proportion of primary studies that considered the test scalability}  
        \label{fig: rq3_whether_scalability}
    \end{subfigure}
    \hfill
    \begin{subfigure}[b]{0.49\textwidth}
        \centering
        \includegraphics[width=0.8\textwidth]{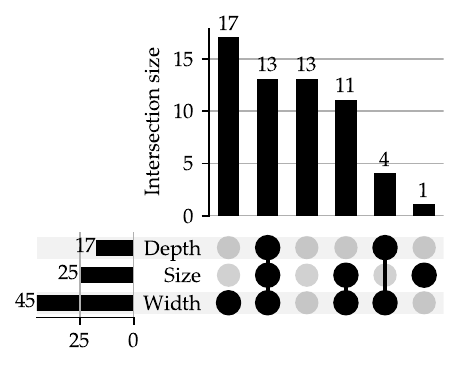}
        \caption{Upset plot about the primary studies reporting the three metrics of involved quantum circuits}  
        \label{fig: rq3_scalability_upsetplot}
    \end{subfigure}

    \caption{Quantity statistics for primary studies in terms of the scalability issue}
    \Description{Two subfigures, where the left is a pie chart that showcases whether the primary studies considered the test scalability, while the right is an upset plot displaying the primary studies reporting the circuit depth, size, and depth of the involved CUTs.}
    \label{fig: rq3_scalability_paper_numbers}
\end{figure}

Towards empirical studies on QST, test scalability can be effectively examined by incorporating a scalable high-level quantum program that can generate quantum circuits of varying sizes. Taking an example of the QFT showcased in Figure~\ref{fig: PUT_versus_CUT}, its functionality depends on the total number of input qubits (denoted as \texttt{n}). Different values of \texttt{n} in turn lead to varying circuit sizes and depths. 
 
With respect to the primary studies, according to Figure~\ref{fig: rq3_whether_scalability}, only 25 out of~\NumberOfPrimaryStudies discussed the test scalability. This result suggests that the scalability is recognized within the community, but has not been covered in the majority of existing studies. 

As we move on to Figure~\ref{fig: rq3_scalability_upsetplot}, it showcases the status of primary studies reporting the three typical metrics to measure the circuit scalability, i.e., circuit width, size, and depth. Forty-five primary studies reported the width of CUTs, while the sizes and depths were less discussed. This may result from a great concern that the increase in the qubit number exponentially raises the computational complexity, as the representation of $n$-qubit evolution requires a $2^n \times 2^n$ matrix stored in the classical simulator. Still, we observed that 13 primary studies reported all three metrics simultaneously, second only to the number of studies that reported the circuit width alone. This finding encourages considering multiple metrics beyond a single one, as meaningful for a more comprehensive assessment of the scalability. For example, in physical quantum hardware, the assessment merely relying on the circuit width (i.e., the mostly discussed one) may exaggerate the execution overhead for circuits in which multiple qubits are mostly operated in parallel, while including the depth can, in turn, quantify the sequential structure more accurately.

\subsubsection{Scale of Employed Quantum Circuits}
\label{sec: rq3: circuit_scale}

\begin{figure}[!t]
    \centering
    \begin{subfigure}[b]{0.485\textwidth}
        \centering
        \includegraphics[width=\textwidth]{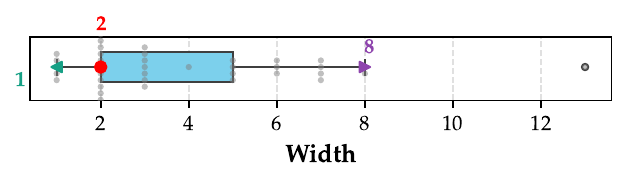}
        \caption{Minimum circuit widths}  
        \label{fig: rq3_scalability_Qubit_min}
    \end{subfigure}
    \hfill
    \begin{subfigure}[b]{0.485\textwidth}
        \centering
        \includegraphics[width=\textwidth]{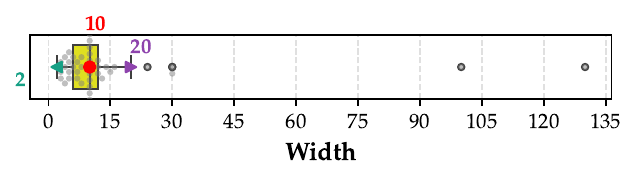}
        \caption{Maximum circuit widths}  
        \label{fig: rq3_scalability_Qubit_max}
    \end{subfigure}

    \begin{subfigure}[b]{0.485\textwidth}
        \centering
        \includegraphics[width=\textwidth]{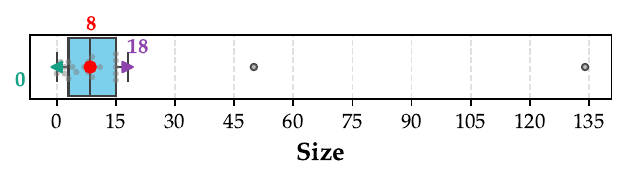}
        \caption{Minimum circuit sizes}  
        \label{fig: rq3_scalability_Gate_min}
    \end{subfigure}
    \hfill
    \begin{subfigure}[b]{0.485\textwidth}
        \centering
        \includegraphics[width=\textwidth]{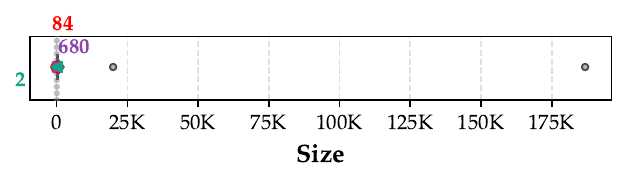}
        \caption{Maximum circuit sizes}  
        \label{fig: rq3_scalability_Gate_max}
    \end{subfigure}

    \begin{subfigure}[b]{0.485\textwidth}
        \centering
        \includegraphics[width=\textwidth]{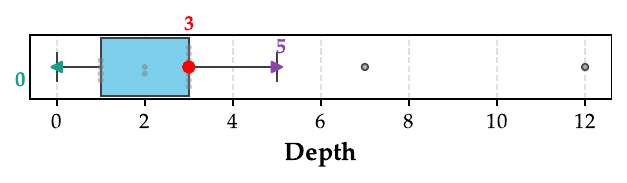}
        \caption{Minimum circuit depths}  
        \label{fig: rq3_scalability_Depth_min}
    \end{subfigure}
    \hfill
    \begin{subfigure}[b]{0.485\textwidth}
        \centering
        \includegraphics[width=\textwidth]{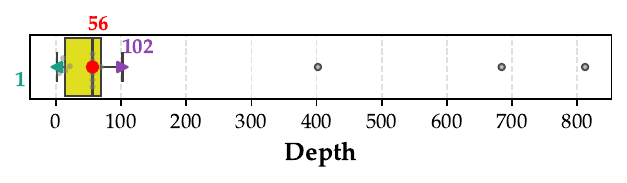}
        \caption{Maximum circuit depths}  
        \label{fig: rq3_scalability_Depth_max}
    \end{subfigure}
    
    \vspace{0.3em}
    \begin{subfigure}[b]{0.8\textwidth}
        \centering
        \includegraphics[width=0.6\textwidth]{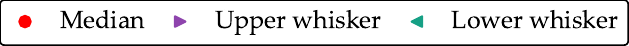}
    \end{subfigure}
    \caption{Complexity measures for CUTs involved in primary studies}
    \Description{A composite figure including six boxplots, where the boxplots respectively depict the minimum and maximum numbers for circuit widths, sizes, and depths of the CUTs employed in primary studies}
    \label{fig: rq3_scalability_metrics}
\end{figure}

In this part, we focus on the concrete scales of the CUTs included in the primary studies. Figure~\ref{fig: rq3_scalability_metrics} presents six boxplots showing the distributions of the minimum and maximum values for each of the three metrics reported across the primary studies. 

Considering the circuit width, the median values for the minimum and maximum are 2 and 10, respectively. On the one hand, this indicates that a moderate number of studies still investigated relatively simple quantum circuits. Nonetheless, a small circuit scale does not necessarily imply low complexity of the corresponding program, such as the high-level programs adopted in~\cite{li2025preparation, long2022testing}, where the circuit scale depends on the classical test inputs without the changes in the code itself. On the other hand, by considering both the median of the maximum circuit width and the upper whisker reaching 20, it is suggested that performing QST with more than 10 qubits remains feasible. In addition, we observed two extreme cases for the maximum qubit counts, i.e., 100~\cite{\PaperThirtySeven} and 130~\cite{\PaperFiftyFour} qubits upon classical simulation. In fact, the former~\cite{\PaperThirtySeven} did not actually run the 100-qubit Bernstein-Vazirani Algorithm for fault detection but classical analysis, and the latter~\cite{\PaperFiftyFour} claimed to employ a dedicated backend using decision diagrams as a data structure. 

According to both Figures~\ref{fig: rq3_scalability_Gate_max} and~\ref{fig: rq3_scalability_Depth_max}, the medium values for the maximum circuit sizes and depths are 84 and 56, respectively, while their upper whiskers reach 680 and 102, which conveys that deep or large-size circuits have been explored for QST. 
Because gate actions on disjoint sets of qubits can be executed in parallel, this result follows the intuition that the total gate count should typically be larger than the circuit depths. 
Especially, the maximum value shown in Figure~\ref{fig: rq3_scalability_Gate_max} is contributed by~\cite{\PaperEightyFour}, where this study executed a 10-qubit Harrow-Hassidim-Lloyd Algorithm with the number of total gates originally as 186,798. In addition to reporting the total number of quantum gates, we identified three studies~\cite{\PaperFifteen, \PaperThirtySeven, \PaperSixtySeven} that listed the number of specific quantum gates, such as single-qubit gates, CNOT gates, non-Clifford gates, etc. Some specific gates indeed play a special role in measuring the scalability of circuits and resulting complexity for QST, where, for example, circuits composed of only Clifford gates are promised to be simulated efficiently in polynomial time and perfectly based on stabilizer codes\footnote{This statement pertains to the Gottesman-Knill theorem, and we offer the literature~\cite{gottesman1998heisenberg, aaronson2004improved} for readers of interest.}. 

\highlight{With respect to quantifying the scale of abstract unitary operation in an unbiased manner, which could involve the modularized design of high-level quantum programs, two studies~\cite{li2025preparation,\PaperSeventy} mentioned \revise{evaluating the size and depth of transpiled circuits, where these synthesized operators can be decomposed and converted into basic gates.} Otherwise, a large-dimensional synthesized operator could be recognized with only one size and depth, thereby causing underestimation of complexity.}

\subsubsection{Summary of RQ3}
\text{}

\Subtitle{Takeaway 3-1} Only 42.5\% of the primary studies considered the scalability in empirical studies, thereby calling for more studies to be concerned with this issue. Forty-five studies reported the circuit width (a.k.a. the number of qubits) for CUTs, indicating a great attention to this metric.
Moreover, thirteen studies provided a summary including all of the circuit width, size, and depth.

\Subtitle{Takeaway 3-2} Regarding the referable scale of adopted quantum circuits, the medium value and the upper bound without outliers for the maximum qubit count turn out to be 10 and 20, respectively. The medium values for maximum circuit width and maximum circuit depth have been extended to 84 and 56.

\section{\texorpdfstring{\titlecap{\GroupTest}}{Testing Process Setups}}
\label{sec: process}

\subsection{Overview of Testing Process}\label{sec:testing_overview}
\begin{figure}
    \centering
    \includegraphics[width=0.75\columnwidth]{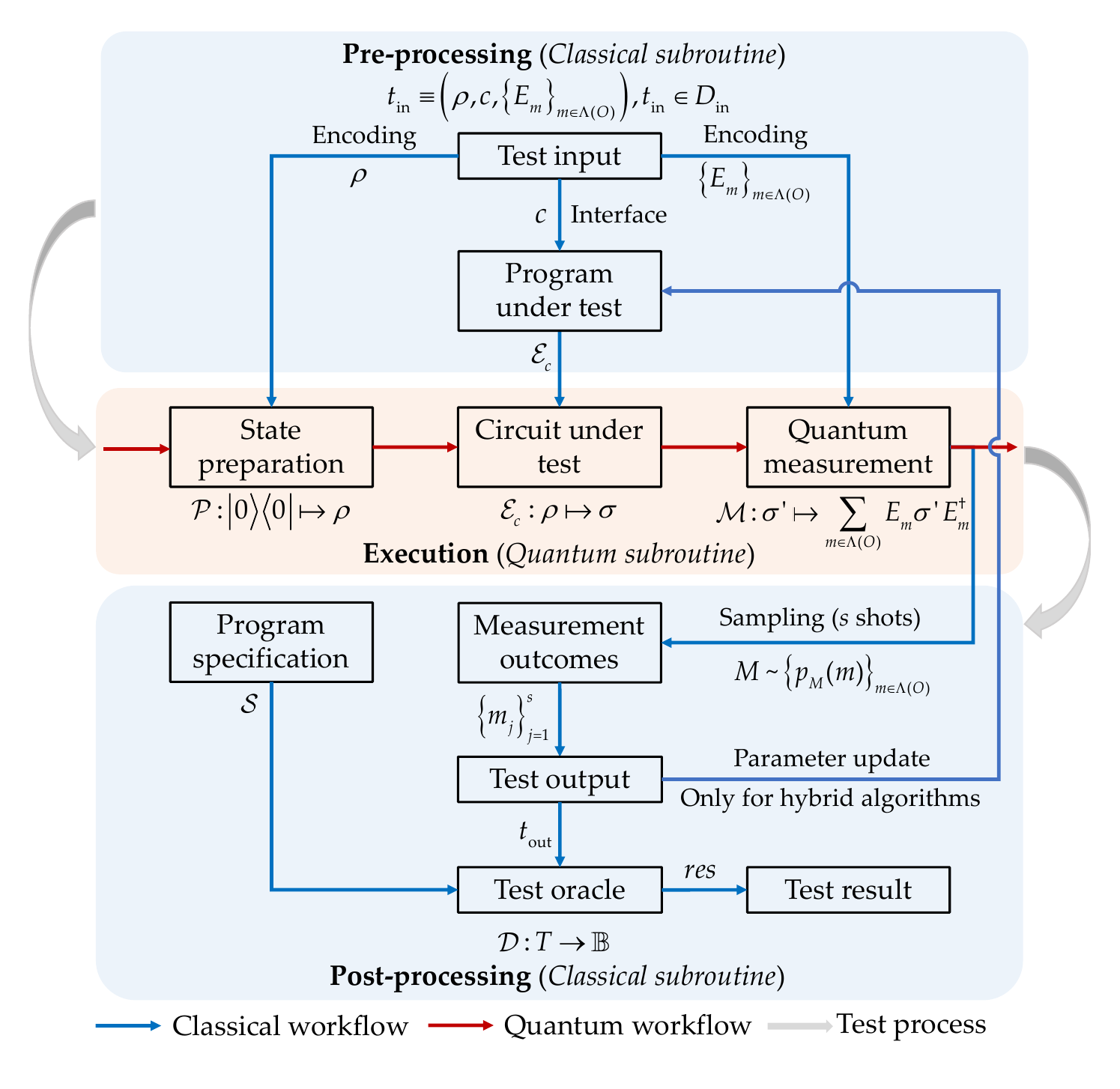}
    \caption{Key steps in a universal test process for QST}
    \Description{A figure displaying a framework of a universal test process, which includes the pre-processing, execution, and post-processing.}
    \label{fig: test_process}
\end{figure}

To begin with, we introduce a universal framework for executing a test process. Figure~\ref{fig: test_process} illustrates key steps in QST, and the discussion about quantum states is assumed within the space $\mathcal{L}(\mathcal{H})$.

\begin{definition}[\textit{Test input}]
    \label{def: test_input}
    The test input $t_{\text{in}}$ of a test case $t$ for QST can be denoted as a triple $t_{\text{in}}\equiv \left(\rho, c, \{E_m\}_{m \in \Lambda(O)}\right)$, where $\Lambda(O)$ is denoted as the set of all candidate measurement outcomes dependent on the selected observable $O$\footnote{The composite mapping can be formally defined as $\Lambda\equiv \Lambda_{\text{tran}} \circ \Lambda_{\text{spec}}$, where $\Lambda_{\text{spec}}$ aims to obtain the spectrum of an observable $O$, while $\Lambda_{\text{tran}}$ maps from the spectrum to the numbers obeying to the programming conventions. For example, the spectrum for Pauli-Z measurement is $\{+1,-1\}$, where their eigenstates are $\ket{0}$ for $+1$ and $\ket{1}$ for $-1$. Therefore, the measurement outcomes conventionally in programming are $\Lambda(Z)=\{0,1\}$, since they match the classical computers that operate on binary data.}.
    The density operator $\rho\in \mathcal{L}(\mathcal{H})$ refers to the initial quantum state of the CUT. The tuple $c\in \left\{\left(c_1, c_2, \cdots,c_l\right)\in \prod_{j\in[l]} C_j\left| c_j\in C^*_j\left(\left(c_k\right)_{k\neq j}\right),\forall j\in[l]\right.\right\}$\footnote{For $l \in \mathbb{N}^+$, the set $[l]$ is defined as $\{1,2,\cdots, l\}$.}
    includes $l$ classical arguments that determine the quantum channel for a CUT, where $C_j$ as a component of the Cartesian product is the maximum independent input domain for the classical argument $c_j$, and $C^*_j\left(\left(c_k\right)_{k\neq j}\right)$ forms the possible constrained input domain for $c_j$ determined by other relevant classical arguments. The collection of $|\Lambda(O)|$ POVMs $\{E_m\}_{m \in \Lambda(O)} \subset \left\{E\in \mathcal{L}(\mathcal{H})| E>0 \right\}$ serves for the measurement operation.
\end{definition}

We define the test input $t_{\text{in}}$ in Definition~\ref{def: test_input}. This definition is intended to be universal and does not imply that the three components are indispensable across all test scenarios. 
Analogous to CST, the classical input $c$ with $l$ arguments is provided for the PUT via its interface. It is worth noting that dependencies may exist among arguments. For instance, if the qubit count $c_1$ fully decides on the dimension of the vector $c_2$ for a linear operation, a possible dependency as $\text{dim}(c_2)=c_1$ can be created. 
In contrast, both the initial quantum state $\rho$ and measurement operation $\{E_m\}_{m \in \Lambda(O)}$ should be encoded into the quantum circuits for test execution, where the former and the latter lie prior and subsequent to the CUT, respectively. There are several feasible approaches to encoding for $\rho$, such as using the mathematical representation of an initial state or employing basic gates with low overhead. The encoding of $\{E_m\}_{m \in \Lambda(O)}$ may rely on extra operations from the default measurement basis (e.g., Pauli-Z measurement in Qiskit), such as transforming the Pauli-Z measurement (e.g., $E_0=\ket{0}\bra{0}, E_1=\ket{1}\bra{1}\}$) to the Pauli-X measurement (e.g., $E_0=\ket{+}\bra{+}, E_1=\ket{-}\bra{-}$), which can be realized by adding a Hadamard gate to each output qubit.

\begin{EnvRevise}
To instantiate Definition~\ref{def: test_input}, we take an integer comparator used in~\cite{li2025preparation} as an example to elaborate on the test input design. Its program specification for pure-state inputs can be denoted as
$$
\ket{0}_n\ket{y}_n\mapsto
\begin{cases}
\ket{\mathbf{1}(y\ge L)}_n\ket{y}_n, & geq = 1  \\
\ket{\mathbf{1}(y< L)}_n\ket{y}_n, & geq = 0
\end{cases},
$$
where $n$ denotes the number of qubits in each quantum register; $y$ and $L$ are two non-negative integers to be compared; the predicate $\mathbf{1}(A)$ equals $1$ if the event $A$ holds and $0$ otherwise; and the Boolean variable $geq$ specifies the comparison mode. 
Then, the initial quantum state only includes the $n$-qubit quantum register used to store $y$, i.e., $\rho= \ket{y}_n\bra{y}_n$. Based on the functionality, the quantum register is initialized as a computational basis state, so a valid instance of 4 qubits could be: $\ket{0101}\bra{0101}$ in a bitstring form, or $\ket{5}_4\bra{5}_4$ in a decimal form.
The classical input tuple can be written as $c=\left(n, L, geq\right)$. Despite the data form: $n\in\mathbb{N^{+}}$, $L\in\mathbb{Z}$ and $geq\in\mathbb{B}$, an implicit constraint to align with the algorithmic principle is that an $n$-qubit quantum register is capable of accurately encoding the classical data $L$, implying $0 \le L \le 2^n -1$. Based on this, we can offer an instance of $c$ as $(4, 3, 1)$. 
To directly extract the comparison outcome, we perform a Pauli-Z measurement on the ancilla register, whose corresponding POVM is given by $\left\{\ket{z}_n\bra{z}_n\right\}_{z=0}^{2^n-1}$. Based on the test input instance given above, we can get a deterministic bitstring $0001$ (i.e., $5 \ge 3$ holds) in the $4$-bit classical register if the test passes.
\end{EnvRevise}

\begin{definition}[\textit{Test oracle}]
    \label{def: test oracle}
    A test oracle $\mathcal{D}: T\rightarrow\mathbb{B}$ is a partial function from the executed test suite to a Boolean set that corresponds to a passed and failed test.
\end{definition}
As is shown in Definition~\ref{def: test oracle}, \revise{a test oracle for QST is primarily used to determine whether a test within the given test suite $T$ passes or fails~\cite{li2026dynamic}}\footnote{We remind that the terminology ``oracle'' varies in QC and SE. The ``oracle'' in QC generally refers to a black-box quantum subroutine implementing a classical function $\mathcal{O}_f:\mathbb{B}^{n_{\text{in}}}\rightarrow\mathbb{B}^{n_{\text{out}}}$, where $n_{\text{in}}$ and $n_{\text{out}}$ indicate the number of input and output qubits involved in the oracle operation. In our paper, ``test oracle'' is merely specified as an SE concept.}. The test oracle works in reliance on the program specification based on test requirements and test outputs. Especially, test outputs $t_{\text{out}}$ for QST should be associated with a group of measurement outcomes $\{m_j\}_{j=1}^s$ upon $s$ shots, such as estimations for 
the probability distribution $\{p_{M}(m)\}_{m\in\Lambda(O)}$ and the expectation of the observable $\braket{O}$. In SE, a test oracle hardly ensures no false negative and no false positive simultaneously~\cite{barr2014oracle}. A false negative occurs when a failed test is mistakenly reported as passed, while a false positive is the opposite, i.e., reporting a passed test as failed. By contrast, the \textit{ground truth} is considered an ideal test oracle that always produces correct results, but it is difficult to obtain in practice. Therefore, an effective test oracle should approximate the ground truth as closely as possible.
 
\subsection{\texorpdfstring{\RQSection{4}}{RQ4}}\label{sec: rq4}
A test input serves as a critical component of a test case to be executed. In a systemic view, a test input determines how a PUT is triggered and then affects what response comes from the PUT. For QST, the formation of test inputs should account for qubit evolution corresponding to the functionality of quantum programs and possible hybrid classical-quantum subroutines regarding the program structure, which manifest observable differences from CST. To this end, this RQ overviews how test inputs were considered in the primary studies and provides a comprehensive summary of test inputs applicable to QST.

\subsubsection{Considerations for Input Types and Sizes}

\begin{figure}[!t]
    \centering
    \begin{subfigure}[b]{0.98\columnwidth}
        \centering
        \includegraphics[width=0.6\columnwidth]{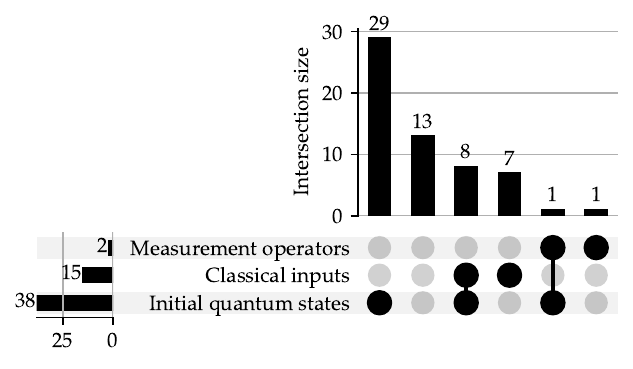}
        \caption{The number of primary studies in terms of the formed test inputs}  
        \label{fig: rq4_input_type_distribution}
    \end{subfigure}
 
    \begin{subfigure}[b]{0.98\columnwidth}
        \centering
        \includegraphics[width=0.7\columnwidth]{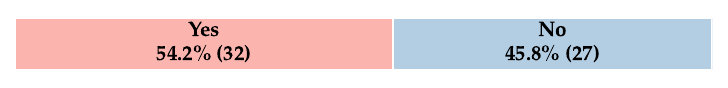}
        \caption{Proportion for the studies reporting the size of employed test suites}  
        \label{fig: rq4_whether_counts}
    \end{subfigure}

    \caption{Quantity statistics for the test cases adopted in the primary studies}  
    \Description{A composite figure including two subplots: the top is an upset plot about how primary studies considered measurement operators, classical inputs, and initial quantum states as parts of test inputs, while the bottom displays whether the primary studies reported the size of test suites.}
    \label{fig: rq4_baselines}
\end{figure}

Figure~\ref{fig: rq4_input_type_distribution} shows which component of $t_{\text{in}}$ was considered in the primary studies\revise{, where the intersection size of the UpSet plot denotes the number of samples corresponding to each set combination indicated by the black dots within the same column.}
In this RQ, we only recognize these components being claimed as test inputs in corresponding studies. We found that the initial quantum states were the most frequently employed, appearing in 38 studies, and 29 studies used initial quantum states as the unique test input type. Their popularity could stem from being the first defined input type for QST in the context of testing quantum circuits~\cite{ali2021assessing}. In comparison, only 15 studies discussed classical inputs. This is because classical inputs are excluded from the low-level quantum programs that have earned popularity with many QST studies. 
To be inspiring, two studies included the measurement operators in the test inputs. This consideration is reasonable, since any measurement outcome is relative to the chosen measurement basis, rather than invariant. 

From the perspective of the Cartesian product, the number of input types included in the test case determines the \revise{\textit{size of the test input domain}, such as the number of valid values for an enumerable variable}. Hence, more test cases are usually required in a test suite for an input domain with a large \revise{size} to ensure test sufficiency. Figure~\ref{fig: rq4_whether_counts} is introduced to show the convention of reporting test suite sizes. However, only 54.2\% of the primary studies explicitly reported or indirectly indicated the sizes of their test suites, highlighting the need for greater attention to this issue in future research.

\subsubsection{Test Input Types and Properties}

\begin{table}[!t]
    \footnotesize
    \centering
    \caption{Test input types and test input properties of test cases}
    \label{tab: rq4_input_type_names}
    \resizebox{.98\columnwidth}{!}{
        
\begin{tabular}{>{\centering\arraybackslash}p{0.22\columnwidth}
p{0.4\columnwidth}
p{0.25\columnwidth}
>{\centering\arraybackslash}p{0.02\columnwidth}}
    \toprule[1pt]
    \multicolumn{1}{c}{\textbf{Test input types}} & \multicolumn{1}{c}{\textbf{Test input properties}} & \multicolumn{1}{c}{\textbf{Primary studies}} & \multicolumn{1}{c}{\textbf{\#}} \\
    \cmidrule(lr){1-1} \cmidrule(lr){2-2} \cmidrule(lr){3-3} \cmidrule(lr){4-4}  
    Initial quantum states & Computational basis states & \cite{\PaperOne, \PaperThree, \PaperFive, \PaperTwelve, \PaperSeventeen, \PaperTwenty, \PaperTwentyOne, \PaperTwentyFour, \PaperThirtyEight, \PaperFiftyTwo, \PaperFiftyNine, \PaperSixty, \PaperSixtySeven, \PaperSeventyOne, \PaperEightyThree, \PaperEightyFour, \PaperOneHundredAndSix, \PaperOneHundredAndEight, \PaperOneHundredAndThirteen, \PaperOneHundredAndFifteen, \PaperOneHundredAndSixteen} & 21 \\ 
     & Fully separable states & \cite{\PaperThirteen, \PaperThirtyFive, \PaperFortyFour, \PaperFiftyTwo, \PaperEightySix, \PaperOneHundredAndThree, \PaperOneHundredAndTwentyTwo} & 7 \\ 
     & Default state & \cite{\PaperOneHundredAndEight, \PaperOneHundredAndTwelve, \PaperOneHundredAndSeventeen, \PaperOneHundredAndTwentyOne} & 4 \\ 
     & States prepared by specific circuits & \cite{\PaperSix, \PaperOneHundredAndFive, \PaperOneHundredAndNineteen} & 3 \\ 
     & Mixed states & \cite{\PaperFortySix, \PaperSeventyOne} & 2 \\ 
     & Superposition states & \cite{\PaperSeventyOne, \PaperOneHundredAndEight} & 2 \\ 
     & State vector by a quantum constraint solver & \cite{\PaperFortyFive} & 1 \\ 
     & Pauli eigenstates & \cite{\PaperSeventy} & 1 \\ 
     & Eigenvector of a unitary operation & \cite{\PaperOneHundredAndSeventeen} & 1 \\ 
    \cmidrule(lr){1-1} \cmidrule(lr){2-2} \cmidrule(lr){3-3} \cmidrule(lr){4-4} 
    Classical inputs & Numbers & \cite{\PaperTwelve, \PaperTwentySeven, \PaperFortySix, \PaperSeventyOne, \PaperOneHundredAndEight} & 5 \\ 
     & Images & \cite{\PaperThirtySix, \PaperEightyOne, \PaperOneHundredAndSeven, \PaperOneHundredAndTwentyThree} & 4 \\ 
     & Vectors & \cite{\PaperFourteen, \PaperFortySix, \PaperOneHundredAndTwentyOne} & 3 \\ 
     & Matrices & \cite{\PaperFortyFour, \PaperFortySix} & 2 \\ 
     & Classical arguments for oracle & \cite{\PaperThirtyThree} & 1 \\ 
    \cmidrule(lr){1-1} \cmidrule(lr){2-2} \cmidrule(lr){3-3} \cmidrule(lr){4-4} 
    Measurement operators & Helstrom measurement operators & \cite{\PaperThirtySeven} & 1 \\ 
     & Pauli strings & \cite{\PaperThirtyNine} & 1 \\
    \bottomrule[1pt]
\end{tabular}
        
    }
\end{table}

Table~\ref{tab: rq4_input_type_names} presents the details of test input types along with their properties manifested in the primary studies.

Regarding the initial quantum states, the majority of studies (21 in total) selected the computational basis states as (part of) test inputs. This trend is understandable, since computational basis states constitute the quantum analogue of classical bit strings, making it straightforward to adapt CST techniques to promote QST. In addition, quantum-specific properties of initial quantum states have been explored, including fully separable states, mixed states, superposition states, and Pauli eigenstates. A great number of studies considering fully separable states may be due to their easy preparation through basic single-qubit gates like the parametric rotation gates, where the state preparation in a view of the Bloch sphere was involved in three studies~\cite{\PaperThirteen, \PaperThirtyFive, \PaperFiftyTwo}. 
Also, some studies introduced specific approaches to prepare intended quantum states, such as the custom circuit patterns for state initialization (e.g., the inverse QFT generator adopted in~\cite{\PaperSix, \PaperOneHundredAndFive}) and the constraint solver to calculate the state vectors of input states~\cite{xia2025quantum}. The default state in the table indicates that all input qubits exist in $\ket{0}$, which was claimed or implicated as the test input in 4 studies. Although this state is deemed a special instance of computational basis states, we emphasize it since the default state is the only valid initial quantum state for many searching and optimization algorithms, such as Grover Search studied by~\cite{\PaperOneHundredAndEight}. Finally, taking an example of the eigenvectors used for Quantum Phase Estimation~\cite{\PaperOneHundredAndSeventeen}, we should become aware that the selection of valid initial states might be constrained by other inputs, according to the program functionality.
 
As for classical inputs, the relevant data formats demonstrate diversity. Common numerical inputs with various dimensions, including numbers, vectors, and matrices, have been used in the primary studies. For example, studies such as~\cite {long2024testing, li2025preparation} treated qubit counts as a type of input numbers, enabling PUTs to be easily scalable. Images were specific to the four studies on testing quantum machine learning models for image classification, which, in terms of test inputs, show strong consistency with testing deep learning models in CSE for the same task. Classical inputs for oracle operations are highlighted, because they constitute specialized test arguments for some \textit{oracle-based quantum programs}, such as Grove Search in~\cite{\PaperThirtyThree} whose test inputs specify unstructured databases under query. 

In addition, we can gain valuable insights into quantum information principles from the two studies that use measurement operators as test inputs. Chen et al.~\cite{\PaperThirtySeven} investigated the generation of test patterns defined by a pair of initial quantum states and measurement operators. They explored Helstrom measurement~\cite{barnett2009quantum} to minimize the error of distinguishing output states of a correct program from those of a fault program. Muqeet et al.~\cite{\PaperThirtyNine} used Pauli strings as the measurement basis for the only test input type, motivated by search and optimization tasks that allow only the default initial state. Because the Pauli family exhibits specific commutation properties~\cite{reggio2024fast}, employing Pauli strings as test inputs allows the test suite to be reduced by removing operationally equivalent test cases.

\revise{In summary, despite the diversity of test input types, the construct validity of an empirical evaluation depends on whether the chosen inputs are aligned with the intended program functionality.
For example, it is difficult to interpret the behavior of an integer comparator shown in Section~\ref{sec:testing_overview} if the quantum register is prepared in a state other than the computational basis. Under such an input, the PUT may execute without crashing, yet the resulting outputs may no longer support meaningful correctness assessment, which compromises the credibility of the evaluation.}

\subsubsection{Summary of RQ4}
\text{}

\Subtitle{Takeaway 4-1} Initial quantum states are the most used test inputs, followed by classical inputs and measurement operators. 
Despite the potential impact of test input formation on test suite size, only 54.2\% of the primary studies reported or implied test suite sizes.

\Subtitle{Takeaway 4-2} The majority of primary studies selected computational basis states as the initial quantum states. Besides, the formation of initial states considered quantum-specific properties (e.g., separability and superposition) and state preparation approaches (e.g., using specific circuits and constraint solvers). The classical inputs cover various common classical data (e.g., numbers and matrices) as well as a quantum-specific component (i.e., oracle). Meanwhile, two studies borrowed principles of quantum information to explore measurement operators as test inputs.

\subsection{\texorpdfstring{\RQSection{5}}{RQ5}}\label{sec: rq5}
The test oracle problem is a fundamental issue rooted in SE. For QST, the inherent properties of quantum measurement make the test oracle problem particularly challenging. On the one hand, quantum measurements introduce extra randomness into the test results, while the ground truth should be deterministic instead of probabilistic. On the other hand, quantum measurement is an inherently irreversible process that maps a quantum state in a high-dimensional Hilbert space to a finite set of classical outcomes, resulting in significant information loss and making it difficult to reconstruct the original quantum state from the measurement results. Out of both concerns, this RQ reviews the test oracles employed in primary studies to summarize feasible solutions to test output analysis. 

\subsubsection{Program Specification}
Before zooming into the test oracle as an executable approach, we should revisit the program specification, as it documents the testing objectives and guides the construction of appropriate test oracles. 
To judge the correctness of quantum programs, recent studies~\cite{ali2021assessing, long2024testing, paltenghi2024survey} have proposed formal program specifications for QST, and all of these existing program specifications only focus on the mathematical relation between inputs and expected outputs. 

In a general view of the input-output relation, we first classify the program specification for QST into the three categories, inspired by the definitions of test oracles for CST~\cite{barr2014oracle}.
\begin{itemize}[leftmargin=*]
    \item \textit{Specified program specification}: Refers to a formal specification that describes the expected behavior of a PUT, such as a mathematical formula to explain the mapping from input to output quantum states.
    \item \textit{Derived program specification}: Indicates information derived from various artifacts (e.g., documentation and system executions), properties of the PUT, or variants of it.
    \item \textit{Implicit program specification}: Means general true implicit knowledge to distinguish between a correct and incorrect behavior, regardless of the PUT, such as a resulting program crash, which definitely indicates a failed test. 
\end{itemize}
Based on the above categories, we found that the program specifications proposed in these recent studies are categorized as specified, demonstrating their importance in QST. Then, we reformulate and renew specified program specifications with different output types through mappings, where $D_{\text{in}}$ is the test input domain: 
\begin{itemize}[leftmargin=*]
    \item \textit{Probability distribution}: Involves a mapping to the probability space for test outputs, i.e., $\mathcal{S}_{\text{pd}}: D_{\text{in}} \rightarrow (\Lambda(O),2^{\Lambda(O)}, p_M)$. For example, we can use a tuple $\{(m, p_M(m))\}_{m\in\Lambda(O)}$ to denote the expected probability distribution yielded by a test input $t_{\text{in}}$.
    \item \textit{Quantum state}: Describes a mapping to the expected output quantum states, where $\mathcal{S}_{\text{qs}}: D_{\text{in}}  \rightarrow  \mathcal{H}$ refers to documenting state vectors, while $\mathcal{S}_{\text{qs}}: D_{\text{in}} \rightarrow  \mathcal{L}(\mathcal{H})$ can accommodate general quantum states in a density operator form.
    \item \textit{Classical outcome}: Includes a mapping to classical numbers, i.e., $\mathcal{S}_{\text{co}}: D_{\text{in}}  \rightarrow  \mathbb{R}$, which is consistent with many classical programs for numerical computation. This specification still happens in QST, such as a PUT for estimating the ground energy of a hydrogen molecule.
    \item \textit{Quantum operator}: Represents the mapping itself for qubit evolution involved in the CUT, thereby $\mathcal{S}_{\text{qo}}\in \mathrm{SU}(2^n)$ in the form of a unitary operator for $n$ qubits or $\mathcal{S}_{\text{qo}}\in \mathcal{L}(\mathcal{L}(\mathcal{H}))$ through a general representation of a quantum channel. Note that this specification can provide structural details of PUT rather than a shallow black-box model. Inspired by~\cite{paltenghi2024survey}, we can design the specification as a sequence of unitary matrices corresponding to required quantum gates or subroutines.
\end{itemize}

\begin{table}[!t]
    \small
    \centering
    \caption{Program specifications and corresponding types for output analysis}
    \label{tab: rq5_program_specification}
    \resizebox{.98\columnwidth}{!}{
        
\begin{tabular}{>{\centering\arraybackslash}p{0.25\columnwidth}
p{0.3\columnwidth}
p{0.42\columnwidth}
c}
    \toprule[1pt]
    \multicolumn{1}{c}{\textbf{Program specifications}} & \multicolumn{1}{c}{\textbf{Types for output analysis}} & \multicolumn{1}{c}{\textbf{Primary studies}} & \multicolumn{1}{c}{\textbf{\#}} \\
    \cmidrule(lr){1-1} \cmidrule(lr){2-2} \cmidrule(lr){3-3} \cmidrule(lr){4-4}  
    Specified & Probability distribution & \cite{\PaperOne, \PaperThree, \PaperFive, \PaperTwelve, \PaperSeventeen, \PaperTwenty, \PaperTwentyOne, \PaperThirtyOne, \PaperThirtyNine, \PaperFiftyNine, \PaperSixty, \PaperSeventyOne, \PaperSeventyThree, \PaperSeventyEight, \PaperEightyFour, \PaperOneHundredAndFourteen} & 16 \\ 
     & Quantum state & \cite{\PaperFourteen, \PaperThirtyEight, \PaperFortyThree, \PaperFortySix, \PaperSeventyOne, \PaperEightyThree, \PaperOneHundredAndSix, \PaperOneHundredAndEight, \PaperOneHundredAndTwelve, \PaperOneHundredAndFourteen, \PaperOneHundredAndFifteen, \PaperOneHundredAndSeventeen, \PaperOneHundredAndTwentyOne} & 13 \\ 
     & Classical outcome & \cite{\PaperFifteen, \PaperThirtySix, \PaperSeventyOne, \PaperEightyOne, \PaperOneHundredAndSix, \PaperOneHundredAndSeven, \PaperOneHundredAndEight} & 7 \\ 
     & Quantum operator & \cite{\PaperThirtySeven, \PaperEightyThree, \PaperOneHundredAndFourteen, \PaperOneHundredAndSeventeen} & 4 \\ 
    \cmidrule(lr){1-1} \cmidrule(lr){2-2} \cmidrule(lr){3-3} \cmidrule(lr){4-4} 
    Derived & General property & \cite{\PaperThirtyThree, \PaperFortyOne, \PaperFortyFour, \PaperFiftyFour, \PaperSeventy, \PaperSeventyFour, \PaperNinetyFour, \PaperOneHundredAndNineteen, \PaperOneHundredAndTwentyTwo} & 9 \\ 
     & Original program output & \cite{\PaperSix, \PaperThirteen, \PaperSeventyFour, \PaperOneHundredAndNine, \PaperOneHundredAndSixteen} & 5 \\ 
     & Metamorphic relation & \cite{\PaperTwentyFour, \PaperTwentySeven} & 2 \\ 
    \cmidrule(lr){1-1} \cmidrule(lr){2-2} \cmidrule(lr){3-3} \cmidrule(lr){4-4} 
    Implicit & Program error & \cite{\PaperSeventyFour} & 1 \\
    \cmidrule(lr){1-4} 
    
    \multicolumn{4}{p{1.1\columnwidth}}{\textbf{Total number 
    of primary studies for each program specification:} Specified (32), Derived (15), Implicit (1)}\\
    \bottomrule[1pt]
\end{tabular}
        
    }
\end{table}

Next, we review the program specification used by primary studies and annotate data according to our proposed definitions mentioned above. From Table~\ref{tab: rq5_program_specification}, specified program specifications were considered in 32 studies, which was the most used specification type compared to the derived and implicit ones. Within the specified program specifications, the probability distribution and quantum state were the top two output types in terms of study counts. We infer that the popularity of both output types could relate to the great impact of the two pioneer publications~\cite{ali2021assessing, long2024testing} proposing such program specifications. The adoption of classical outcomes should rely on the functionalities of PUTs, and we found that most of the involved PUTs were expected to yield deterministic outputs, such as oracle-based quantum programs~\cite{\PaperSeventyOne, \PaperOneHundredAndSix, \PaperOneHundredAndEight} and quantum machine learning models for classification~\cite{\PaperThirtySix, \PaperEightyOne, \PaperOneHundredAndSeven}. As for the quantum operators with the least attention currently, the two studies~\cite{\PaperThirtySeven, \PaperEightyThree} employed unitary matrices to represent the expected programs, while the other two~\cite{\PaperOneHundredAndFourteen, \PaperOneHundredAndSeventeen} suggested the Choi-matrix representation for the expected quantum channel.

Regarding the derived program specifications, we preliminarily identified three types for output analysis. General properties were employed in 9 studies, where they indicated the invariants designed for quantum-specific properties, like checking superposition and entanglement of quantum states~(e.g.,~\cite{huang2019statistical}) or examining equivalence of two quantum circuits (e.g.,~\cite{long2024equivalence}). 
Aside from merely checking the given properties of the quantum states involved in PUTs, we found that two studies~\cite{\PaperFortyFour, \PaperSeventyFour} employed program properties for fault detection. For example, assertions about the inequality between total counts and successful counts were adopted in~\cite{\PaperSeventyFour}, and the study~\cite{tan2025hornbro} used the assertions in the product of components under check, both of which aimed to capture output-related properties or partial functionalities, rather than establishing full output correctness.
Although metamorphic relations can be regarded as a special subset of program properties, we distinguish them from ``general properties'' because metamorphic relations describe relationships among multiple related test inputs and their corresponding outputs, and are specifically designed to support correctness checking in the absence of fully specified program specifications. From the result of the literature review, metamorphic relations gained the least attention within the derived specifications, as there were only two studies on metamorphic testing of quantum programs. In addition, five studies treated the original program outputs (i.e., the outputs produced by a fully known bugless program to which the buggy variants correspond) as the program specification, which could be a temporarily feasible scheme specific to certain controlled experiments without easily acquired output expectations.

Finally, only one study~\cite{\PaperSeventyFour} on mutation testing of quantum programs discussed program errors. One possible explanation for the limited focus on implicit specifications is that most work targeted triggering bugs of quantum subroutines (i.e., CUTs), while most of such errors in this study were revealed by classical mutants.

\subsubsection{Test Oracle}
For QST, a test oracle depends on a predefined criterion to judge whether the PUT follows the given program specification. It should be emphasized that the test oracle could take significant efforts to conduct such a comparison, which, however, appears trivial in CST. 
For example, we intend to test a quantum program whose program specification specifies an expected quantum state, but the test output we can offer for the test oracle is a group of measurement outcomes rather than the original state, such that the test oracle may compromise to compare the one group of measurement outcomes produced from the PUTs and the other group derived from the program specification.
In this situation, a valid test oracle must rely on an additional \textit{evaluation mechanism} to effectively distinguish between the two groups, such as hypothesis testing. The need for such an evaluation mechanism primarily arises from the unique properties of quantum measurements.

Then, we overview candidate test oracles suitable for QST. Only two test oracles tailored to QST have been claimed (i.e., wrong output oracle and output probability oracle~\cite{ali2021assessing}) in existing work, whereas both kinds still struggle to cover all the underlying techniques adopted in the primary studies. Motivated by the fact that current literature lacks a unified terminology or clear boundaries for test oracles, we summarize and extend the list of test oracles for QST as follows:
\begin{itemize}[leftmargin=*]
    \item \textit{Wrong Output Oracle (WOO)}: Given one or a group of test outputs, and a collection of valid outputs with deterministic values provided by the program specification, the test passes iff the evaluation mechanism confirms that each test output falls within the given collection; otherwise, the test fails. Owing to the focus on the individual output, WOO almost behaves like the test oracles used to test deterministic classical software.
    \item \textit{Output Probability Oracle (OPO)}: Given the probability distribution estimated by a group of measurement outcomes, and the expected probability distribution from the program specification, the test passes iff the evaluation mechanism regards the two distributions as statistically consistent; otherwise, the test fails. OPO is specifically designed for scenarios where the focus is on the overall distribution rather than on individual outcomes.
    \item \textit{Property-Based Oracle (PBO)}: Given a set of invariants or properties that the program output should satisfy, the test passes iff the evaluation mechanism verifies that all of these properties are satisfied; otherwise, the test fails. To this end, PBO can be applied to test scenarios aside from fault detection, such as checking output properties.
    \item \textit{Dominant Output Oracle (DOO)}: Given a group of measurement outcomes, and a collection of valid outputs with deterministic values specified by the program specification, the test passes iff the evaluation mechanism detects the presence of a certain outcome that dominates other outcomes and falls within the given collection; otherwise, the test fails. From the perspective of the trade-off between WOO and OPO, DOO allows a limited number of measurement outcomes that deviate from the valid outputs and does not require consistency across the overall probability distributions.
    \item \textit{Quantum State Oracle (QSO)}: Given a derived or reconstructed output quantum state of the CUT and an expected quantum state given by the program specification, the test passes iff the evaluation mechanism recognizes the equivalence of the two quantum states; otherwise, the test fails. Different from the above four test oracles, QSO may depend on multiple test inputs (e.g., multiple collections of measurement operators for quantum state tomography) within one test case to reconstruct the quantum state.
    \item \textit{Quantum Operation Oracle (QOO)}: Given a derived or reconstructed quantum operation of the CUT and an expected quantum operation given by the program specification, the test passes iff the evaluation mechanism identifies the consistent equivalence of each pair of components within the two quantum operations; otherwise, the test fails. Similar to QSO, QOO could need multiple test inputs to recover quantum information as well, but differently, QOO imposes stricter requirements on the internal structure of the CUT rather than merely on the relationships between inputs and outputs. 
\end{itemize}
\revise{Regarding the dependency of the six test oracles on specified, derived, and implicit program specifications, we observe that many oracle implementations rely on specified program specifications, because these specifications explicitly characterize the correct behavior required for oracle construction. In contrast, PBO is more flexible: it can operate with derived and implicit specifications, because the required invariants or properties can often be constructed without full knowledge of the expected outputs.}

\begin{table}[!t]
    \small
    \centering
    \caption{Test oracles along with their evaluation mechanisms}
    \label{tab: rq5_oracle_type}
    \resizebox{.98\columnwidth}{!}{
        
\begin{tabular}{>{\centering\arraybackslash}p{0.15\columnwidth}
p{0.4\columnwidth}
p{0.4\columnwidth}
c}
    \toprule[1pt]
    \multicolumn{1}{c}{\textbf{Test oracles}} & \multicolumn{1}{c}{\textbf{Evaluation mechanisms}} & \multicolumn{1}{c}{\textbf{Primary studies}} & \multicolumn{1}{c}{\textbf{\#}} \\
    \cmidrule(lr){1-1} \cmidrule(lr){2-2} \cmidrule(lr){3-3} \cmidrule(lr){4-4}  
    WOO & Original outcome & \cite{\PaperOne, \PaperThree, \PaperFive, \PaperSeventeen, \PaperTwenty, \PaperTwentyOne, \PaperFortyThree, \PaperFiftyNine, \PaperSixty, \PaperSeventyOne, \PaperSeventyThree, \PaperSeventyEight} & 12 \\ 
     & Inverse test & \cite{\PaperSeventyOne, \PaperOneHundredAndSix, \PaperOneHundredAndEight, \PaperOneHundredAndTwelve} & 4 \\ 
     & Swap test & \cite{\PaperFiftyTwo, \PaperOneHundredAndTwelve} & 2 \\ 
    \cmidrule(lr){1-1} \cmidrule(lr){2-2} \cmidrule(lr){3-3} \cmidrule(lr){4-4} 
    OPO & Pearson's chi-squared test & \cite{\PaperFive, \PaperSeventeen, \PaperTwenty, \PaperTwentyOne, \PaperThirtyOne, \PaperFortyThree, \PaperFiftyTwo, \PaperSixty, \PaperSeventyThree, \PaperSeventyEight, \PaperEightyFour, \PaperOneHundredAndTwelve, \PaperOneHundredAndThirteen, \PaperOneHundredAndFourteen, \PaperOneHundredAndSeventeen} & 15 \\ 
     & Wilcoxon signed rank test & \cite{\PaperThree, \PaperFiftyNine} & 2 \\ 
     & Expectation value & \cite{\PaperThirtyNine} & 1 \\ 
     & Mann-Whitney U test & \cite{\PaperFortySix} & 1 \\ 
     & Jensen-Shannon divergence & \cite{\PaperEightyThree} & 1 \\ 
     & G-test & \cite{\PaperOneHundredAndTwelve} & 1 \\ 
     & Multinomial test & \cite{\PaperOneHundredAndTwelve} & 1 \\ 
     & Monte Carlo Pearson's chi-squared test & \cite{\PaperOneHundredAndTwelve} & 1 \\ 
     & Monte Carlo G-test & \cite{\PaperOneHundredAndTwelve} & 1 \\ 
     & Monte Carlo multionmial test & \cite{\PaperOneHundredAndTwelve} & 1 \\ 
    \cmidrule(lr){1-1} \cmidrule(lr){2-2} \cmidrule(lr){3-3} \cmidrule(lr){4-4} 
    PBO & Original outcome & \cite{\PaperTwentyFour, \PaperTwentySeven, \PaperNinetyFour} & 3 \\ 
     & Pearson's chi-squared test & \cite{\PaperTwelve, \PaperThirteen} & 2 \\ 
     & Fisher's exact test & \cite{\PaperThirtyThree, \PaperOneHundredAndNineteen} & 2 \\ 
     & T test & \cite{\PaperThirteen} & 1 \\ 
     & Swap test & \cite{\PaperSeventy} & 1 \\ 
     & Trace-related distance & \cite{\PaperSeventy} & 1 \\ 
     & Holm-Bonferroni correction & \cite{\PaperOneHundredAndNineteen} & 1 \\ 
     & Ancillary qubit & \cite{\PaperOneHundredAndTwentyTwo} & 1 \\ 
    \cmidrule(lr){1-1} \cmidrule(lr){2-2} \cmidrule(lr){3-3} \cmidrule(lr){4-4} 
    DOO & Frequency-based criterion & \cite{\PaperFifteen, \PaperSeventyOne, \PaperSeventyEight} & 3 \\ 
    \cmidrule(lr){1-1} \cmidrule(lr){2-2} \cmidrule(lr){3-3} \cmidrule(lr){4-4} 
    QSO & Quantum fidelity & \cite{\PaperOneHundredAndFourteen, \PaperOneHundredAndFifteen, \PaperOneHundredAndSeventeen} & 3 \\ 
     & Quantum state tomography & \cite{\PaperOneHundredAndFourteen, \PaperOneHundredAndSeventeen} & 2 \\ 
     & Original outcome & \cite{\PaperOneHundredAndTwelve} & 1 \\ 
     & Bloch vector representation & \cite{\PaperOneHundredAndFifteen} & 1 \\ 
    \cmidrule(lr){1-1} \cmidrule(lr){2-2} \cmidrule(lr){3-3} \cmidrule(lr){4-4} 
    QOO & Quantum process tomography & \cite{\PaperOneHundredAndFourteen, \PaperOneHundredAndSeventeen} & 2 \\ 
     & Quantum fidelity & \cite{\PaperOneHundredAndFourteen, \PaperOneHundredAndSeventeen} & 2 \\
    \cmidrule(lr){1-4} 
    
    \multicolumn{4}{p{1.1\columnwidth}}{\textbf{Total number 
    of primary studies for each test oracle:} OPO (20), WOO (16), PBO (9), QSO (4), DOO (3), QOO (2)}\\
    \bottomrule[1pt]
\end{tabular}
        
    }
\end{table}

Table~\ref{tab: rq5_oracle_type} lists the test oracles along with their evaluation mechanisms employed in primary studies. Since we identify and summarize four new test oracles that have not been claimed in previous work, the identification of test oracle types is based on the definitions introduced above. This means that even if a certain test oracle was not explicitly defined or clarified in one study, it may still fall within one of our defined types. 

We found that, among the six test oracles,  the two proposed by the existing work~\cite{ali2021assessing} (i.e., OPO and WOO) were used more frequently than the other four. As for the family of OPOs, the evaluation mechanisms can be roughly categorized into hypothesis tests (e.g., Wilcoxon signed rank test) and distance measures (e.g., Jensen-Shannon divergence). Among all the evaluation mechanisms for OPOs, the majority of studies (i.e., 15) adopted Pearson's chi-squared test, which belongs to the \textit{nonparametric hypothesis tests}. 
Still, we found the employment of \textit{parametric hypothesis tests}, like G-test and multinomial test in~\cite{miranskyy2025on}.
The two parametric hypothesis tests are established upon assumptions like the counts of measurement outcomes following a multinomial distribution, and can also be categorized as \textit{one-sample statistical tests} that allow comparing test samples against the theoretically specified probability distribution, making them applicable to cases with small shot counts.
In addition, Miranskyy et al.~\cite{miranskyy2025on} introduced Monte Carlo variants of statistical tests to improve their robustness by repeatedly sampling from the expected distribution, thereby mitigating the sensitivity issues and the tendency of some tests to latch onto positive cases. 

Regarding the family of WOOs, most of the studies (i.e., 12) explicitly adopted the original outcomes from the PUT for comparison. This evaluation mechanism is simple and cost-efficient, but its effectiveness is confined to testing quantum programs with deterministic expected outputs. It can be observed that several primary studies proposed evaluation mechanisms for WOOs via extra quantum operations, such that WOOs can be extended to test more general quantum programs. 
Swap test newly introduces an ancilla qubit to store the comparison result between the expected and test output states, and then WOO can work on the sole ancilla qubit. 
Excluding the ancilla qubits but increasing the circuit depth, inverse test appends an inverse implementation of the quantum operation relevant to the expected CUT and checks whether the final quantum state returns the default or initial state, which indicate a passed test\footnote{To be objective, despite the ability of fault detection, the WOO using inverse testing could not always serve as a conventionally defined test oracle like those in CSE. As a side effect, the original outcomes of CUTs cannot be exposed to testers, because the inverse implementation alters the output quantum state. Furthermore, the comparison against the program specification is hardly explained in intuition.}. 

The evaluation mechanisms for PBOs show overlaps with those for OPOs and WOOs, such as the swap test and Pearson's chi-squared test, indicating that some evaluation mechanisms manifest generalization across test oracles. Especially, the study~\cite{\PaperOneHundredAndNineteen} used Holm-Bonferroni correction to confine the family-wise error rate when applying multiple statistical tests consecutively.

DOOs were considered in only three studies, and the evaluation mechanisms consistently regarded the values of measurement outcomes with the highest frequency (or probability) as dominant. 
For example, given ten measurement outcomes in decimal: 1, 2, 2, 3, 3, 3, 3, 4, 6, and 6, the value 3 is the dominant one.
The study~\cite{\PaperFifteen} introduced the DOO to ensure the reliability of testing single-output circuits upon noisy backends. The remaining two studies~\cite{\PaperSeventyOne, \PaperSeventyEight} were motivated by special quantum algorithms, where they adopted the DOO\footnote{The study~\cite{\PaperSeventyEight} classified such a test oracle under WOO. In our paper, however, we follow our proposed definition and refer to it as DOO to clearly distinguish it from the general WOO.} for the \textit{output-dominant algorithms}, the term that is defined by~\cite{\PaperSeventyEight} and indicates quantum algorithms that only care about the output with the highest probability (e.g., Grover Search and Variational Quantum Eigensolver). 

Quantum fidelity and quantum tomography were engaged in both QSO and QOO, where quantum tomography recovers quantum states or operations from the classical measurement outcomes, and quantum fidelity measures the distinction between recovered and expected quantum information. The study~\cite{miranskyy2025on} proposed the \textit{state-vector test} that straightforwardly leverages the output quantum state originally from the CUT, and the feasibility of the QSO using such an evaluation mechanism is only allowed by the ideal simulator built on state vectors. Instead of full quantum tomography with high overheads, the study~\cite{oldfield2025bloch} proposed a Bloch vector representation of output quantum states, thereby focusing on reconstructing individual qubits instead of joint output states.

\revise{Overall, when employing a test oracle, we suggest an evaluation on possible impact factors to analyze its effectiveness and cost, such as evaluation mechanisms and number of shots (more details can be found in Section~\ref{sec: shots}). For instance, a study~\cite{li2026dynamic} reported that the effectiveness of OPO with distance measures, such as the Jensen-Shannon divergence, strongly depends on the configured threshold. Moreover, the required shots for OPO would exponentially increase with qubit counts of the CUT, whereas such a trend may not hold true for WOO. Owing to their significance, failing to consider these impact factors threatens the validity of empirical evaluations.}
\subsubsection{Summary of RQ5}
\text{}

\Subtitle{Takeaway 5-1} We defined and identified specified, derived, and implicit program specifications employed in primary studies. The majority of the primary studies considered specified program specifications. Within this type, probability distributions and quantum states were the top two output types in terms of the number of papers adopting them.

\Subtitle{Takeaway 5-2} We summarized and listed six test oracles used in the primary studies, where the four (i.e., property-based oracle, dominant output oracle, quantum state oracle, and quantum operation oracle) were proposed in this paper, and the two (i.e., output probability oracle and wrong output oracle) were previously proposed by existing studies. The output probability oracle and the wrong output oracle were the two test oracles most commonly employed. Besides, primary studies adopted diverse evaluation mechanisms to execute particular test oracles, such as hypothesis tests, statistical measures, extra quantum operations, and quantum tomography techniques.

\section{\texorpdfstring{\titlecap{\GroupEval}}{Approach Evaluation}}
\label{sec: evaluation}

\subsection{\texorpdfstring{\RQSection{6}}{RQ6}}\label{sec: rq6}
In SE practices, evaluation metrics pose a significant impact on \textit{construct validity} of empirical studies, this terminology that refers to the adequacy of concept definitions and the representativeness of the indicators used for those concepts~\cite{sjoberg2022construct}. Different metrics could correspond to multiple perspectives of one concept, and it would introduce undesirable bias if the employed metrics cannot cover crucial perspectives of the concept. In the context of QST, reusing well-defined and appropriately applied metrics from existing studies can ensure robust construct validity for future research. To support subsequent studies in construct validity with methodological rigor, this RQ seeks to review how primary studies considered evaluation metrics and which metrics were used for particular SE problems.

\subsubsection{Considerations of Cost-effectiveness Analysis}
The concern about the cost-effectiveness trade-off also exists in QST, as both CST and QST share the purpose of software quality assurance. Recent studies~\cite{zhang2025quantum, zhang2025empirical} on quantum optimization for SE, within the scope of QSE, have underscored the trade-off between cost and effectiveness, as well as the importance of conducting empirical cost-effectiveness analyses of proposed approaches. These findings motivate our paper to investigate whether the QST studies have taken cost and effectiveness into account in terms of evaluation metrics.

\begin{figure}[!t]
    \begin{subfigure}[b]{0.98\columnwidth}
        \centering
        \includegraphics[width=0.25\columnwidth]{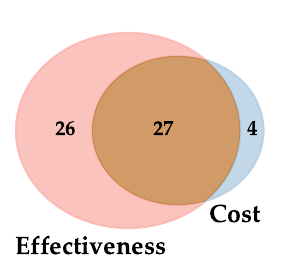}
    \end{subfigure}
    \caption{The number of primary studies adopting effectiveness or cost metrics}  
    \label{fig: rq6_baselines}
    \Description{A Venn diagram showing the consideration of effectiveness and cost metrics for primary studies}
\end{figure}

Figure~\ref{fig: rq6_baselines} displays a Venn diagram about the number of primary studies, where we remind that cost and effectiveness target the implementation of test approaches rather than the test process itself. There are 45.8\% (27 out of \NumberOfPrimaryStudies) primary studies that considered both cost and effectiveness metrics, the proportion exceeding the number of studies that focused solely on either objective. This reveals that cost-effectiveness analysis in existing QST studies has, to a certain extent, reached a consensus, which helps to build a comprehensive construct of empirical studies. For comparison within the two objectives, we found that effectiveness metrics were involved in many more studies than cost metrics, implying a research tendency to emphasize effectiveness over cost in approach evaluation. This is understandable, since a test approach, for instance, is of little value if it substantially compromises its fault-detection capability, even if it incurs very low execution cost.

\subsubsection{Evaluation Metrics for SE Problems}
\begin{table}[!t]
    \small
    \centering
    \caption{Effectiveness and cost metrics used for each SE problem, where for each case, only metrics that rank within the top five (ties included) in frequency are displayed}
    \label{tab: rq6_metrics_names}
    \resizebox{.98\columnwidth}{!}{
        
\begin{tabular}{c  p{0.46\columnwidth} p{0.38\columnwidth}}
    \toprule[1pt]
    \multicolumn{1}{c}{\textbf{SE problems}} & \multicolumn{1}{c}{\textbf{Effectiveness (\#)}} & \multicolumn{1}{c}{\textbf{Cost (\#)}} \\
    \cmidrule(lr){1-1} \cmidrule(lr){2-2} \cmidrule(lr){3-3}  
    Software testing & Percentage of failed tests (15),\newline Mutation score (10),\newline Accuracy (6),\newline Precision (5),\newline Recall (5). & Overall execution time (15),\newline Circuit simulation time (2),\newline Number of gates (2),\newline Number of shots (2),\newline Circuit depth (2),\newline Number of iterations (2). \\ 
    \cmidrule(lr){1-1} \cmidrule(lr){2-2} \cmidrule(lr){3-3} 
Runtime assertion & Success rate in identifying errors (1),\newline Reduction rate for the number of lines (1),\newline F1 score (1),\newline Fidelity of quantum states (1),\newline False positives (1),\newline Success rate of program execution (1). & Number of gates (3),\newline Number of qubits (3),\newline Number of measurement operators (2),\newline Number of shots (1),\newline Overall execution time (1),\newline Circuit depth (1). \\ 
    \cmidrule(lr){1-1} \cmidrule(lr){2-2} \cmidrule(lr){3-3} 
Fault localization & Localization rate (1),\newline Probability of successfully locating a bug (1),\newline Mutation score (1),\newline Percentage of statements before finding faults (1). & Number of gates (1),\newline QPU execution time (1). \\ 
    \cmidrule(lr){1-1} \cmidrule(lr){2-2} \cmidrule(lr){3-3} 
Program repair & Repair rate (3),\newline Distance between the repaired and correct programs (1),\newline Approximation accuracy (1),\newline Number of successful runs (1),\newline Number of equivalent patches (1),\newline Error of similarity (1),\newline Error of unitarity (1). & Overall execution time (2),\newline Repair time (1),\newline Number of gates (1),\newline Test generation time (1),\newline Localization time (1). \\
    \bottomrule[1pt]
\end{tabular}
        
    }
\end{table}

In this part, we delve into metrics appropriate for specific SE problems relevant to QST. Table~\ref{tab: rq6_metrics_names} presents the top five (including ties) effectiveness and cost metrics for each SE problem, where the complete list of metrics can be seen in our artifact~\cite{li_2026_18159893}. During the data annotation, we consolidated metrics with similar semantics or complementary meanings to reduce the number of candidates, such as both ``number of failed tests'' and ``percentage of passed tests'' being normalized into a unified metric ``percentage of failed tests''.

To begin with, we discuss software testing, as the majority of the primary studies fall within this scope. The two most used effectiveness metrics are the percentage of failed tests and mutation score, where both metrics are explicitly associated with the quantity of identified faults, while the latter is tailored for mutation analysis. 
The other three (i.e., accuracy, precision, and recall) are generally responsible for testing-related classification tasks, such as testing quantum machine learning models for image classification (e.g.,~\cite{shi2025quantest}) and evaluation of test oracles' ability to give correct test results (e.g.,~\cite{\PaperThirtyNine}). 
\revise{It is important to ensure that evaluation metrics are aligned with the research objectives, such as the fact that great values of the percentage of failed tests or mutation score cannot explicitly reflect the test oracle effectiveness, since simply using such metrics do not account for whether the reported failures correspond to actual faults or are incorrectly flagged due to inaccuracy of the test oracle (e.g., false positives).}

The majority of studies employed execution time for software testing within QST. We found that two studies~\cite{\PaperThree, \PaperSeventeen} reported circuit simulation time separate from the overall cost of the test process, which suggests a potential need to independently investigate components within the whole test process. 
\revise{The diversity in measuring time cost also highlights the importance of clearly reporting how such metrics are derived from test execution, thereby enabling fair, accurate, and reproducible comparisons across studies.}
In addition, we identified quantum-specific metrics to measure the test cost, in terms of aspects like quantum gates, shots, and depth. Different from those metrics investigated in Section~\ref{sec: rq3: circuit_scale}, the three in this RQ are treated as dependent variables to quantify complexities of the entire quantum circuits or other operations than CUTs, where these complexities result from executing the proposed test approach. 

As for the other three SE problems, whose metric pools are relatively shallow due to the scarcity of corresponding primary studies, we extracted some problem-specific metrics (e.g., localization and repair rate respectively for fault localization and program repair) along with approach-specific metrics (e.g., percentage of statements before finding faults~\cite{\PaperOneHundredAndNine} and error of similarity~\cite{li2024automatic}). Most of such exclusive metrics are used to evaluate the effectiveness of proposed approaches, while the cost metrics are more likely to be adapted to general approaches or problems, such as execution time and circuit complexity, which have already been discussed in the prior part of software testing. 

In the end, we offer an overview that almost all metrics, specific to QC but not limited to certain approaches and problems, are relevant to cost. Especially, from the perspective of test implementation upon a quantum computing platform, we should distinguish circuit simulation time and QPU execution time from the overall execution time, as suggested by a related study~\cite{zhang2025empirical}. Learned from the work on runtime assertion, where the studies~\cite{\PaperFourteen, \PaperOneHundredAndFifteen, \PaperOneHundredAndTwentyOne} introduced additional quantum components to construct assertions, we should also be aware of the testing-induced overhead in the quantum circuit, in addition to the cost of the CUT subroutines.

\subsubsection{Summary of RQ6}
\text{}

\Subtitle{Takeaway 6-1} There are 45.8\% of primary studies that discussed both effectiveness and cost metrics in their empirical studies. For comparison in terms of the counts of primary studies, effectiveness metrics have attracted more attention than cost metrics.

\Subtitle{Takeaway 6-2} We found that several metrics used in the primary studies were tailored to a specific SE problem or a particular proposed approach. Metrics pertaining to detected faults and execution time were commonly used in primary studies. Besides, we observed that most general metrics, without restriction to certain approaches or problems and with specificity to QC, predominantly demonstrate the relevance to cost, such as execution time for quantum subroutines and complexity measures for quantum circuits.

\subsection{\texorpdfstring{\RQSection{7}}{RQ7}}\label{sec: rq7}
In SE research, the presence of a valid baseline gives a common basis of comparison, facilitating the repeatability and improvement of a wide variety of SE experiments~\cite{gay2010baseline}. Regarding QST, a research area in its infancy, baselines can provide a starting goal for comparably evaluating the newly developed testing approaches. Given the importance, this RQ investigates how and what baselines were considered in the primary studies. Also, we further delve into details like the motivations for including specific baselines and the statistical rigor of the comparative evaluations among approaches.
 
\subsubsection{Considerations for Baselines}
\begin{figure}[!t]
    \centering
    \begin{subfigure}[b]{0.54\columnwidth}
        \centering
        \includegraphics[width=0.75\columnwidth]{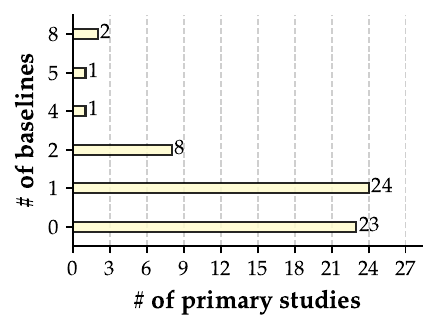}
        \caption{Number of the baselines adopted in each primary study}  
        \label{fig: rq7_number_of_baselines}
    \end{subfigure}
    \hfill
    \begin{subfigure}[b]{0.44\columnwidth}
        \centering
        \includegraphics[width=0.7\columnwidth]{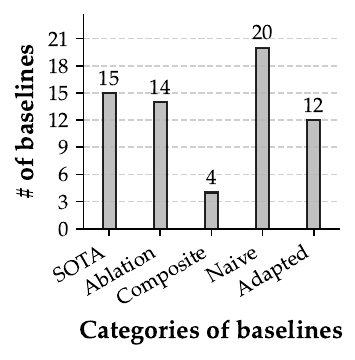}
        \caption{Distribution of the baseline categories}  
        \label{fig: rq7_inclusion_motivation}
    \end{subfigure}
    \Description{A composite figure, where the left histogram illustrates the number of baselines adopted in the primary studies, and the right bar chart visualizes the distribution of baseline categories.}
    \caption{Quantity statistics for the baselines adopted in the primary studies}  
    \label{fig: rq7_baselines}
\end{figure}
Observed from Figure~\ref{fig: rq7_number_of_baselines}, there are 36 (2+1+1+8+24) primary studies incorporating at least one baseline in their experiments, where two-thirds of studies (24 out of 36) included only one baseline. 
\revise{Among the 23 primary studies without any baselines, we found that some merely presented a proof-of-concept demonstration to preliminarily show that their approaches could work (like~\cite{\PaperTwentySeven, \PaperThirtyFive}); some pioneered a research track within QST, which had been mostly unexplored (like~\cite{\PaperTwelve} for statistical assertions and~\cite{\PaperFiftyOne} for program repair); and some of the others focused on tooling usability rather than approach advantages (such as~\cite{\PaperOne, \PaperThree}).}
\revise{Generally, one plausible explanation for such a still-developing practice of baseline adoption} is the scarcity of established approaches within each specific track, in spite of sharing the scope of QST. For instance, only two studies~\cite{\PaperFive, \PaperSixty} investigated combinatorial testing, implicating a shallow pool of viable QST baselines in this track. Borrowing baselines from other tracks with different testing assumptions and purposes, such as mutation testing or fuzz testing, may not be methodologically sound.

Motivated by this current state, we aim to identify how existing primary studies determined their starting point for baseline inclusion, which in turn can help guide future empirical research. In the following, we define five baseline categories based on the primary studies:
\begin{itemize}[leftmargin=*]
    \item \textit{SOTA (State Of The Art) baselines}: Indicate the newest and most advanced approaches, typically drawn from the latest relevant studies in a particular field. They typically serve as strong comparative baselines to demonstrate the advantage of the newly developed approach.
    \item \textit{Ablation baselines}: Refer to variants introduced within the primary study itself for conducting \textit{ablation studies}~\cite{meyes2019ablation}, which aim to isolate the contribution of specific components included in the proposed approach.
    \item \textit{Naive baselines}: Stand for the simplest approaches within the same track, such as fully random or heuristic-free approaches. Outperforming a naive baseline is a prerequisite for justifying the introduction of more advanced but complex approaches.
    \item \textit{Adapted baselines}: Specify the approaches or techniques adapted from multiple domains or communities, owing to the interdisciplinarity of QSE, but not being employed in existing studies within the same track.
    \item \textit{Composite baselines}: Denote synthesized approaches that combine components from SOTA, naive, or adapted methods within the corresponding tracks. Composite baselines could provide extensive comparative evaluation for relatively sophisticated techniques.
\end{itemize}

Figure~\ref{fig: rq7_inclusion_motivation} visualizes the baseline distribution based on the above-defined categories, where the counts are not derived from a deduplicated set of baselines but explicitly collected from each study. We found that the most frequently used type was naive baselines with 20 samples, suggesting that a non-negligible proportion of existing QST studies in certain tracks may still be preliminary in nature. In addition, the numbers of SOTA, ablation, and adapted baselines are observed to approximate, which could reveal the scattered diversity of current baselines. Four composite baselines merely appeared in one study~\cite{tan2025hornbro} on program repair. The motivation for including such baselines is reasonable and inspirational, because program repair is associated with multiple phases like fault localization, patch generation, and test execution, such that it is natural to integrate competitive approaches designed for each phase.

\subsubsection{Specific Baselines for Software Engineering Problems and Defined Categories}

\begin{table}[!t]
    \small
    \centering
    \caption{Baselines adopted in the primary studies, which could be considered in future studies}
    \label{tab: rq7_baseline_names}
    \resizebox{.98\columnwidth}{!}{
        
\begin{tabular}{c  c  p{0.7\columnwidth}}
    \toprule[1pt]
    \multicolumn{1}{c}{\textbf{SE problems}} & \multicolumn{1}{c}{\textbf{Categories}} & \multicolumn{1}{c}{\textbf{Baselines (Correspondings primary studies)}} \\
    \cmidrule(lr){1-1} \cmidrule(lr){2-2} \cmidrule(lr){3-3}  
    Software testing & SOTA & QuraTest (\cite{\PaperFortyFive, \PaperOneHundredAndFive}), SimBAy (\cite{\PaperThirtySix}), ETO (\cite{\PaperThirtyNine}), Quito (\cite{\PaperFortyFive}), Pure-state test cases (\cite{\PaperFortySix}), Pearson's chi-squared test (\cite{\PaperOneHundredAndTwelve}), QMutPy (\cite{\PaperOneHundredAndThirteen}), Muskit (\cite{\PaperOneHundredAndThirteen}), QSharpCheck (\cite{\PaperOneHundredAndNineteen}) \\ 
     & Naive & Random testing (\cite{\PaperFive, \PaperFiftyTwo, \PaperSixty, \PaperOneHundredAndSeven}), Random search (\cite{\PaperSeventeen, \PaperTwenty, \PaperTwentyOne, \PaperFortyThree}), Random state vector generator (\cite{\PaperFortyFive, \PaperOneHundredAndThree}), Random circuit generator (\cite{\PaperSix}), Random coherent noise (\cite{\PaperThirtySix}), SAMPLE (\cite{\PaperThirtyEight}), Random quantum circuit generator (\cite{\PaperFortyFive}), Noise prediction model based on linear regression (\cite{\PaperSixtySeven}) \\ 
     & Adapted & Adaptive random testing (\cite{\PaperFiftyTwo}), Tradifuzz (\cite{\PaperEightySix}), Swap test (\cite{\PaperOneHundredAndTwelve}), Statevector test (\cite{\PaperOneHundredAndTwelve}), G-test (\cite{\PaperOneHundredAndTwelve}), Multinomial test (\cite{\PaperOneHundredAndTwelve}), Monte Carlo Pearson's chi-squared test (\cite{\PaperOneHundredAndTwelve}), Monte Carlo G-test (\cite{\PaperOneHundredAndTwelve}), Monte Carlo multinomial test (\cite{\PaperOneHundredAndTwelve}) \\ 
    \cmidrule(lr){1-1} \cmidrule(lr){2-2} \cmidrule(lr){3-3} 
    Runtime assertion & SOTA & QECA (\cite{\PaperFourteen, \PaperOneHundredAndTwentyOne}), Stat (\cite{\PaperFourteen}), Proq (\cite{\PaperOneHundredAndFifteen}) \\ 
    \cmidrule(lr){1-1} \cmidrule(lr){2-2} \cmidrule(lr){3-3} 
    Fault localization & Naive & Naive linear search (\cite{\PaperThirtyOne, \PaperEightyFour}), Naive binary search (\cite{\PaperThirtyOne, \PaperEightyFour}) \\ 
     & Adapted & Spectrum-based fault localization (\cite{\PaperOneHundredAndNine}) \\ 
    \cmidrule(lr){1-1} \cmidrule(lr){2-2} \cmidrule(lr){3-3} 
    Program repair & SOTA & LLM-QAPR (\cite{\PaperFortyFour}) \\ 
     & Composite & Syn-QFST (\cite{\PaperFortyFour}), Syn-QSD (\cite{\PaperFortyFour}), Quito (\cite{\PaperFortyFour}), QuSBT (\cite{\PaperFortyFour}) \\     & Naive & Basic-QAPR (\cite{\PaperFortyFour}) \\     & Adapted & GenProg (\cite{\PaperEightyThree}), TBar (\cite{\PaperEightyThree}) \\
    \cmidrule(lr){1-3}
\multicolumn{3}{p{\columnwidth}}{\textbf{Number of deduplicated baselines:} \newline
    Software testing (26): SOTA (9), Naive (8), Adapted (9)\newline
    Runtime assertion (3): SOTA (3)\newline
    Fault localization (3): Naive (2), Adapted (1)\newline
    Program repair (8): SOTA (1), Composite (4), Naive (1), Adapted (2)}\\
    \bottomrule[1pt]
\end{tabular}
        
    }
\end{table}

To instantiate the baselines previously employed, Table~\ref{tab: rq7_baseline_names} summarizes the baselines corresponding to the defined categories and the SE problems. We do not discuss baselines for ablation studies here, as such baselines are usually tailored to specific approaches and are rarely reusable for evaluating other approaches.

From the perspective of SE problems, 26 baselines are identified for software testing, and their category distribution for this SE problem shows an almost uniform pattern. In comparison, the employed baselines for runtime assertion and fault localization are relatively few. Despite only 3 studies on program repair (referred to Figure~\ref{fig: bib_se_problems}), 8 different baselines have been explored. This count should be largely contributed to by the composite baselines, ascribed to the inherent complexity of program repair. For example, QuSBT is an SOTA search-based approach for test case generation~\cite{wang2022qusbt}, and QuSBT was incorporated in~\cite{tan2025hornbro} as a baseline by replacing only the test-case generation stage within the discussed pipeline of program repair.

In the view of baseline categories, the largest number of SOTA baselines were used for software testing across the four problems, with 9 in total. Their target tracks include test case generation (e.g., QuraTest~\cite{ye2024quratest}, Quito~\cite{wang2022quito} and Pure-state test cases), test oracle problem (e.g., ETO and Pearson's chi-squared test), and mutation testing (e.g., QMutPy~\cite{fortunato2022qmutpy} and Muskit~\cite{mendiluze2022muskit}). The adapted baselines are borrowed from approaches or techniques of fields including CSE (e.g., adaptive random testing~\cite{chen2004adaptive} and spectrum-based fault localization~\cite{abreu2007accuracy}), QC (e.g., swap test~\cite{buhrman2001quantum}), and statistics (e.g., Monte Carlo multinomial test~\cite{bourguignon2007selection}). The naive baselines still gain attention in SE problems, except for runtime assertion. The above findings convey that, although the latest SOTA baselines are expected to be the most competitive for the subsequent approaches, comparing with naive baselines remains necessary for most of the discussed SE problems. For example, several systematic empirical studies on testing classical traditional software~\cite{arcuri2011adaptive} and classical intelligent software~\cite{mazouni2024policy} have argued that, under the circumstances of evaluating generalization, SOTA baselines widely followed by papers in high-tier venues may underperform simpler or naive baselines in consideration of cost and effectiveness.

\subsubsection{Statistical Tests for Contrastive Analysis}
The uncertainty of quantum measurements enables running the same QST experiments to produce varied test results under a certain metric. Analogously, research on \textbf{Search-Based Software Engineering (SBSE)} has addressed how to evaluate randomized heuristic algorithms convincingly. The SBSE community commonly accepts statistical tests to assess whether there is enough empirical evidence to claim a difference between experimental results from two or more randomized algorithms~\cite{arcuri2011practical}. Since statistical tests can provide a theoretical support for assessing the significance of experimental differences, especially for those that are not notable, we study how statistical tests are adopted in QST for approach comparison.

\begin{table}[!t]
    \small
    \centering
    \caption{Statistical approaches used for the comparison with baselines}
    \label{tab: rq7_baseline_comparison}
    \resizebox{.98\columnwidth}{!}{
        
\begin{tabular}{p{0.36\columnwidth}  p{0.62\columnwidth}}
    \toprule[1pt]
    \multicolumn{1}{c}{\textbf{Statistical tests (\#)}} & \multicolumn{1}{c}{\textbf{Associated statistics (\#)}} \\
    \cmidrule(lr){1-1} \cmidrule(lr){2-2}  
    Mann-Whitney U test (8) & $p$-value (8), Vargha and Delaney's $\hat{A}_{12}$ statistics (7), Magnitude (2) \\ 
    Fisher's exact test (2) & Odds ratio (2), $p$-value (1) \\ 
    Kruskal-Wallis test (1) & $p$-value (1) \\ 
    Spearman's rank correlation test (1) & Spearman rank coefficient (1) \\ 
    Wilcoxon signed-rank test (1) & $p$-value (1), Cliff's $\delta$ (1) \\
    \cmidrule(lr){1-2}
    \multicolumn{2}{p{\columnwidth}}{\textbf{Total number of primary studies:} 9}\\
    \bottomrule[1pt]
\end{tabular}
        
    }
    {\justify
        {\selectfont
            ``\#'' indicates the number of corresponding primary studies.
        }
    \par}
\end{table}

Observing the statistical tests illustrated in Table~\ref{tab: rq7_baseline_comparison}, merely 9 primary studies utilized statistical tests for approach comparison. All the employed statistical tests belong to nonparametric hypothesis tests that make minimal assumptions about the underlying distribution of the studied data. Mann-Whitney U test was the most frequently used in 8 primary studies, where its popularity manifests the consistency with the SBSE community~\cite{arcuri2011practical}. Regarding the usage for empirical studies, both Mann-Whitney U test and Wilcoxon signed-rank test can be employed to compare two test approaches, while Kruskal-Wallis test aims at the comparison among three approaches. Given that Fisher’s exact test is well-suited for binary outcomes, particularly in small-sample settings, two studies~\cite{\PaperFive, \PaperSixty} applied it to compare the proportions of passed versus failed test suites produced by combinatorial testing and random testing. Compared to simply identifying the difference between approaches, Spearman's rank correlation test is inclined for assessing their associations, such as~\cite{oldfield2025faster} that employed this test to reveal the correlation for a group of experimental metrics. 

About the statistics, $p$-value was the most involved. This is because $p$-value plays a necessary role in determining whether to reject or accept the null hypothesis $\text{H}_0$ in the involved statistical tests, where $\text{H}_0$ is usually defined to state no difference being observed. Also, the primary studies adopted the effect size measures, including Vargha and Delaney's $\hat{A}_{12}$ statistics, odds ratio, along with Cliff's $\delta$, to quantify the strength of statistical significance and even indicate which method tended to outperform the other. There are two studies~\cite{\PaperFortyThree, \PaperSixty} using magnitudes to further depict effect sizes across discrete categories. For example, the magnitude for $\hat{A}_{12}$ statistics is defined to indicate negligible, small, medium, or large differences, with each level corresponding to a distinct and non-overlapping range of $\hat{A}_{12}$.

\subsubsection{Summary of RQ7}
\text{}

\Subtitle{Takeaway 7-1} Around 61.0\% (36 out of~\NumberOfPrimaryStudies) primary studies considered at least one baseline. Among the adopted baselines, the most used are naive approaches with the simplest mechanisms. Besides, we identified moderate numbers of SOTA baselines (15), ablation studies (14), and approaches adapted from various communities (12), as well as a small number (i.e., 4) of synthesized approaches serving as composite baselines. 

\Subtitle{Takeaway 7-2} Software testing incorporates the largest number of SOTA baselines among the four fine-grained SE problems, and the tracks where these SOTA baselines are applicable manifest diversity. The composite baselines have been adopted only for program repair. Besides, the naive approaches have still been considered moderately in SE problems except for runtime assertion, indicating their non-trivial role in contrastive analysis.

\Subtitle{Takeaway 7-3} Only 9 primary studies explored 5 nonparametric hypothesis tests for experimental contrastive analysis, where Mann-Whitney U test was the most frequently used. The statistics associated with the employed statistical tests primarily included $p$-value, effect size measures, and magnitude of effect sizes.

\section{\texorpdfstring{\titlecap{\GroupExp}}{Experimental Configurations And Resources}}
\label{sec: settings}

\subsection{\texorpdfstring{\RQSection{8}}{RQ8}}\label{sec: rq8}
As one of the vital indicators to analyze the threats to validity of empirical studies, \textit{conclusion validity} is related to the degree to which statistical conclusions about the relationships among variables are reasonable. The statistical repetitions have been widely discussed in the QSE community~\cite{zhang2025empirical}. One concern arises from the probabilistic nature of QC, which could bring uncertainty to the results of experiments. However, with practical considerations for computational cost, we cannot rely on an extremely large number of statistical repetitions to mitigate randomness. 

\begin{figure}
    \centering
    \includegraphics[width=0.78\textwidth]{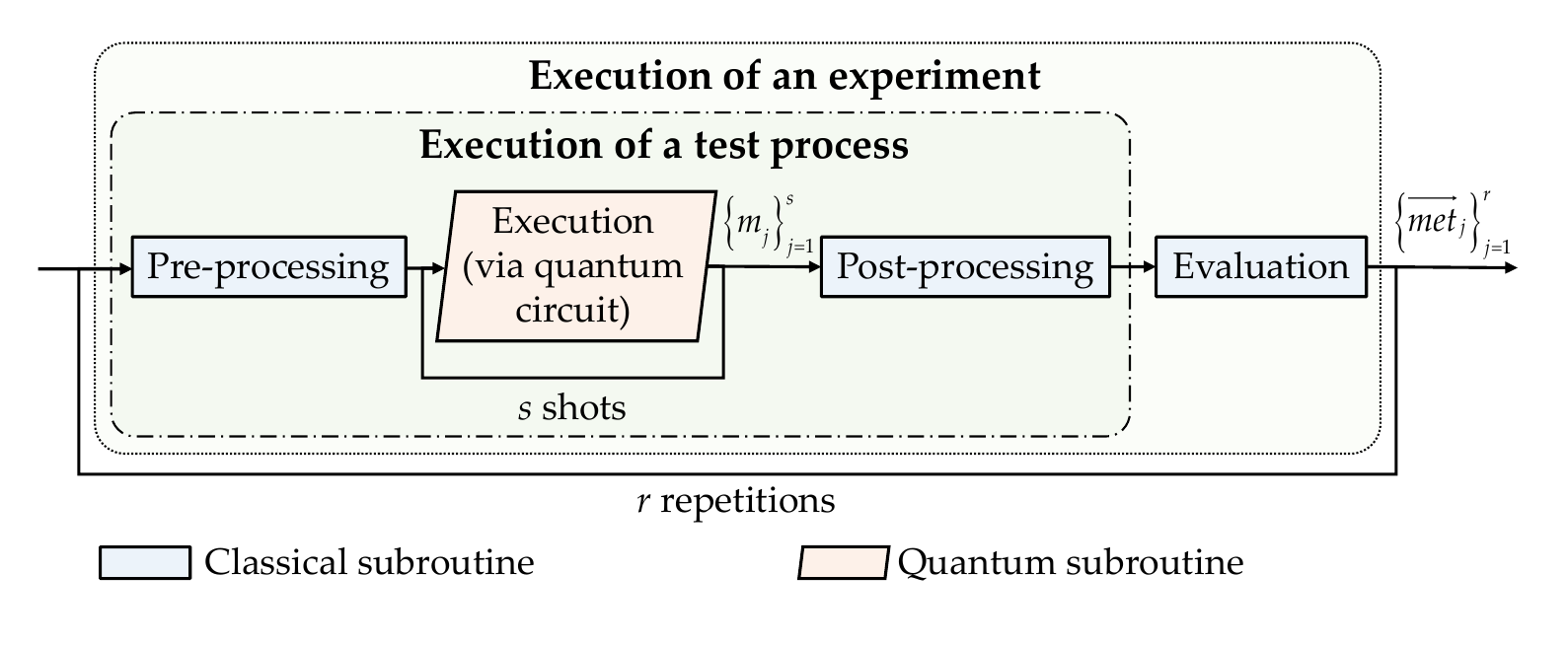}
    \caption{The roles of statistical repetitions (i.e., shots and experimental repetitions) in QST experiments}
    \Description{A framework that can differentiate shots from experimental repetitions}
    \label{fig: repeats}
\end{figure}

Hence, we intend to revisit primary studies and present an overview of current schemes to the experimental configurations of statistical repetitions. More particularly, this RQ zooms into two important experimental configurations, i.e., number of shots (denoted as $s$) and number of experimental repetitions (signified as $r$), which statistically impact the conclusion validity of QST experiments. Figure~\ref{fig: repeats} illustrates their heterogeneous roles in experiments. From the perspective of the test process, the shots indicate $s$ measurement outcomes $\{m_{j}\}_{j=1}^{s}$ output from the quantum circuit and serve as one source of statistical variation in the evaluation. Meanwhile, in view of the overall experiment, the inclusion of experimental repetitions is performed to ensure that the metric vectors $\{\overrightarrow{met_j}\}_{j=1}^r$ (i.e., potential multiple metrics), obtained from $r$ implementations of the same test process, are statistically significant. In a nutshell, the statistical uncertainty observed in the evaluation metrics is propagated by the probabilistic nature of quantum measurements and any randomized strategies used in the test process (e.g., random selection of test cases). 
 
\subsubsection{Shots}\label{sec: shots}
\begin{figure}[!t]
    \centering
    \begin{subfigure}[b]{0.49\textwidth}
        \centering
        \includegraphics[width=0.72\textwidth]{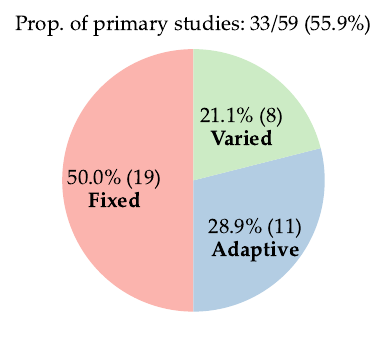}
        \caption{Distribution for shot configuration categories}  
        \label{fig: rq8_shot_config_types}
    \end{subfigure}
    \hfill
    \begin{subfigure}[b]{0.49\textwidth}
        \centering
        \includegraphics[width=0.85\textwidth]{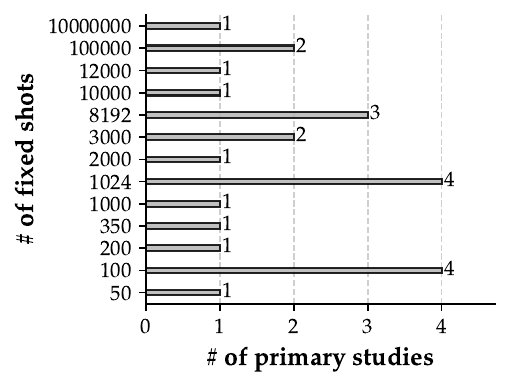}
        \caption{Number of shots for the fixed configurations}  
        \label{fig: rq8_fixed_shots}
    \end{subfigure}
    \caption{Visualization for shot counts configured in the primary studies}
    \Description{The figure consists of two subfigures illustrating shot configurations used in the primary studies. 
    The left subfigure shows the distribution of different shot configuration categories.
    The right subfigure presents the distribution of the number of shots used in experiments with fixed shot configurations.}
    \label{fig: rq8}
\end{figure}

Based on the primary studies, we observed that some experiments employed specific shot configurations rather than maintaining a fixed number of shots throughout the evaluation. Accordingly, we define the following categories to provide a more comprehensive overview of shot configuration strategies:
\begin{itemize}[leftmargin=*]
    \item \textit{Fixed configuration of shots}: Indicates the configuration that adopts a fixed number of shots throughout one experiment.
    \item \textit{Varied configuration of shots}: Denotes the configuration that attempts multiple shot counts in one experiment, where the shot count is treated as a configurable independent variable.
    \item \textit{Adaptive configuration of shots}: Refers to the configuration whose number of shots in one experiment is adaptive to available arguments or predefined criteria. Unlike the varied configuration of shots, the shot count for the adaptive configuration of shots is deemed a dependent variable.
\end{itemize}

Figure~\ref{fig: rq8_shot_config_types} displays the distribution of our defined categories for shot configurations, where 33 primary studies reported how shots were configured in experiments. Half as well as the majority of studies explicitly employed fixed numbers of shots, while varied and adaptive shots were still considered in smaller portions of the literature.

Observed from Figure~\ref{fig: rq8_fixed_shots} that shows specific shot counts for the fixed configurations, we found that the range spans from a small number (i.e., 50~\cite{\PaperFiftyTwo}) to a rather great number (i.e., 10,000,000~\cite{\PaperOneHundredAndTwelve}). There are 13 values for fixed shots, and each value is considered by no more than four studies, indicating a scattered distribution. Hence, it appears unsound to suggest a common configuration for fixed shots based on the primary studies. 

\begin{table}[!t]
    \small
    \centering
    \caption{Specific configuration schemes for adaptive and varied shots}
    \label{tab: rq8_varied_and_adaptive_shots}
    \resizebox{.98\textwidth}{!}{
        
\begin{tabular}{c  p{0.75\columnwidth} p{0.2\columnwidth}}
    \toprule[1pt]
    \multicolumn{1}{c}{\textbf{Categories}} & \multicolumn{1}{c}{\textbf{Configuration schemes}} & \multicolumn{1}{c}{\textbf{Primary studies}} \\
    \cmidrule(lr){1-1} \cmidrule(lr){2-2} \cmidrule(lr){3-3}  
    Adaptive & The shot count determined by the number of possible outputs for a specific input multiplied by 100 & \cite{\PaperFive, \PaperSeventeen, \PaperTwenty, \PaperTwentyOne, \PaperFiftyTwo} \\ 
     & The shot count configured as $10^4$ for programs with 4 or 5 qubits and $10^5$ for programs with 6 or 7 qubits & \cite{\PaperSix} \\ 
     & The minimum shots calculated by the selected algorithms and also associated with the testing scenarios & \cite{\PaperThirtyEight} \\ 
     & The shot count determined by the rank of program specification multiplied by 10 & \cite{\PaperFortyThree} \\ 
     & The minimum shots determined by an estimated lower bound & \cite{\PaperSeventy} \\ 
     & The minimum shots determined by statistics of Pearson's chi-squared test and increased by 100 per step up to 100000 in accordance with the early determination & \cite{\PaperEightyFour} \\ 
     & The minimum shots determined by the Quantum Chernoff bound & \cite{\PaperOneHundredAndTwelve} \\ 
    \cmidrule(lr){1-1} \cmidrule(lr){2-2} \cmidrule(lr){3-3} 
    Varied & From $10^1$ to $10^5$ & \cite{\PaperThirtySix} \\ 
     & From $2^3$ to $2^{10}$ & \cite{\PaperFortySix} \\     & Proportional to the lower bound ranging from 0.05 to 1.0 & \cite{\PaperSeventy} \\     & From $10^0$ to $10^3$ & \cite{\PaperOneHundredAndSix} \\     & From $10^2$ to $10^5$ & \cite{\PaperOneHundredAndSeven} \\     & From $10^0$ to $10^4$ & \cite{\PaperOneHundredAndTwelve} \\     & From $10^0$ to $10^5$ & \cite{\PaperOneHundredAndSeventeen} \\     & From 12 to 3200 & \cite{\PaperOneHundredAndNineteen} \\
    \bottomrule[1pt]
\end{tabular}
        
    }
\end{table}

Table~\ref{tab: rq8_varied_and_adaptive_shots} lists the configuration schemes for adaptive and varied shots in detail. Considering configurations for varied shots first, we found that many of the primary studies varied their shot counts exponentially, such as from $10^0$ to $10^5$ in~\cite{\PaperOneHundredAndSeventeen}. Relative to linear changes, exponentially increasing the number of shots covers a far wider range, and thus more easily reveals threshold cases, like the minimum shot count needed for convergence of specific metrics in probability. Furthermore, we concluded that the motivations for these studies adopting varied shots were mainly to present empirical results of the shots' impact on test results or to evaluate the sensitivity of the proposed approaches to shots. Such considerations are methodologically valuable, as they strengthen the \textit{internal validity} of empirical studies. 

As for adaptive configurations, we found that multiplying the number of possible outputs for a given input by 100 was the most popular scheme, adopted by 5 of the discussed 11 studies. Generally, the adaptive configuration schemes can be roughly divided into two sorts. The first sort determines the actual shot counts heuristically, including 7 studies~\cite{\PaperFive, \PaperSeventeen, \PaperTwenty, \PaperTwentyOne, \PaperFiftyTwo, \PaperSix, \PaperFortyThree}. It can be observed that 6 studies had the shot counts proportional to the number of possible outputs or the \textit{rank of the program specification}. As a novel concept proposed in~\cite{oldfield2025faster}, the latter refers specifically to the number of outcomes with non-zero probability under the program specification, which differs from the number of possible outcomes determined by the qubit count (i.e., $n$ qubits corresponding to $2^n$ possible outcomes). Compared with the first sort, the other sort depends on the bounds obtained from rigorous mathematical derivations or strong statistical evidence. However, this sort of scheme can hardly recommend a highly precise or even deterministic shot count, but can provide an asymptotic or probabilistic guarantee on the range of shots. Besides, for the four studies~\cite{\PaperThirtyEight, \PaperSeventy, \PaperEightyFour, \PaperOneHundredAndTwelve} within the second sort, we observed that they all intended to derive the minimum shots required for a sufficient and reliable evaluation. Since the bound derivation is approach-specific and involves sophisticated mathematical principles, we suggest that readers interested in further details refer to the corresponding studies.
 
\subsubsection{Experimental Repetitions}
\begin{figure}[t!]
    \centering
    \includegraphics[width=0.78\textwidth]{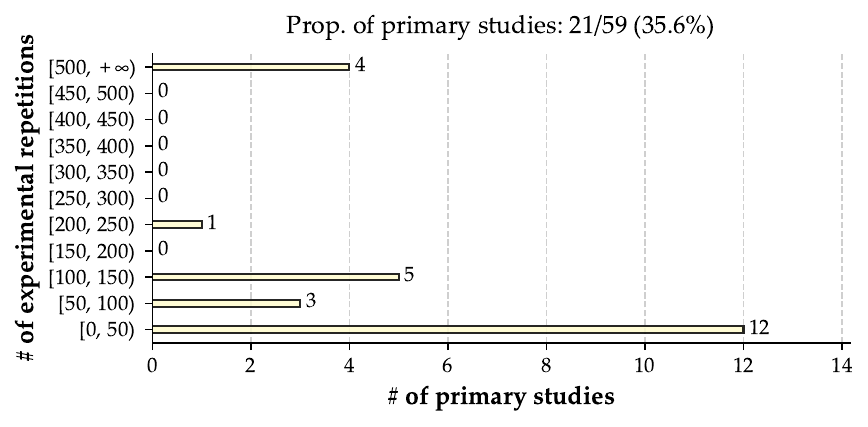}
    \caption{Histogram about experimental repetitions configured in experiments} 
    \Description{A histogram illustrates the number of experimental repetitions adopted in primary studies.}
    \label{fig: rq8_number_of_repetitions}
\end{figure}

Figure~\ref{fig: rq8_number_of_repetitions} presents a histogram pertaining to the number of experimental repetitions employed in primary studies, where we found that only 35.6\% of primary studies mentioned this configuration. Concerning the studies offering valid configurations, 57.1\% (12 out of 21) of these studies set the experimental repetitions below 50. We should note that although prior work in CSE recommends running each randomized algorithm at least 1,000 times for a reliable evaluation~\cite{arcuri2011practical}, applying the same criterion to QST is considerably more challenging. We found that only two studies~\cite{\PaperSixtySeven, \PaperOneHundredAndTwelve} configured 1,000 repetitions, the highest number among all of these involved primary studies. One explanation is the concern with cost, since numerous shots incur execution costs from running quantum circuits repeatedly, and the repetitions across the entire experiment further multiply the overall cost. 

In addition, we observed that three studies~\cite{\PaperThirtySix, \PaperFortyThree, \PaperOneHundredAndTwelve} included multiple configurations of experimental repetitions. For instance, the study~\cite{\PaperFortyThree} set 30 and 100 repetitions, respectively, for two different experiments, where the larger number was motivated by the fact that the maximum number of objective-function evaluations only grew quadratically with the problem size. By adjusting the number of repetitions to the (estimated) experimental complexity, the trade-off between execution cost and result reliability can be achieved to some extent.

\subsubsection{Summary of RQ8}
\text{}

\Subtitle{Takeaway 8-1} The fixed configuration of shots is considered in the majority of primary studies (i.e., 50.0\%), while the varied and adaptive configurations are also involved. The choices of fixed shots were scattered, making it hard to identify a common preference. For the varied configuration, many studies varied their shot counts exponentially. As for the adaptive configuration, we found that the adjustment of shot counts in these related studies followed either heuristic intuitions that provided specific counts or mathematical arguments that derived bounds on the minimum required shots.
 
\Subtitle{Takeaway 8-2} Only 35.6\% of primary studies reported experimental repetitions, suggesting that more attention is required for this configuration. Besides, more than half of the discussed primary studies (i.e., 12 out of 21) employed below 50 experimental repetitions, and the maximum repetitions contributed by only two studies were 1,000, which could showcase latent inconsistency between considerations for QST and conventions for classical SBSE. 

\subsection{\texorpdfstring{\RQSection{9}}{RQ9}}\label{sec: rq9}
The actual implementation of QST is strongly constrained by the properties of the underlying backends that should be associated with the hardware resources. Recently, the most common backends discussed in the community include classical ideal simulator, classical noisy simulator, and quantum physical hardware. In the current NISQ era, both the noisy simulator and physical hardware should account for the potential impact of quantum noise, which demonstrates a significant distinction from CST.
Given the significant attention to backend selection in QSE research~\cite{murillo2025quantum}, we propose this RQ to investigate the choices of primary studies regarding this issue.

\subsubsection{Classical Simulator versus Quantum Hardware}

\begin{figure}[!t]
    \centering
    \begin{subfigure}[b]{0.485\textwidth}
        \centering
        \includegraphics[width=0.98\textwidth]{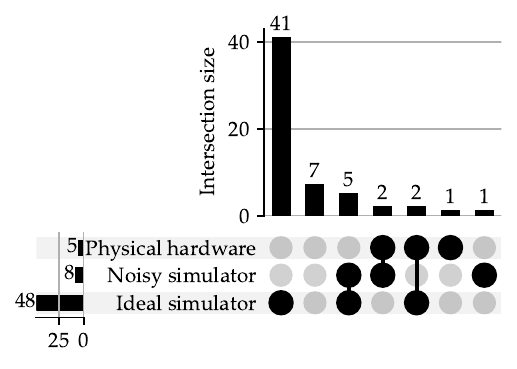}
        \caption{Upset plot of the backend selection}  
        \label{fig: rq9_number_of_backends}
    \end{subfigure}
    \begin{subfigure}[b]{0.485\textwidth}
        \centering
        \includegraphics[width=0.98\textwidth]{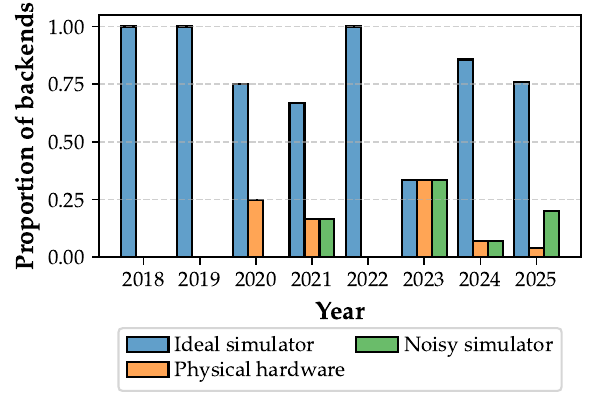}
        \caption{Proportion of each backend over the years}  
        \label{fig: rq9_backend_temporal_trend}
    \end{subfigure}
    \caption{Execution Backends adopted in the primary studies for running the CUTs}  
    \label{fig: rq9}
    \Description{The figure contains two subfigures illustrating execution backends used in the primary studies. 
    The left subfigure is an upset plot showing the combinations and frequencies of three backends across studies. 
    The right subfigure depicts the temporal distribution of different execution backends, presenting the proportion of each backend adopted in studies over time.}
\end{figure}

Figure~\ref{fig: rq9_number_of_backends} shows an upset plot of the backends selected in the primary studies. Until now, 81.4\% (48 of~\NumberOfPrimaryStudies) of the primary studies have implemented QST experiments on ideal simulators, while physical hardware and noisy simulation have been seldom explored. We noticed that 41 studies merely discussed ideal simulators, which is still understandable. The ideal simulation approximates a post-NISQ quantum computer that is fully fault-tolerant, and enables the evaluation to focus on the identification of code-level faults. Besides, accessing current quantum computers while ensuring their expected effectiveness leads to significantly high overhead~\cite{wang2021qdiff, preskill2018quantum}, which could hinder comprehensive empirical studies based on the physical hardware. We found that 7 (5+2) studies considered both ideal simulations and noisy environments, providing a balance between evaluating the applicability of test approaches on current NISQ devices and conducting empirical studies with the cost accounted for. Apart from the cases involving noise-free environments, 4 (2+1+1) primary studies~\cite{\PaperFifteen, \PaperSixtySeven, \PaperNinetyFour, \PaperThirtyNine} only considered the noisy backends (i.e., noisy simulators or physical hardware). Among the four, two studies~\cite{\PaperFifteen, \PaperSixtySeven} especially examined the impact of noise on CUTs, rather than the fault-detection purpose, which implies potential testing-related issues specific to QSE. 

Considering the evolving focus of the QSE community and the progress of real quantum hardware, we also display the temporal trend of backend selection in Figure~\ref {fig: rq9_backend_temporal_trend}. The proportions in this figure are relative to the frequency of backends adopted by the primary studies released in a given year. The ideal simulation significantly dominates other backends in years except 2023. Since 2020, physical hardware has been considered in QST empirical studies, marking the feasibility of employing NISQ devices for QST research. 
For each of the recent years from 2023 to 2025, both the noisy simulator and the physical hardware have been explored, indicating the community's continued interest in the noise-aware scenarios. 

\subsubsection{Shot-based Simulator versus Shot-independent Simulator}
\begin{figure}[!t]
    \centering
    \includegraphics[width=0.9\textwidth]{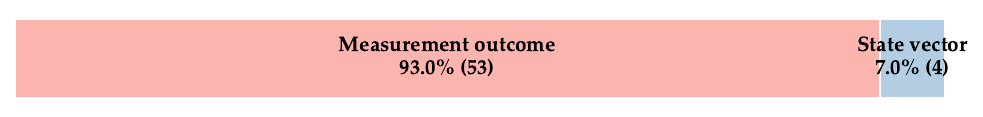}
    \caption{The choices of ideal classical simulation, where the number enclosed by the parentheses indicates the number of corresponding primary studies}
    \Description{This figure shows the choices of ideal classical simulators, where one results in measurement outcomes while the other merely returns state vectors.}
    \label{fig: rq9_backend_output}
\end{figure}
    
Nowadays, quantum software development kits provide diverse ideal simulators, enabling practitioners to simulate quantum systems for heterogeneous purposes. For example, Qiskit supports the numerical simulation completely built on state vectors and unitary matrices, aside from imitating real quantum systems based on quantum measurements\footnote{See the documentation of Qiskit Aer: \url{https://qiskit.github.io/qiskit-aer/apidocs/aer_provider.html}.}. For backends that do not model quantum measurements, their outcomes are independent of shots and therefore cannot fully reproduce the behavior of a realistic quantum system, owing to a violation of the postulate of quantum mechanics. 
\revise{However, such backends indeed provide convenience for practitioners to fully access the original data stored in quantum circuits through classical and noise-free simulation.}
In light of the above discussion, we are interested in whether the primary studies adopted shot-independent backends for QST, and, if so, the specific purposes for \revise{why} 
such backends were employed.

Figure~\ref{fig: rq9_backend_output} demonstrates the adoption of different backends, where a study may consider multiple backend types. Almost all primary studies (i.e., 93.0\%) used the backends that produce measurement outcomes, reflecting a research mainstream that focuses on testing quantum programs on realistic quantum devices or their practical simulations. Only four studies~\cite{\PaperFourteen, \PaperFiftyFour, \PaperOneHundredAndTwelve, \PaperOneHundredAndTwentyOne} included classical ideal simulators based on state vectors, while the other types of ideal simulators, such as the unitary-based backend, were not involved. We identified that the three studies~\cite{\PaperFourteen, \PaperFiftyFour, \PaperOneHundredAndTwentyOne} investigated runtime assertion for debugging, and the other one~\cite{\PaperOneHundredAndTwelve} studied fault detection of quantum programs by comparing state vectors.

\revise{The above results suggest an emerging use of shot-independent simulators in the current NISQ era, despite a relatively limited number of studies. One implication is that researchers should explicitly justify how their simulator choice aligns with testing objectives, for example, by clarifying whether state-vector simulation is employed solely to verify program correctness within a small scale, rather than to evaluate techniques for possible execution on quantum hardware. Furthermore, it would be misleading to directly compare two sets of test results obtained from shot-based and shot-independent simulators without accounting for differences in their execution models and research objectives.}

\subsubsection{Summary of RQ9}
\text{}

\Subtitle{Takeaway 9-1} Ideal simulators overall dominate the backends used for QST, even in terms of most years. Noisy simulators and physical hardware were employed in only a few studies to assess the performance of test approaches on current NISQ devices. Besides, we found that some studies simultaneously considered noise-free and noisy backends in their experiments.

\Subtitle{Takeaway 9-2} Almost all the primary studies (93.0\%) employed shot-based backends that produced measurement outcomes. Only four studies considered the classical ideal simulator based on state vectors, where such shot-independent backends have been employed for runtime assertion and software testing.

\subsection{\texorpdfstring{\RQSection{10}}{RQ10}}\label{sec: rq10}
Open-source tooling plays a crucial role in SE research, as it can enhance the traceability and reproducibility of scientific studies and benefit reusability for baseline comparison and software iteration.
In this RQ, we present the available tooling identified in primary studies, focusing on the sources of PUTs used for evidence generation and the artifacts provided to support the generated evidence.

\subsubsection{Available Sources of PUTs}
\begin{table}[!t]
    \small
    \centering
    \caption{Available sources of object programs mentioned in the primary studies}
    \label{tab: rq10_available_program_sources}
    \resizebox{.98\textwidth}{!}{
        
\begin{tabular}{p{{0.25\columnwidth}}  p{{0.49\columnwidth}} p{{0.215\columnwidth}} c}
    \toprule[1pt]
    \multicolumn{1}{c}{\textbf{Program sources}} & \multicolumn{1}{c}{\textbf{Repository links}} & \multicolumn{1}{c}{\textbf{Primary studies}} & \multicolumn{1}{c}{\textbf{\#}} \\
    \cmidrule(lr){1-1} \cmidrule(lr){2-2} \cmidrule(lr){3-3} \cmidrule(lr){4-4}  
    Bugs4Q~\cite{zhao2021bugs4q} & {\url{https://github.com/Z-928/Bugs4Q}} & \cite{\PaperFortyFour ,\PaperFortyFive ,\PaperFiftyOne ,\PaperEightyThree ,\PaperEightyFour ,\PaperOneHundredAndNine} & 6 \\ 
    MQT bench~\cite{quetschlich2023mqt} & {\url{https://www.cda.cit.tum.de/mqtbench/}} & \cite{\PaperThirtyNine ,\PaperFiftyFour ,\PaperSeventyEight ,\PaperOneHundredAndTwelve} & 4 \\ 
    Qbugs~\cite{campos2021qbugs} & N/A & \cite{\PaperFortyFour ,\PaperEightyThree ,\PaperEightyFour} & 3 \\ 
    VeriQBench~\cite{chen2022VeriQBench} & {\revise{\url{https://github.com/Veri-Q/Benchmark}}} & \cite{\PaperSix ,\PaperFiftyTwo} & 2 \\ 
    O'reilly~\cite{johnston2019programming} & {\url{https://oreilly-qc.github.io/}} & \cite{\PaperTwentySeven ,\PaperFiftyNine} & 2 \\ 
    Qiskit aqua library & {\url{https://github.com/qiskit-community/qiskit-aqua#migration-guide}} & \cite{\PaperSeventyFour ,\PaperOneHundredAndNine} & 2 \\ 
    RevLib~\cite{wille2008revlib} & {\url{https://www.revlib.org/}} & \cite{\PaperFifteen} & 1 \\ 
    Qiskit textbook & {\url{https://qiskit.org/textbook/ch-algorithms/shor.html}} & \cite{\PaperTwentySeven} & 1 \\ 
    Rieffel et al.~\cite{rieffel2011quantum} & N/A & \cite{\PaperFortyThree} & 1 \\ 
    Qiskit circuit library & {\url{https://github.com/Qiskit/qiskit/tree/stable/1.2/qiskit/circuit/library}} & \cite{\PaperFortySix} & 1 \\ 
    Quanto~\cite{pointing2024optimizing} & N/A & \cite{\PaperSeventy} & 1 \\ 
    Juan Carlos et al.~\cite{garcia2011equivalent} & N/A & \cite{\PaperSeventy} & 1 \\ 
    Quantum algorithm zoo & {\url{https://quantumalgorithmzoo.org/}} & \cite{\PaperSeventyThree} & 1 \\ 
    Quantum code repository & {\url{https://quantumcomputinguk.org/code-repository}} & \cite{\PaperSeventyThree} & 1 \\ 
    QASMBench~\cite{li2023qasmbench} & {\url{https://github.com/pnnl/QASMBench}} & \cite{\PaperEightyFour} & 1 \\
    \cmidrule(lr){1-4} 
    \multicolumn{4}{l}{\textbf{Number of sources}: 15; \textbf{Number of repositories}: 11} \\   
    \bottomrule[1pt]
\end{tabular}
        
    }
    {\justify
        {\selectfont
            ``N/A'' indicates the source did not provide an available repository link, but the programs can still be obtained from the literature.
        }
    \par}
\end{table}

According to Table~\ref{tab: rq10_available_program_sources}, we identified 15 available PUT sources from the primary studies, and \revise{11} open-source repositories with accessible links corresponding to these sources. The presence of these repositories enables future studies to access and reuse the included PUTs more conveniently. 

In addition, we discovered that the two most popular sources were Bugs4Q and MQT bench, which were used in 6 and 4 primary studies, respectively. MQT bench is an evolving software repository comprising around 70,000 benchmark circuits with qubit counts ranging from 2 to 130, thereby offering abundant candidate PUTs written by OpenQASM. In comparison, Bugs4Q is one of the two benchmarks (i.e., the other is Qbugs) on real-world buggy programs, and it includes Qiskit bugs from popular programming platforms\footnote{We found two archived versions of Bugs4Q. The previous version (i.e., the one exactly mentioned by primary studies) corresponding to the conference publication~\cite{zhao2021bugs4q} incorporates 36 Qiskit bugs from GitHub, while the new version (\url{ https://github.com/Z-928/Bugs4Q-Framework}) for the journal publication~\cite{zhao2023bugs4q} includes 42 Qiskit bugs from three platforms (GitHub, Stack Overflow, and Stack Exchange).}. Even so, the adoption of PUT sources remains scattered, so we cannot draw conclusions about widely preferred sources or suggest universal ones for future empirical studies.

\subsubsection{Available Artifacts of Primary Studies}
\begin{table}[!t]
    \small
    \centering
    \caption{Available artifacts (up to~\SnowballingEndTime) provided by the primary studies}
    \label{tab: rq10_available_artifacts}
    \resizebox{.98\textwidth}{!}{
        
\begin{tabular}{c p{0.95\columnwidth}}
    \toprule[1pt]
    
    \multicolumn{2}{l}{\textbf{Software testing:}} \\
    \cite{\PaperOne} & \textit{GitHub}: {\url{https://github.com/Simula-COMPLEX/muskit}}, \textit{Zenodo}: {\url{https://zenodo.org/records/5288917}} \\
    \cite{\PaperThree} & \textit{GitHub}: {\url{https://github.com/Simula-COMPLEX/quito}}, \textit{Zenodo}: {\url{https://zenodo.org/records/5288665}} \\
    \cite{\PaperFive} & \textit{GitHub}: {\url{https://github.com/Simula-COMPLEX/qucat-tool}} \\
    \cite{\PaperThirteen} & \textit{GitHub}: {\url{https://github.com/ShahinHonarvar/QSharpCheck}} \\
    \cite{\PaperSeventeen} & \textit{GitHub}: {\url{https://github.com/Simula-COMPLEX/qusbt-tool}} \\
    \cite{\PaperTwenty} & \textit{GitHub}: {\url{https://github.com/Simula-COMPLEX/MutTG-paper}} \\
    \cite{\PaperTwentyOne} & \textit{GitHub}: {\url{https://github.com/Simula-COMPLEX/qusbt/}} \\
    \cite{\PaperTwentyFour} & \textit{GitHub}: {\url{https://github.com/LuisLlana/metamorphic_testing/}} \\
    \cite{\PaperTwentySeven} & \textit{GitHub}: {\url{https://github.com/biromiro/feup-gulbenkian-qc-mt}} \\
    \cite{\PaperThirtyThree} & \textit{Figshare}: {\url{https://figshare.com/articles/software/Delta_Debugging_for_Property-Based_Regression_Testing_of_Quantum_Programs/25075154?file=44241593}}, \textit{GitHub}: {\url{https://github.com/GabrielPontolillo/ddregression}} \\
    \cite{\PaperThirtySix} & \textit{GitHub}: {\url{https://github.com/am0x00/QuanTest}} \\
    \cite{\PaperThirtySeven} & \textit{GitHub}: {\url{https://github.com/cccorn/Q-ATPG}} \\
    \cite{\PaperThirtyEight} & \textit{Zenodo}: {\url{https://zenodo.org/records/13370788}} \\
    \cite{\PaperThirtyNine} & \textit{GitHub}: {\url{https://github.com/AsmarMuqeet/QOPS}} \\
    \cite{\PaperFortyThree} & \textit{Zenodo}: {\url{https://zenodo.org/records/11191215}} \\
    \cite{\PaperFortySix} & \textit{Zenodo}: {\url{https://doi.org/10.5281/zenodo.15462299}} \\
    \cite{\PaperFiftyNine} & \textit{Personal Website}: {\url{https://simula-complex.github.io/Quantum-Software-Engineering/ICST21.html}} \\
    \cite{\PaperSixty} & \textit{GitHub}: {\url{https://github.com/Simula-COMPLEX/qucat-paper}} \\
    \cite{\PaperSixtySeven} & \textit{GitHub}: {\url{https://github.com/WindFrank/QuantumDataAndProgram}} \\
    \cite{\PaperSeventy} & \textit{GitHub}: {\url{https://github.com/MgcosA/EvaluationCodeOfQuantumRelationChecking/}} \\
    \cite{\PaperSeventyFour} & \textit{GitHub}: {\url{https://github.com/danielfobooss/mutpy}} \\
    \cite{\PaperSeventyEight} & \textit{GitHub}: {\url{https://github.com/EnautMendi/Quantum-Circuit-Mutants-Empirical-Evaluation}} \\
    \cite{\PaperNinetyFour} & \textit{Figshare}: {\url{https://figshare.com/articles/software/QuCheck_with_QOIN/28772231}} \\
    \cite{\PaperOneHundredAndSix} & \textit{GitHub}: {\url{https://github.com/MgcosA/Code_of_Testing_Oracle_Quantum_Program_Article}} \\
    \cite{\PaperOneHundredAndThirteen} & \textit{GitHub}: {\url{https://github.com/sinugarc/QCRMut}} \\
    \cite{\PaperOneHundredAndFourteen} & \textit{GitHub}: {\url{https://github.com/csiro/QUT/tree/main}} \\
    \cite{\PaperOneHundredAndSixteen} & \textit{GitHub}: {\url{https://github.com/Ahmik-Virani/Differentiating-Quantum-Bug-From-Noise-Statistical-Approach}} \\
    \cite{\PaperOneHundredAndNineteen} & \textit{Figshare}: {\url{https://figshare.com/articles/software/QuCheck_A_Property-based_Testing_Framework_for_Quantum_Programs_in_Qiskit/27919539?file=53357996}} \\
    \cmidrule(lr){1-2}

    \multicolumn{2}{l}{\textbf{Runtime assertion:}} \\
    \cite{\PaperFortyOne} & \textit{OSF}: {\url{https://osf.io/k6ygp/overview}} \\
    \cite{\PaperFiftyFour} & \textit{GitHub}: {\url{https://github.com/munich-quantum-toolkit/debugger}} \\
    \cite{\PaperOneHundredAndFifteen} & \textit{Zenodo}: {\url{https://zenodo.org/records/15708438}} \\
    \cite{\PaperOneHundredAndTwentyTwo} & \textit{GitHub}: {\url{https://github.com/revilooliver/Quantum-Circuits-for-Dynamic-Runtime-Assertions-in-Quantum-Computation}}, \textit{Zenodo}: {\url{https://zenodo.org/records/3597507}} \\
    \cmidrule(lr){1-2}

    \multicolumn{2}{l}{\textbf{Program repair:}} \\
    \cite{\PaperFortyFour} & \textit{Zenodo}: {\url{https://zenodo.org/records/14288140}} \\
    \cmidrule(lr){1-2}

    \multicolumn{2}{l}{\textbf{Platforms: } \textit{GitHub} (24), \textit{Zenodo} (8), \textit{Figshare} (3), \textit{OSF} (1), \textit{Personal Website} (1)} \\
    \multicolumn{2}{l}{\textbf{Proportion of available artifacts}: 55.9\% (33/59)} \\
    \bottomrule[1pt]
\end{tabular}

    }
\end{table}

Table~\ref{tab: rq10_available_artifacts} lists the artifact links along with their released platforms of the primary studies. There are beyond half of the primary studies (i.e., 55.9\%) having their artifacts publicly available, indicating that the consensus of open-source practices is gradually taking shape within the QSE community. However, the available artifact for fault localization remains absent, thereby creating an intensive need for future studies to contribute artifacts to this SE problem.

Regarding the platforms used for artifact release, there are four public platforms having been explored: GitHub, Zenodo, Figshare, and OSF (i.e., Open Science Framework).
GitHub is the most commonly employed in the primary studies (24 studies), followed by Zenodo (8 studies). Besides, four studies~\cite{\PaperOne, \PaperThree, \PaperThirtyThree, \PaperOneHundredAndTwentyTwo} provide multiple repository links in their publications. Each of the four encompasses a development-oriented GitHub repository that offers convenience for collaborative development and maintenance, along with a link to the platform that promises long-term archiving with digital object identifiers (e.g., Zenodo and Figshare). 
The consideration of both development and archiving suggests a desirable practice for artifact release of empirical studies.

\subsubsection{Summary of RQ10}
\text{}

\Subtitle{Takeaway 10-1} We found 15 sources for PUTs provided by the primary studies. \revise{Eleven} sources offer available repositories, which are convenient to reuse the included PUTs. Besides, both Bugs4Q and Qbugs are benchmarks collecting real-world buggy programs based on quantum software development kits.

\Subtitle{Takeaway 10-2} Over half of the primary studies (i.e., 55.9\%) made their research artifacts publicly available, implying the open-source practices being gradually shaped and acknowledged within the community. GitHub is the most popular platform for releasing artifacts. Even so, four studies took into account both GitHub along a platform like Zenodo and Figshare for long-term archives.


\section{Threats to Validity}
\label{sec: threats}

Several aspects could threaten the validity of our paper, as discussed below.

\Subtitle{Internal validity} 
The phase to include and analyze literature cannot avoid introducing unexpected uncertainty and subjectivity, although the double check was performed by the authors.

\Subtitle{External validity} 
We acknowledge that some findings can only represent the current state of QST studies, given the evolving practices in the QSE community and the advancement of quantum computing technology. For example, if quantum computers have more usable qubits and stronger fault-tolerant capabilities, more QST experiments would be conducted on physical hardware. Even so, these findings remain valuable, supporting the reasonability of near-term work and offering motivations for longer-term research.

\Subtitle{Construct validity} 
\revise{One potential threat to construct validity lies in the choice of keywords used for the literature search. In particular, the heavy reliance on a single keyword ``test*'' to identify SE activity within our scope may narrow the search query. However, this threat could be mitigated by the subsequent snowballing search. Compared with keyword search, only a limited number of additional papers were identified during snowballing search (\PaperNumSnowballingTotal versus~\PaperNumKeywordTotal), suggesting that the query of keyword search was relatively comprehensive.}
\highlight{The extraction of data items could threaten the construct validity \revise{as well}. We cannot ensure that the coded data (e.g., consolidation and categorization) is semantically identical to the original data reported in the primary studies, due to inconsistencies in similar terminology and variation in writing style.
\revise{Also, incomplete and ambiguous reports in primary studies impact the quality of evidence synthesis in our paper.}
\revise{To mitigate above two threats and enhance transparency, we provide a unified and standardized overview of the primary studies through well-defined categories.}}

\Subtitle{Conclusion validity} 
The conclusion validity may be threatened by reliance on the valid metadata from a moderate but not very extensive pool of primary studies (i.e.,~\NumberOfPrimaryStudies). We confirm that the QST empirical studies are still in their infancy, but the notable increase in related studies over the years underscores the need to conduct this paper in the current context. Moreover, the inclusion of preprint papers (up to 23.7\% of the primary studies) via snowballing could bring another threat. Some of them do not undergo peer review and the available versions may iterate over time, thereby affecting the reliability of the synthesized evidence and consequently threatening the validity of the conclusions. To mitigate this threat, the qualitative analysis (e.g., the categorization into our defined labels) throughout this paper was carefully conducted by the authors after evaluating the reasonability of statements in the primary studies with the preprint papers included. In the content of each RQ, we strive to opt only representative, undisputed, and dependable metadata as examples for further analysis.

\section{Discussions}
\label{sec: discussion}

In this section, we present insights into the current state of research and outline feasible future directions, based on the findings from the ten RQs. We also highlight key distinctions between QST and CST, as well as between SE practices and QC objectives.
\revise{In the following four parts from Section~\ref{sec: requirements} to Section~\ref{sec: research_artifacts}, we discuss four high-level aspects in detail, including test requirements, test oracles, test scalability, and open-source tooling, which should attract more attention in future studies. Actionable recommendations corresponding to empirical results of RQs are summarized in Section~\ref{sec: recommendation}.}

\subsection{Requirements-driven Testing}\label{sec: requirements}
In CSE, it has been argued that software testing may involve a requirements gap, i.e., omissions in the translation of requirements into design and implementation~\cite{huizinga2007automated}. A similar concern arises for QST, where testing practice is still in its early stages, and guidance on quantum software requirements is limited. Despite the scarcity of studies on requirements engineering for quantum software systems~\cite{yue2023towards, sepulveda2024systematic}, empirical QST studies should be grounded in plausible requirements from real-world testing scenarios. In this section, we discuss two aspects in which requirements can guide testing decisions: the objects under test and the execution backends.

\subsubsection{Functionality and Structure of PUTs}
In CSE, functional testing and structural testing are two runtime techniques used to verify program correctness with respect to software requirements. QST should consider both aspects and, moreover, clarify whether the focus is on program-level objects or circuit-level objects.

\Subtitle{Functional testing} 
Most primary studies treated PUTs as black-box models to conduct functional testing, because quantum measurement makes code instrumentation difficult. Still, it is important to distinguish the functionality of the whole program from that of the quantum circuit alone. For example (RQ1), quantum neural networks are tested in existing studies either as subroutines with fixed circuits or as models trained on data. In the former case, test requirements mainly concern the unitary operation implemented by the circuit, whereas in the latter case, they evaluate model performance associated with algorithmic functionality. In addition, as discussed in RQ4, functional testing should consider input types that align with the required functionality. Compared to testing low-level quantum programs that target quantum circuits (CUTs), testing high-level quantum programs should consider classical test inputs, since these programs are often scalable and hybrid, combining classical and quantum subroutines. 
\revise{Beyond classical inputs, recent studies have further explored extending the test input domain suitable for low- and high-level PUTs. In particular, two studies treated measurement operators as test inputs~\cite{\PaperThirtySeven, \PaperThirtyNine}, which provide an alternative means of designing effective test cases, especially in scenarios where preparing initial quantum states falls outside the PUT's defined functionality.} In comparison, four studies~\cite{\PaperSeventyOne, \PaperOneHundredAndSix, \PaperOneHundredAndEight, \PaperOneHundredAndTwelve} that adopted an inverse test apply an additional basis transformation relative to the default measurement basis, but the transformed measurement operators mainly serve fault detection and are not explicitly treated as test inputs. Therefore, testing requirements should be made explicit when implementing the testing process. For example, if measurement is specified as part of the test inputs, measurement operators should not be arbitrarily altered during testing. 

\Subtitle{Structural testing} 
In comparison, structural testing has received limited attention in existing work. According to recent studies~\cite{long2024testing, zhao2024towards}, structural testing of quantum programs or quantum software systems should be encouraged, because it can further assess whether control flow, data flow, and quantum circuit topology are implemented as required. In structural testing, evaluating a CUT solely by mathematical equivalence to an expected quantum circuit is not desirable, since the CUT may still contain unintended or structurally inconsistent gate sequences that a black-box view would overlook.
For future empirical studies on structural testing at the circuit level, we suggest specifying expected structures using quantum operators (as discussed in RQ5) to depict the intended sequence of basic gates or gate blocks (e.g., the oracle operation in Grover Search).
Additionally, white-box and grey-box techniques remain feasible for structural testing of high-level quantum programs, because classical subroutines in a hybrid PUT are not subject to the constraints imposed by the postulates of quantum mechanics.



\subsubsection{Backend Selection}
Backend selection remains debated in the community, and one concern is the rationale for using ideal simulators. Based on RQ9, we argue that ideal simulation should not be dismissed, as it is used in the majority of primary studies. In practice, backend selection is not a black-and-white decision and should be guided by specific software requirements.

\Subtitle{Code-level testing} 
\revise{If the test requirement is to identify faults or bugs at the code level and the simulation costs are affordable by classical computers, classical simulation should be acceptable for empirical studies that evaluate testing methods in early-stage QST.
In comparison, access to current NISQ devices via cloud platforms is expensive, and results from specific hardware are hard to reproduce, both of which deviate from common software engineering practices.
Appropriate adoption of ideal simulation still makes sense, resulting in a smaller execution cost than noisy simulation. 
Moreover, quantum computers are steadily progressing toward fault tolerance~\cite{preskill2018quantum}, thereby mitigating the adverse effects of quantum noise as hardware continues to advance.
Even so, code-level testing should also pay attention to the underlying use of quantum hardware, since simulating large-scale CUTs is challenging for classical computers in theory.
Early-stage QST research is encouraged to explore how hardware characteristics beyond noise, such as quantum measurement, qubit connectivity, and native gate sets, influence test results, whether on classical simulators or physical hardware.
This, in turn, provides a methodological foundation for future research and applications, once practical quantum hardware becomes sufficiently reliable and usable.
}

 

\Subtitle{System-level testing} 
A vital distinction between CST and QST is that executing quantum programs is tightly coupled with specific hardware systems; backend properties should therefore be accounted for during testing. When testing depends on physical properties of the execution system, relying on ideal simulators may deviate from such requirements. Even for tasks that aim to provide practical evidence beyond methodological validation, such as investigating the impact of noise~\cite {\PaperFifteen, \PaperSixtySeven} (RQ9), evaluation on physical hardware is necessary. This is because the significance of these tasks lies in the limitations of current NISQ devices, and noisy simulators cannot fully capture time-varying hardware attributes (e.g., drift, crosstalk, and thermal fluctuation).
Moreover, it is important to distinguish noisy simulators from actual quantum hardware. A noisy simulator emulates the logical behavior of an NISQ device by injecting mathematically modeled noise, yet it does not reproduce the genuine physical dynamics of real quantum systems.





\subsection{Test Oracle Problem}
The test oracle problem arises in almost every empirical study on QST and can substantially affect experimental evaluations. For example, oracle-induced false positives may inflate the number of reported faults, thereby confounding effectiveness-related metrics.
We note that, while CST research has extensively studied oracle absence (e.g., oracle generation~\cite{padgham2013model, dinella2022toga} and metamorphic testing~\cite{segura2016survey, chen2018metamorphic}), empirical QST studies face additional challenges rooted in quantum semantics and measurement. Therefore, before directly transferring CST-style solutions, it is important to first address several QC-specific issues that determine whether an oracle is well-defined and interpretable for a given specification. In this subsection, we start from program specifications and discuss how specification--oracle alignment should be handled and reported in empirical QST studies.

\subsubsection{Gap Between Program Specifications and Test Oracles}
A first concern is whether the adopted test oracle and its evaluation procedure align with the program specification. This concern is motivated by RQ5: specifications in primary studies imply output types beyond probability distributions, and some test oracle types are only meaningful under specific program functionalities. Therefore, empirical QST studies should make the specification assumptions explicit and justify that the oracle implements the intended notion of correctness.


\Subtitle{Output type}
To illustrate specification--oracle consistency, consider a specification that requires an expected output quantum state. A test oracle based solely on measurement outcomes may lose effectiveness when an incorrect output state differs from the expected one only by a relative phase, e.g., $\alpha\ket{0}+\beta\ket{1}$ and $\alpha\ket{0}+\beta e^{i\theta}\ket{1}$ ($\theta\neq2k\pi,k\in\mathbb{Z}$), which yield identical probabilities $p_{M}(0)=|\alpha|^2$ and $p_{M}(1)=|\beta|^2$ under Pauli-Z measurement. In such cases, studies should clarify whether the specification is phase-sensitive and, if so, use a test oracle that reflects state-level relations. At the same time, QC techniques such as full quantum tomography can incur prohibitive overhead. Therefore, we suggest that future empirical studies adopt or develop specification-aware test oracles that (i) match the specified output type as closely as possible and (ii) report the resulting trade-off between test oracle effectiveness and its evaluation cost.

\Subtitle{Program functionality}
Primary studies also imply functional differences among quantum programs. For example, oracle-based quantum programs discussed in~\cite{long2025a} are expected to yield deterministic outcomes, whereas output-dominant algorithms defined in~\cite{\PaperSeventyEight} only care about the most probable outcome rather than the entire distribution. This indicates that an OPO, which focuses on comparing probability distributions, is neither universally effective nor universally efficient, even though it is used in the majority of primary studies (RQ5). 
\revise{Instead, DOO naturally matches the functional testing of output-dominant algorithms.
On the one hand, for such algorithms, functional correctness is defined in terms of measurement outcomes rather than requiring identical final quantum states; it is sufficient to obtain the correct dominant output with high probability.
On the other hand, identifying the dominant output typically requires fewer shots than reconstructing the full output probability distribution, thereby benefiting the efficiency of test oracle execution.}
Hence, future empirical QST studies should explicitly state the expected functionality of the PUT and justify how the chosen oracle matches that functionality.

\subsubsection{Unavailability of Fully Specified Program Specifications}
Not all quantum programs admit fully specified and tractable program specifications. For example, the \textbf{Quantum Approximate Optimization Algorithm (QAOA)} explores hybrid classical--quantum workflows to solve combinatorial optimization problems that are difficult for classical computing alone. In contrast to foundational subroutines such as QFT, whose expected states can be explicitly derived and efficiently computed classically, QAOA is produced through an iterative optimization process, and its expected solution may be intractable for classical algorithms---consistent with its goal of exploring quantum advantage. This makes it difficult to define complete, executable specifications that can directly serve as test oracles.

\Subtitle{Pseudo-oracle}
\revise{As} one possible solution\revise{, a} \textit{pseudo-oracle} is an independently produced program intended to satisfy the same specification as the original program~\cite{davis1981pseudo}. A test passes only if no inconsistency is detected between the original program and the independently produced program when executed with the same test input. This idea is related to fault-tolerant software reliability techniques such as \textit{N-version programming} with voting, which has also been discussed for quantum software systems~\cite{saito2024towards}. In light of RQ5, using a derived specification based on program outputs partially follows the pseudo-oracle principle. However, directly relying on a bug-free version to generate the ``reference output'' before seeding defects is discouraged, because such a version is typically unavailable in real-world scenarios. For empirical studies, a more practical alternative is to use multiple independently developed implementations, such as a classical counterpart of the quantum program under test, and to explicitly report the assumptions under which this counterpart serves as a reference.

\Subtitle{Property-based specification}
As another alternative, several studies (RQ5) used assertions over program properties for fault detection.
Metamorphic testing designs a set of test inputs and metamorphic relations that specify expected relationships among the corresponding outputs.
Compared to CST, metamorphic testing for QST remains limited, with only two primary studies identified~\cite{\PaperTwentyFour, \PaperTwentySeven}. This highlights a need for future studies to further explore metamorphic testing for complex quantum programs (e.g., programs with hybrid subroutines). Inspired by~\cite{\PaperSeventyFour,tan2025hornbro}, using a subset of properties rather than a complete specification can still be effective in triggering failures, since certain properties (e.g., the relative ordering of measurement probabilities rather than their exact values) are often easier to analyze in practice. In summary, we suggest that future empirical studies systematically explore property-based specifications to alleviate the oracle problem for fault detection. More importantly, they should explicitly report such partial specifications and distinguish them from formal, complete specifications. Although it is impossible to obtain a complete set of properties~\cite{chen2018metamorphic}, we recommend including a critical analysis of how the selected properties relate to the intended faulty behaviors of the PUTs and what kinds of faults they may fail to expose.

\subsection{Test Scalability}
In QST, the scalability of a testing approach concerns whether it can remain effective and cost-efficient as the complexity of quantum programs increases, a common situation in real-world software development.
In this part, we discuss test scalability from two perspectives: complexity measures and test execution. The first perspective concerns how empirical studies characterize the complexity of PUTs and what is considered ``complex enough'' for testing. The second perspective concerns what challenges a testing approach must address when applied to such complex programs. Based on the primary studies, we outline directions for future work, with an emphasis on cost-effectiveness when testing larger-scale circuits.

\subsubsection{Complexity Measures of PUTs}
As shown in RQ3, current studies primarily measured PUT complexity through the corresponding CUTs, using circuit width, size, and depth.
However, empirical studies can benefit from a broader set of measures that better reflect both software development and program execution. This, in turn, can guide PUT selection under the rationale that more complex programs are often associated with higher quality risks and are therefore both critical and challenging to test.

\Subtitle{Complexity for development} 
As discussed by Zhao~\cite{zhao2021some}, the size of quantum software can be measured at multiple development levels, including code, design, and specification (e.g., the number of connectors between quantum and classical components at the design level). \highlight{Based on Figure~\ref{fig: PUT_versus_CUT}, it is worth noting that the development complexity of \revise{PUTs, especially for the high-level ones,} is not always consistent with the complexity of the involved CUTs. For instance, a high-level quantum program with many lines of code may exhibit high development complexity, while an individual test case may yield only small-scale circuits, such as \texttt{n=1} for QFT in that figure.} Therefore, future empirical studies should be aware of this gap when reporting complexity, especially when the evaluation is dominated by circuit-level measures. Moreover, it is meaningful to investigate how complexity measures shared by classical and quantum programs relate to quantum circuit complexity, thereby clarifying the purpose and focus of empirical studies in QST.

\Subtitle{Complexity for execution} 
In existing work, execution complexity largely depends on circuit scale. Given the emphasis on qubit counts, future empirical studies on test scalability should include PUTs with sufficiently large qubit counts, as suggested by the distributions observed in RQ3 (e.g., not being limited to very small qubit counts and encouraging no fewer than 10 qubits). In addition, reporting both circuit size and depth enables a more complete characterization of circuit complexity, and studies may consider exploring larger sizes and depths (e.g., size $\ge 84$ and depth $\ge 56$) when feasible. Finally, when testing high-level quantum programs that use abstract representations of quantum operations (e.g., a QFT block synthesized into basic gates), adopting metrics computed on decomposed circuits, as suggested by~\cite{long2024equivalence, li2025preparation}, can provide a more comparable and less biased estimation of execution complexity.

\subsubsection{Challenges of Testing Large-scale Circuits}
Testing large-scale circuits faces high-dimensional input and output spaces. Here, we discuss two practical challenges in empirical studies: test cases and statistical repetitions, corresponding to test inputs and outputs, respectively.

\Subtitle{Test cases} 
As the input domain grows, exhaustive testing becomes infeasible, and achieving adequate input coverage becomes increasingly costly. For example, for a family of CUTs with increasing qubit counts, executing a single test case often incurs substantially higher simulation overhead, while contributing only marginally to input coverage due to the exponential growth of the input space. One direction is to design test inputs that are more likely to trigger faults. Motivated by the gap observed in RQ4, we further suggest that future empirical studies explicitly report the test suite size and discuss its impact on evaluation results, following common SE practice~\cite{chen2020revisiting}. Without deliberate test-case design, using test suites of the same size across CUTs of different scales can introduce threats to validity. Using mutation score as an illustrative example (RQ6), mutants involving larger qubit counts may be inherently harder to kill when only a limited number of test cases originally designed for smaller circuits are applied. In such settings, observed effectiveness may conflate the capability of the approach with the adequacy of the test suite size.

\Subtitle{Statistical repetitions}
Shot counts in QST have received increasing attention in recent studies~\cite{miranskyy2025cost, ye2025measurement}, yet the relationship between the shots required for reliable inference and the qubit counts of CUTs has not been systematically studied. By the intuition behind the \textit{curse of dimensionality}~\cite{bellman1966dynamic}, reliable conclusions from statistical analyses may require shot counts to grow rapidly with the output-space dimension determined by the number of qubits. 
\revise{In theory, this suggests that using a fixed number of shots may not always work for experiments on CUTs of varying sizes, when it is necessary to capture features such as probability distributions; otherwise, the statistical significance may be compromised.}
At the same time, resource costs (e.g., simulator memory and circuit execution time) limit the feasibility of using very large numbers of shots. Therefore, future empirical studies may consider adaptive shot configurations (RQ8), informed by both theoretical considerations and practical constraints. In addition, when designing QST experiments, it is important to examine the joint effect of shot count and experimental repetitions: if shot counts are insufficient, outcome variance may increase, which in turn may require more repetitions to achieve statistically meaningful comparisons.
\revise{We suggest that when it is challenging to secure a theoretically sufficient shot count, the number of experimental repetitions can be moderately increased within affordable cost constraints.}

\subsection{Open-source Tooling}~\label{sec: research_artifacts}
Several top-tier SE venues increasingly emphasize open-source tooling and research artifacts, as reflected by initiatives such as the Artifact Evaluation track at ICSE and the replicated computational results report suggested by TOSEM.
Based on our literature review and analysis, open-source tooling is closely tied to the design, execution, and evaluation of empirical studies in QST.
Therefore, we encourage future studies to release tools as open source to improve transparency, reproducibility, and reuse within the QSE community. In the following two parts, we provide suggestions regarding public benchmarks and research artifacts.

\subsubsection{Available Benchmarks}
Here, we highlight two types of benchmarks that are particularly useful for empirical QST studies: (i) benchmarks of scalable high-level quantum programs for controlled experiments, and (ii) real-world quantum programs for evaluating industrial relevance.

\Subtitle{High-level quantum programs} 
Based on RQ10, the PUT benchmarks used by at least two studies include MQT Bench and VeriQBench. However, these benchmarks mainly consist of assembly-based low-level quantum programs and were primarily created for QC research rather than being tailored to QST. Therefore, there is a need for a public benchmark that targets scalable high-level quantum programs, given their importance in developer-oriented quantum software development. Drawing inspiration from the Software-artifact Infrastructure Repository\footnote{SIR link: \url{https://sir.csc.ncsu.edu/portal/index.php}} (SIR)~\cite{do2005supporting}, QST-specific benchmarks should include not only a pool of programs but also basic documentation and exemplary test cases to improve readability and reusability.

\revise{\Subtitle{Approximation to real-world bugs}}
Combining RQ1 and RQ10, we found that Bugs4Q is currently the only publicly available repository that incorporates real-world buggy programs along with their fixes. However, only a limited subset of these programs is suitable for QST, so it is unrealistic to expect near-term studies to rely exclusively on real-world quantum programs. In the long term, future work can expand the set of PUTs derived from real-world quantum software development. Moreover, mutation operators are crucial for controlled empirical studies in QST: a sufficiently rich and representative set of operators can approximate a broad spectrum of bug types that may arise in practice. 
\revise{As a suggestion, when designing mutation operators, refer to the real-world bug types summarized in systematic empirical studies, like~\cite{paltenghi2022bugs, bensoussan2025taxonomy}. It would be more helpful to measure how well the approximation matches real-world scenarios (e.g., the coverage of bug types) to support a systematic discussion of potential threats to validity.}
\revise{In addition, motivated} by the limited availability observed in RQ2, we suggest extending mutation operators for classical and hybrid \revise{target when testing high-level quantum programs. The insufficiency of the two types of mutation targets can be attributed to the current focus on testing low-level quantum programs without modular design.}

\subsubsection{Research Artifacts}
\highlight{RQ10 indicates that open-source artifacts are increasingly recognized in the QSE community. Future empirical studies on QST can follow this practice, and it is better to provide long-term archived repositories. Beyond availability, high-quality artifacts should be reusable and functional, consistent with ACM artifact review and badging requirements\footnote{Artifact Review and Badging Version 1.1: \url{https://www.acm.org/publications/policies/artifact-review-and-badging-current}}. For example, when prior work is used as a baseline, releasing executable artifacts can reduce the burden on follow-up studies and help avoid methodological bias introduced during re-implementation and reproduction.}

\begin{EnvRevise}
\subsection{Actionable Recommendations for Future Studies}\label{sec: recommendation}
\begin{table}[!t]
    \small
    \renewcommand{\arraystretch}{0.92}
    \centering
    \caption{List of actionable recommendations}
    \label{tab: recommendation_list}
    \resizebox{.98\columnwidth}{!}{
        \revise{\begin{tabular}{p{0.11\columnwidth}  c  p{0.85\columnwidth}}
    \toprule[1pt]
    \multicolumn{1}{c}{\textbf{Aspect}} 
    & \textbf{RQ} 
    & \multicolumn{1}{c }{\textbf{Recommendation}} \\
    \cmidrule(lr){1-1} 
    \cmidrule(lr){2-2}
    \cmidrule(lr){3-3}
    Design preference
    & \hyperref[sec: rq1]{1} 
    & When evaluating testing approaches on general quantum algorithms and subroutines, the Quantum Fourier Transform deserves the prioritized consideration.\\ \cmidrule(lr){2-2} \cmidrule(lr){3-3}
    & \hyperref[sec: rq6]{6}
    & Utilize both cost- and effectiveness-related metrics to identify how the test approach achieves a balance between the two mutually contradicting aspects. 
    \\ \cmidrule(lr){2-2} \cmidrule(lr){3-3}
    & \hyperref[sec: rq7]{7} 
    & Regarding baseline adoption, it is necessary to consider naive approaches like random testing and random search, and then demonstrate that the proposed approach can outperform them.\\ \cmidrule(lr){2-2} \cmidrule(lr){3-3}
    & \hyperref[sec: rq9]{9} 
    & Using an ideal simulation is allowed in cases without explicitly benchmarking quantum hardware. Meanwhile, evaluation on multiple backends, including noise simulation and physical hardware, could strengthen the approach's applicability in the NISQ era.\\ \cmidrule(lr){2-2} \cmidrule(lr){3-3}
    \cmidrule(lr){1-1} 
    \cmidrule(lr){2-2}
    \cmidrule(lr){3-3}
    Referable quantity  
    & \hyperref[sec: rq1]{1} 
    & To preserve the evaluation diversity on testing general quantum algorithms and subroutines, employ no fewer than 5 programs with heterogeneous functionalities. \\ \cmidrule(lr){2-2} \cmidrule(lr){3-3}
    & \hyperref[sec: rq2]{2} 
    & In terms of sufficiency, the lower bounds for mutant-level variants and version-level variants are 2,200 and 19, respectively.
    \\ \cmidrule(lr){2-2} \cmidrule(lr){3-3}
    & \hyperref[sec: rq3]{3}
    & When assessing test scalability, execute the CUT with at least 10 qubits, and it is still feasible to explore up to 20 qubits in QST in the NISQ era. 
    \\ \cmidrule(lr){2-2} \cmidrule(lr){3-3}
    & \hyperref[sec: rq8]{8}
    & If high execution overhead, setting experimental repetitions below 50 would be acceptable, while proving that conclusion validity is not severely harmed by this compromise.
    \\ 
    \cmidrule(lr){1-1} 
    \cmidrule(lr){2-2}
    \cmidrule(lr){3-3}
    Reporting practice 
    & \hyperref[sec: rq1]{1}, \hyperref[sec: rq2]{2}, \hyperref[sec: rq7]{7}
    & Regarding testing quantum algorithms and subroutines without real-world bugs, report the design of mutation operators and the generation of buggy variants. Then, analyze the construct validity of corresponding evaluation metrics, such as mutation score for mutant-level variants.
    \\ \cmidrule(lr){2-2} \cmidrule(lr){3-3}
    & \hyperref[sec: rq3]{3}
    & Have a comprehensive report of the PUT complexity, such as disclosing all of the width, size, and depth when quantifying the CUT scale. 
    \\ \cmidrule(lr){2-2} \cmidrule(lr){3-3}
    & \hyperref[sec: rq3]{3}
    & Analyze the scalability of test approaches, such as executing CUTs with different scales. 
    \\ \cmidrule(lr){2-2} \cmidrule(lr){3-3}
    & \hyperref[sec: rq4]{4}, \hyperref[sec: rq5]{5}
    & Elaborate on the design of the test cases, including the generation of test inputs and the derivation of test oracles. Then, justify how the test cases align with the PUTs' characteristics and the testing objectives.
    \\ \cmidrule(lr){2-2} \cmidrule(lr){3-3}
    & \hyperref[sec: rq7]{7}
    & Report how baselines are introduced fairly and appropriately. Besides, discuss the statistical power of comparisons among approaches, for example, using matched statistical tests.
    \\ \cmidrule(lr){2-2} \cmidrule(lr){3-3}
    & \hyperref[sec: rq8]{8}, \hyperref[sec: rq9]{9}
    & Report the configuration of shot counts and experimental repetitions when using a shot-based backend, and suggest a discussion about the rationale or limitation behind this configuration.
    \\ \cmidrule(lr){2-2} \cmidrule(lr){3-3}
    & \hyperref[sec: rq9]{9}
    & Clarify the backends employed, specifying whether they involve classical simulation or quantum hardware, are noise-free or noisy, are shot-based or shot-independent, and other possible factors associated with the test requirements. 
    \\
    \cmidrule(lr){1-1} 
    \cmidrule(lr){2-2}
    \cmidrule(lr){3-3}
    Research gap
    & \hyperref[sec: rq1]{1}, \hyperref[sec: rq2]{2}
    & Further explore techniques for testing quantum machine learning models, and investigate the similarity as well as the difference to practices of testing their counterparts in CSE. \\ \cmidrule(lr){2-2} \cmidrule(lr){3-3}  
    & \hyperref[sec: rq1]{1}, \hyperref[sec: rq2]{2}, \hyperref[sec: rq10]{10} 
    & Construct or enlarge benchmarks for real-world bug types and quantum programs, making them more suitable for QST research.
    \\ \cmidrule(lr){2-2} \cmidrule(lr){3-3}
    & \hyperref[sec: rq2]{2}
    & Extend the pool of mutation operators to match the rich abstractions of high-level quantum programs, for example, by considering variables as classical mutation targets and branches as hybrid mutation targets.
    \\ \cmidrule(lr){2-2} \cmidrule(lr){3-3}
    & \hyperref[sec: rq8]{8} 
    & Further investigate adaptive or varied configurations of shots, which are helpful to evaluate the internal validity of empirical studies.\\   
    \cmidrule(lr){2-2}
    \cmidrule(lr){3-3}
    & \hyperref[sec: rq8]{8} 
    & Explore the statistical power regarding the joint impact of the number of shots and experimental repetitions.\\   
    \cmidrule(lr){1-1} 
    \cmidrule(lr){2-2}
    \cmidrule(lr){3-3}
    Tooling Availability
    & \hyperref[sec: rq10]{10}
    & Make artifacts publicly available on both GitHub for easy access, and long-term archive platforms such as Zenodo or Figshare for persistent storage.
    \\
    \bottomrule[1pt]
\end{tabular}
 }
    }
\end{table}

Based on the statistical evidence in each RQ and the discussions above, we summarize a list of actionable recommendations in Table~\ref{tab: recommendation_list} for future QST research with empirical studies. Five aspects are suggested as key points warranting attention, and each recommendation links to the relevant RQs for traceability. 

``Design preference'' is derived from the conventions adopted by the majority of primary studies and guides specific choices in empirical research design. When applicable and appropriate, adhering to these preferences facilitates more standardized and comparable evaluations across studies.
Also, we consider a ``referable quantity'' that provides more concrete numbers based on our collected statistics, such as the median of a boxplot and the mode of a histogram. 
Next, ``reporting practice'' focuses on guiding research papers to provide a more comprehensive report and analysis, thereby strengthening the reasonability of design, the soundness of observation, and the transparency for reproduction. 
Then, in contrast to ``design preference'' derived from majority practices, ``research gap'' highlights important directions that remain underexplored or insufficiently addressed in existing studies, yet are critical for the long-term advancement of QST. Finally, aside from the design and implementation of empirical studies, ``tooling availability'' offers suggestions for research artifacts. 
\end{EnvRevise}


\section{Related Work}
\label{sec: related}

QSE is an interdisciplinary field that studies how to apply and adapt software engineering principles and practices to the development and maintenance of quantum software and hybrid quantum-classical systems, where the quantum components are executed on quantum hardware or simulators. Several recent reviews and surveys have helped connect QC and SE by consolidating terminology, problem spaces, and research directions~\cite{zhao2020quantum, de2024quantum,murillo2025quantum}. For example, Zhao~\cite{zhao2020quantum} presented a mapping study of QSE from the perspective of the software development life cycle, and Murillo et al.~\cite{murillo2025quantum} reviewed recent QSE studies and summarized challenges expected to be addressed in the coming years.

From the perspective of bringing software engineering practices to QC, Miranskyy et al.~\cite{miranskyy2019testing} were among the first to discuss testing quantum programs. As in \revise{CSE}, software testing has become a central topic in QSE due to the need for software quality assurance. Several studies~\cite{garcia2023quantum, paltenghi2024survey, leite2025testing} surveyed broader testing-related areas than the scope of our work. For instance, Garc{\'\i}a de la Barrera et al.~\cite{garcia2023quantum} discussed testing reversible circuits; Paltenghi et al.~\cite{paltenghi2024survey} covered static analysis in addition to runtime execution; and Leite Ramalho et al.~\cite{leite2025testing} included bug reports from quantum computing platforms. In contrast, our paper focuses on QST, i.e., testing quantum programs that incorporate quantum subroutines, and specifically examines how empirical studies for QST are designed, executed, and reported. This focus is motivated by the rapid growth of QST studies in recent years and aligns with the empirical evaluation practices commonly expected in software testing research.

Moreover, most existing reviews and surveys in QSE emphasize proposed techniques, while empirical study practices are discussed only briefly. Only two studies~\cite{gierisch2025qef, zhang2025empirical} systematically examined experimental practices and provided practical guidance. Gierisch et al.~\cite{gierisch2025qef} proposed an experimental framework for empirical studies of quantum software, including recommendations on experiment descriptors and evaluation metrics. Zhang et al.~\cite{zhang2025empirical} analyzed experimental designs in studies that apply quantum or quantum-inspired optimization to classical SE problems. While these two studies and our work share concerns shaped by QC characteristics (e.g., shot counts and execution backends), our SLR focuses on empirical studies in QST and investigates QST-specific issues such as PUT selection, test input design, and the test oracle problem.

\section{Conclusion}
\label{sec: conclusion}

With the continued advancement of quantum computing and the growing interest in applying software engineering practices to quantum software systems, QST has become important for assuring software quality through runtime execution against program specifications. Empirical studies are widely used to evaluate QST techniques, yet the community still lacks shared conventions on how such studies should be designed, executed, and reported. Motivated by this gap, we presented a methodological analysis of empirical studies in QST to summarize current practice and to provide guidance and directions for future work.

We reviewed~\NumberOfPrimaryStudies primary studies and formulated ten research questions covering four key aspects of QST empirical studies:~\MakeLowercase{\GroupObj},~\MakeLowercase{\GroupTest},~\MakeLowercase{\GroupEval}, and~\MakeLowercase{\GroupExp}. Our findings characterize how existing studies select and construct PUTs, design test inputs and test oracles, quantify effectiveness and efficiency, and configure experimental settings and tool support.

Based on these findings, we highlight several methodological considerations for future empirical QST studies. First, the study design should be aligned with plausible real-world requirements and clearly state whether the goal is code-level fault detection or system-level evaluation under backend effects. Second, the test oracle problem deserves explicit treatment, and oracle choices should be justified against the program specification and the intended notion of correctness. Third, scalability should be addressed with clear complexity measures and with reporting that separates the effect of program scale from confounding factors such as test-suite size, shot counts, and repetitions. Finally, releasing reusable benchmarks and research artifacts can improve transparency, reproducibility, and cumulative progress in QST.
\revise{Moreover, we summarize actionable recommendations from the perspectives of design preference, referable quantity, reporting practice, research gap, and tooling availability.}

\section*{Acknowledgments}
Yuechen Li was partly supported by the China Scholarship Council (CSC) under No. 202506020099. 
\revise{Generative AI, including ChatGPT and DeepSeek, was employed to assist in code generation and textual polishing, where the authors reviewed and examined all the generated code and texts.}

\bibliographystyle{ACM-Reference-Format}
\bibliography{bib/primary_studies, bib/manual, bib/program_sources}

@inproceedings{wohlin2014guidelines,
  title={Guidelines for snowballing in systematic literature studies and a replication in software engineering},
  author={Wohlin, Claes},
  booktitle={Proceedings of the 18th international conference on evaluation and assessment in software engineering},
  pages={1--10},
  year={2014}
}

@article{he2025llm,
  title={LLM-Based Multi-Agent Systems for Software Engineering: Literature Review, Vision, and the Road Ahead},
  author={He, Junda and Treude, Christoph and Lo, David},
  journal={ACM Transactions on Software Engineering and Methodology},
  volume={34},
  number={5},
  pages={1--30},
  year={2025},
  publisher={ACM New York, NY}
}

@article{marcen2024systematic,
  title={A systematic literature review of model-driven engineering using machine learning},
  author={Marc{\'e}n, Ana C and Iglesias, Antonio and Lape{\~n}a, Ra{\'u}l and P{\'e}rez, Francisca and Cetina, Carlos},
  journal={IEEE Transactions on Software Engineering},
  year={2024},
  publisher={IEEE}
}

@inproceedings{du2019deepstellar,
  title={Deepstellar: Model-based quantitative analysis of stateful deep learning systems},
  author={Du, Xiaoning and Xie, Xiaofei and Li, Yi and Ma, Lei and Liu, Yang and Zhao, Jianjun},
  booktitle={Proceedings of the 2019 27th ACM joint meeting on European software engineering conference and symposium on the foundations of software engineering},
  pages={477--487},
  year={2019}
}

@inproceedings{pei2017deepxplore,
  title={Deepxplore: Automated whitebox testing of deep learning systems},
  author={Pei, Kexin and Cao, Yinzhi and Yang, Junfeng and Jana, Suman},
  booktitle={proceedings of the 26th Symposium on Operating Systems Principles},
  pages={1--18},
  year={2017}
}

@article{zhang2020machine,
  title={Machine learning testing: Survey, landscapes and horizons},
  author={Zhang, Jie M and Harman, Mark and Ma, Lei and Liu, Yang},
  journal={IEEE Transactions on Software Engineering},
  volume={48},
  number={1},
  pages={1--36},
  year={2020},
  publisher={IEEE}
}

@article{cong2019quantum,
  title={Quantum convolutional neural networks},
  author={Cong, Iris and Choi, Soonwon and Lukin, Mikhail D},
  journal={Nature Physics},
  volume={15},
  number={12},
  pages={1273--1278},
  year={2019},
  publisher={Nature Publishing Group UK London}
}

@article{mcclean2018barren,
  title={Barren plateaus in quantum neural network training landscapes},
  author={McClean, Jarrod R and Boixo, Sergio and Smelyanskiy, Vadim N and Babbush, Ryan and Neven, Hartmut},
  journal={Nature communications},
  volume={9},
  number={1},
  pages={4812},
  year={2018},
  publisher={Nature Publishing Group UK London}
}

@inproceedings{lago2024threats,
  title={Threats to validity in software engineering--hypocritical paper section or essential analysis?},
  author={Lago, Patricia and Runeson, Per and Song, Qunying and Verdecchia, Roberto},
  booktitle={Proceedings of the 18th ACM/IEEE International Symposium on Empirical Software Engineering and Measurement},
  pages={314--324},
  year={2024}
}

@inproceedings{wright2010validity,
  title={Validity concerns in software engineering research},
  author={Wright, Hyrum K and Kim, Miryung and Perry, Dewayne E},
  booktitle={Proceedings of the FSE/SDP workshop on Future of software engineering research},
  pages={411--414},
  year={2010}
}

@article{zhao2023bugs4q,
  title={Bugs4Q: A benchmark of existing bugs to enable controlled testing and debugging studies for quantum programs},
  author={Zhao, Pengzhan and Miao, Zhongtao and Lan, Shuhan and Zhao, Jianjun},
  journal={Journal of Systems and Software},
  volume={205},
  pages={111805},
  year={2023},
  publisher={Elsevier}
}

@article{murillo2025quantum,
  title={Quantum software engineering: Roadmap and challenges ahead},
  author={Murillo, Juan Manuel and Garcia-Alonso, Jose and Moguel, Enrique and Barzen, Johanna and Leymann, Frank and Ali, Shaukat and Yue, Tao and Arcaini, Paolo and P{\'e}rez-Castillo, Ricardo and Garc{\'\i}a-Rodr{\'\i}guez de Guzm{\'a}n, Ignacio and others},
  journal={ACM Transactions on Software Engineering and Methodology},
  volume={34},
  number={5},
  pages={1--48},
  year={2025},
  publisher={ACM New York, NY}
}

@article{shen1983software,
  title={Software science revisited: A critical analysis of the theory and its empirical support},
  author={Shen, Vincent Yun and Conte, Samuel D. and Dunsmore, Hubert E.},
  journal={IEEE Transactions on Software Engineering},
  number={2},
  pages={155--165},
  year={1983},
  publisher={IEEE}
}

@book{halstead1977elements,
  title={Elements of Software Science (Operating and programming systems series)},
  author={Halstead, Maurice H},
  year={1977},
  publisher={Elsevier Science Inc.}
}

@article{weyuker1988evaluating,
  title={Evaluating software complexity measures},
  author={Weyuker, Elaine J},
  journal={IEEE Transactions on Software Engineering},
  volume={14},
  number={9},
  pages={1357--1365},
  year={1988},
  publisher={IEEE}
}

@inproceedings{zhao2021some,
  title={Some size and structure metrics for quantum software},
  author={Zhao, Jianjun},
  booktitle={2021 IEEE/ACM 2nd International Workshop on Quantum Software Engineering (Q-SE)},
  pages={22--27},
  year={2021},
  organization={IEEE}
}

@article{gottesman1998heisenberg,
  title={The Heisenberg representation of quantum computers},
  author={Gottesman, Daniel},
  journal={arXiv preprint quant-ph/9807006},
  year={1998}
}

@article{aaronson2004improved,
  title={Improved simulation of stabilizer circuits},
  author={Aaronson, Scott and Gottesman, Daniel},
  journal={Physical Review A—Atomic, Molecular, and Optical Physics},
  volume={70},
  number={5},
  pages={052328},
  year={2004},
  publisher={APS}
}

@inproceedings{gay2010baseline,
  title={A baseline method for search-based software engineering},
  author={Gay, Gregory},
  booktitle={Proceedings of the 6th International Conference on Predictive Models in Software Engineering},
  pages={1--11},
  year={2010}
}

@article{meyes2019ablation,
  title={Ablation studies in artificial neural networks},
  author={Meyes, Richard and Lu, Melanie and De Puiseau, Constantin Waubert and Meisen, Tobias},
  journal={arXiv preprint arXiv:1901.08644},
  year={2019}
}

@inproceedings{fortunato2022qmutpy,
  title={QMutPy: A mutation testing tool for quantum algorithms and applications in Qiskit},
  author={Fortunato, Daniel and Campos, Jos{\'e} and Abreu, Rui},
  booktitle={Proceedings of the 31st ACM SIGSOFT International Symposium on Software Testing and Analysis},
  pages={797--800},
  year={2022}
}

@inproceedings{chen2004adaptive,
  title={Adaptive random testing},
  author={Chen, Tsong Yueh and Leung, Hing and Mak, Ieng Kei},
  booktitle={Annual Asian Computing Science Conference},
  pages={320--329},
  year={2004},
  organization={Springer}
}

@inproceedings{abreu2007accuracy,
  title={On the accuracy of spectrum-based fault localization},
  author={Abreu, Rui and Zoeteweij, Peter and Van Gemund, Arjan JC},
  booktitle={Testing: Academic and industrial conference practice and research techniques-MUTATION (TAICPART-MUTATION 2007)},
  pages={89--98},
  year={2007},
  organization={IEEE}
}

@article{bourguignon2007selection,
  title={Selection bias corrections based on the multinomial logit model: Monte Carlo comparisons},
  author={Bourguignon, Fran{\c{c}}ois and Fournier, Martin and Gurgand, Marc},
  journal={Journal of Economic surveys},
  volume={21},
  number={1},
  pages={174--205},
  year={2007},
  publisher={Wiley Online Library}
}

@article{buhrman2001quantum,
  title={Quantum fingerprinting},
  author={Buhrman, Harry and Cleve, Richard and Watrous, John and De Wolf, Ronald},
  journal={Physical review letters},
  volume={87},
  number={16},
  pages={167902},
  year={2001},
  publisher={APS}
}

@inproceedings{mazouni2024policy,
  title={Policy Testing with MDPFuzz (Replicability Study)},
  author={Mazouni, Quentin and Spieker, Helge and Gotlieb, Arnaud and Acher, Mathieu},
  booktitle={Proceedings of the 33rd ACM SIGSOFT International Symposium on Software Testing and Analysis},
  pages={1567--1578},
  year={2024}
}

@inproceedings{arcuri2011adaptive,
  title={Adaptive random testing: An illusion of effectiveness?},
  author={Arcuri, Andrea and Briand, Lionel},
  booktitle={Proceedings of the 2011 International Symposium on Software Testing and Analysis},
  pages={265--275},
  year={2011}
}

@inproceedings{arcuri2011practical,
  title={A practical guide for using statistical tests to assess randomized algorithms in software engineering},
  author={Arcuri, Andrea and Briand, Lionel},
  booktitle={Proceedings of the 33rd international conference on software engineering},
  pages={1--10},
  year={2011}
}

@article{zhao2020quantum,
  title={Quantum software engineering: Landscapes and horizons},
  author={Zhao, Jianjun},
  journal={arXiv preprint arXiv:2007.07047},
  year={2020}
}

@article{de2024quantum,
  title={The quantum frontier of software engineering: A systematic mapping study},
  author={De Stefano, Manuel and Pecorelli, Fabiano and Di Nucci, Dario and Palomba, Fabio and De Lucia, Andrea},
  journal={Information and Software Technology},
  volume={175},
  pages={107525},
  year={2024},
  publisher={Elsevier}
}

@inproceedings{miranskyy2019testing,
  title={On testing quantum programs},
  author={Miranskyy, Andriy and Zhang, Lei},
  booktitle={2019 IEEE/ACM 41st International Conference on Software Engineering: New Ideas and Emerging Results (ICSE-NIER)},
  pages={57--60},
  year={2019},
  organization={IEEE}
}

@article{garcia2023quantum,
  title={Quantum software testing: State of the art},
  author={Garc{\'\i}a de la Barrera, Antonio and Garc{\'\i}a-Rodr{\'\i}guez de Guzm{\'a}n, Ignacio and Polo, Macario and Piattini, Mario},
  journal={Journal of Software: Evolution and Process},
  volume={35},
  number={4},
  pages={e2419},
  year={2023},
  publisher={Wiley Online Library}
}

@article{paltenghi2024survey,
  title={A survey on testing and analysis of quantum software},
  author={Paltenghi, Matteo and Pradel, Michael},
  journal={arXiv preprint arXiv:2410.00650},
  year={2024}
}

@article{leite2025testing,
  title={Testing and debugging quantum programs: The road to 2030},
  author={Leite Ramalho, Neilson Carlos and Amario de Souza, Higor and Lordello Chaim, Marcos},
  journal={ACM Transactions on Software Engineering and Methodology},
  volume={34},
  number={5},
  pages={1--46},
  year={2025},
  publisher={ACM New York, NY}
}

@article{zhang2025empirical,
  title={Empirical Studies on Quantum Optimization for Software Engineering: A Systematic Analysis},
  author={Zhang, Man and Li, Yuechen and Yue, Tao and Cai, Kai-Yuan},
  journal={arXiv preprint arXiv:2510.27113},
  year={2025}
}

@article{gierisch2025qef,
  title={QEF: Reproducible and Exploratory Quantum Software Experiments},
  author={Gierisch, Vincent and Mauerer, Wolfgang},
  journal={arXiv preprint arXiv:2511.04563},
  year={2025}
}

@article{barnett2009quantum,
  title={Quantum state discrimination},
  author={Barnett, Stephen M and Croke, Sarah},
  journal={Advances in Optics and Photonics},
  volume={1},
  number={2},
  pages={238--278},
  year={2009},
  publisher={Optical Society of America}
}

@article{reggio2024fast,
  title={Fast partitioning of pauli strings into commuting families for optimal expectation value measurements of dense operators},
  author={Reggio, Ben and Butt, Nouman and Lytle, Andrew and Draper, Patrick},
  journal={Physical Review A},
  volume={110},
  number={2},
  pages={022606},
  year={2024},
  publisher={APS}
}

@article{barr2014oracle,
  title={The oracle problem in software testing: A survey},
  author={Barr, Earl T and Harman, Mark and McMinn, Phil and Shahbaz, Muzammil and Yoo, Shin},
  journal={IEEE transactions on software engineering},
  volume={41},
  number={5},
  pages={507--525},
  year={2014},
  publisher={IEEE}
}

@book{10.5555/1074100,
author = {Ralston, Anthony and Reilly, Edwin D. and Hemmendinger, David},
title = {Encyclopedia of Computer Science},
year = {2003},
isbn = {0470864125},
publisher = {John Wiley and Sons Ltd.},
address = {GBR}
}

@inproceedings{davis1981pseudo,
  title={Pseudo-oracles for non-testable programs},
  author={Davis, Martin D and Weyuker, Elaine J},
  booktitle={Proceedings of the ACM'81 Conference},
  pages={254--257},
  year={1981}
}

@article{sjoberg2022construct,
  title={Construct validity in software engineering},
  author={Sj{\o}berg, Dag IK and Bergersen, Gunnar Rye},
  journal={IEEE Transactions on Software Engineering},
  volume={49},
  number={3},
  pages={1374--1396},
  year={2022},
  publisher={IEEE}
}

@article{zhou2018cost,
  title={A cost-effective software testing strategy employing online feedback information},
  author={Zhou, Zhi Quan and Sinaga, Arnaldo and Susilo, Willy and Zhao, Lei and Cai, Kai-Yuan},
  journal={Information sciences},
  volume={422},
  pages={318--335},
  year={2018},
  publisher={Elsevier}
}

@article{basili2006comparing,
  title={Comparing the effectiveness of software testing strategies},
  author={Basili, Victor R and Selby, Richard W},
  journal={IEEE transactions on software engineering},
  number={12},
  pages={1278--1296},
  year={2006},
  publisher={IEEE}
}

@article{zhang2025quantum,
  title={Quantum Optimization for Software Engineering: A Survey},
  author={Zhang, Man and Li, Yuechen and Yue, Tao and Cai, Kai-Yuan},
  journal={arXiv preprint arXiv:2506.16878},
  year={2025}
}

@book{huizinga2007automated,
  title={Automated defect prevention: best practices in software management},
  author={Huizinga, Dorota and Kolawa, Adam},
  year={2007},
  publisher={John Wiley \& Sons}
}

@inproceedings{yue2023towards,
  title={Towards quantum software requirements engineering},
  author={Yue, Tao and Ali, Shaukat and Arcaini, Paolo},
  booktitle={2023 IEEE International Conference on Quantum Computing and Engineering (QCE)},
  volume={2},
  pages={161--164},
  year={2023},
  organization={IEEE}
}

@article{sepulveda2024systematic,
  title={Systematic review on requirements engineering in quantum computing: Insights and future directions},
  author={Sepulveda, Samuel and Cravero, Ania and Fonseca, Guillermo and Antonelli, Leandro},
  journal={Electronics},
  volume={13},
  number={15},
  pages={2989},
  year={2024},
  publisher={MDPI}
}

@inproceedings{zhao2024towards,
  title={Towards an architecture description language for hybrid quantum-classical systems},
  author={Zhao, Jianjun},
  booktitle={2024 IEEE International Conference on Quantum Software (QSW)},
  pages={19--23},
  year={2024},
  organization={IEEE}
}

@inproceedings{chen2020revisiting,
  title={Revisiting the relationship between fault detection, test adequacy criteria, and test set size},
  author={Chen, Yiqun T and Gopinath, Rahul and Tadakamalla, Anita and Ernst, Michael D and Holmes, Reid and Fraser, Gordon and Ammann, Paul and Just, Ren{\'e}},
  booktitle={Proceedings of the 35th IEEE/ACM international conference on automated software engineering},
  pages={237--249},
  year={2020}
}

@article{ye2025measurement,
  title={Is Measurement Enough? Rethinking Output Validation in Quantum Program Testing},
  author={Ye, Jiaming and Wu, Xiongfei and Xia, Shangzhou and Zhang, Fuyuan and Zhao, Jianjun},
  journal={arXiv preprint arXiv:2509.16595},
  year={2025}
}

@article{miranskyy2025cost,
  title={The Cost of Certainty: Shot Budgets in Quantum Program Testing},
  author={Miranskyy, Andriy},
  journal={arXiv preprint arXiv:2510.22418},
  year={2025}
}

@article{bellman1966dynamic,
  title={Dynamic programming},
  author={Bellman, Richard},
  journal={science},
  volume={153},
  number={3731},
  pages={34--37},
  year={1966},
  publisher={American Association for the Advancement of Science}
}

@article{padgham2013model,
  title={Model-based test oracle generation for automated unit testing of agent systems},
  author={Padgham, Lin and Zhang, Zhiyong and Thangarajah, John and Miller, Tim},
  journal={IEEE Transactions on Software Engineering},
  volume={39},
  number={9},
  pages={1230--1244},
  year={2013},
  publisher={IEEE}
}

@inproceedings{dinella2022toga,
  title={Toga: A neural method for test oracle generation},
  author={Dinella, Elizabeth and Ryan, Gabriel and Mytkowicz, Todd and Lahiri, Shuvendu K},
  booktitle={Proceedings of the 44th International Conference on Software Engineering},
  pages={2130--2141},
  year={2022}
}

@article{chen2018metamorphic,
  title={Metamorphic testing: A review of challenges and opportunities},
  author={Chen, Tsong Yueh and Kuo, Fei-Ching and Liu, Huai and Poon, Pak-Lok and Towey, Dave and Tse, TH and Zhou, Zhi Quan},
  journal={ACM Computing Surveys (CSUR)},
  volume={51},
  number={1},
  pages={1--27},
  year={2018},
  publisher={ACM New York, NY, USA}
}

@article{segura2016survey,
  title={A survey on metamorphic testing},
  author={Segura, Sergio and Fraser, Gordon and Sanchez, Ana B and Ruiz-Cort{\'e}s, Antonio},
  journal={IEEE Transactions on software engineering},
  volume={42},
  number={9},
  pages={805--824},
  year={2016},
  publisher={IEEE}
}

@inproceedings{saito2024towards,
  title={Towards n-version quantum software systems for reliable classical-quantum computing},
  author={Saito, Shinobu and Endo, Suguru and Suzuki, Yasunari},
  booktitle={2024 IEEE 35th International Symposium on Software Reliability Engineering Workshops (ISSREW)},
  pages={119--120},
  year={2024},
  organization={IEEE}
}

@article{do2005supporting,
  title={Supporting controlled experimentation with testing techniques: An infrastructure and its potential impact},
  author={Do, Hyunsook and Elbaum, Sebastian and Rothermel, Gregg},
  journal={Empirical Software Engineering},
  volume={10},
  number={4},
  pages={405--435},
  year={2005},
  publisher={Springer}
}

@misc{mosca2008quantumalgorithms,
      title={Quantum Algorithms}, 
      author={Michele Mosca},
      year={2008},
      eprint={0808.0369},
      archivePrefix={arXiv},
      primaryClass={quant-ph},
      url={https://arxiv.org/abs/0808.0369}, 
}

@article{abhijith2022quantum,
  title={Quantum Algorithm: Implementations for Beginners},
  author={Abhijith, J and Adedoyin, Adetokunbo and Ambrosiano, John and Anisimov, Petr and Casper, William and Chennupati, Gopinath and Coffrin, Carleton and Djidjev, Hristo and Gunter, David and Karra, Satish and others},
  journal={ACM Transactions on Quantum Computing},
  volume={3},
  number={4},
  pages={1--92},
  year={2022}
}

@article{kak1995quantum,
  title={Quantum neural computing},
  author={Kak, Subhash C},
  journal={Advances in imaging and electron physics},
  volume={94},
  pages={259--313},
  year={1995},
  publisher={Elsevier}
}

@article{liu2018differentiable,
  title={Differentiable learning of quantum circuit born machines},
  author={Liu, Jin-Guo and Wang, Lei},
  journal={Physical Review A},
  volume={98},
  number={6},
  pages={062324},
  year={2018},
  publisher={APS}
}

@article{preskill2018quantum,
  title={Quantum computing in the NISQ era and beyond},
  author={Preskill, John},
  journal={Quantum},
  volume={2},
  pages={79},
  year={2018},
  publisher={Verein zur F{\"o}rderung des Open Access Publizierens in den Quantenwissenschaften}
}

@inproceedings{wang2021qdiff,
  title={QDiff: Differential testing of quantum software stacks},
  author={Wang, Jiyuan and Zhang, Qian and Xu, Guoqing Harry and Kim, Miryung},
  booktitle={2021 36th IEEE/ACM international conference on automated software engineering (ASE)},
  pages={692--704},
  year={2021},
  organization={IEEE}
}

@book{nielsen2010quantum,
  title={Quantum computation and quantum information},
  author={Nielsen, Michael A and Chuang, Isaac L},
  year={2010},
  publisher={Cambridge university press}
}

@article{wood2011tensor,
  title={Tensor networks and graphical calculus for open quantum systems},
  author={Wood, Christopher J and Biamonte, Jacob D and Cory, David G},
  journal={arXiv preprint arXiv:1111.6950},
  year={2011}
}

@article{o2004quantum,
  title={Quantum process tomography of a controlled-NOT gate},
  author={O'Brien, Jeremy L and Pryde, Geoff J and Gilchrist, Alexei and James, Daniel FV and Langford, Nathan K and Ralph, Timothy C and White, Andrew G},
  journal={Physical review letters},
  volume={93},
  number={8},
  pages={080502},
  year={2004},
  publisher={APS}
}

@software{li_2026_18159893,
  author       = {Li, Yuechen and
                  Shao, Minqi and
                  Zhao, Jianjun and
                  Wang, Qichen},
  title        = {Artifact Repository for A Methodological Analysis
                   of Empirical Studies in Quantum Software Testing
                  },
  month        = jan,
  year         = 2026,
  publisher    = {Zenodo},
  version      = {v0.1},
  doi          = {10.5281/zenodo.18159892},
  url          = {https://doi.org/10.5281/zenodo.18159892},
  swhid        = {swh:1:dir:1f7c39e7d99801fa1c78eadb0d44b8bdb3fb7f2d
                   ;origin=https://doi.org/10.5281/zenodo.18159892;vi
                   sit=swh:1:snp:2f19f1452cd30d24a279cd568c9bb7a08628
                   2a86;anchor=swh:1:rel:ab6fb85f7939ef0fac5034b8758f
                   d6d807ea7e1d;path=NahidaNahida-QST-empirical-
                   study-767622b
                  },
}

@article{aleksandrowicz2019qiskit,
  author = {Aleksandrowicz, Gadi and Alexander, Thomas and Barkoutsos, Panagiotis and Bello, Luciano and Ben-Haim, Yael and Bucher, David and Cabrera-Hernández, Francisco Jose and Carballo-Franquis, Jorge and Chen, Adrian and Chen, Chun-Fu and Chow, Jerry M. and Córcoles-Gonzales, Antonio D. and Cross, Abigail J. and Cross, Andrew and Cruz-Benito, Juan and Culver, Chris and De La Puente González, Salvador and De La Torre, Enrique and Ding, Delton and Dumitrescu, Eugene and Duran, Ivan and Eendebak, Pieter and Everitt, Mark and Faro Sertage, Ismael and Frisch, Albert and Fuhrer, Andreas and Gambetta, Jay and Godoy Gago, Borja and Gomez-Mosquera, Juan and Greenberg, Donny and Hamamura, Ikko and Havlicek, Vojtech and Hellmers, Joe and Herok, tukasz and Horii, Hiroshi and Hu, Shaohan and Imamichi, Takashi and Itoko, Toshinari and Javadi-Abhari, Ali and Kanazawa, Naoki and Karazeev, Anton and Krsulich, Kevin and Liu, Peng and Luh, Yang and Maeng, Yunho and Marques, Manoel and Martín-Fernández, Francisco Jose and McClure, Douglas T. and McKay, David and Meesala, Srujan and Mezzacapo, Antonio and Moll, Nikolaj and Moreda Rodríguez, Diego and Nannicini, Giacomo and Nation, Paul and Ollitrault, Pauline and O'Riordan, Lee James and Paik, Hanhee and Pérez, Jesús and Phan, Anna and Pistoia, Marco and Prutyanov, Viktor and Reuter, Max and Rice, Julia and Rodríguez Davila, Abdón and Putra Rudy, Raymond Harry and Ryu, Mingi and Sathaye, Ninad and Schnabel, Chris and Schoute, Eddie and Setia, Kanav and Shi, Yunong and Silva, Adenilton and Siraichi, Yukio and Sivarajah, Seyon and Smolin, John A. and Soeken, Mathias and Takahashi, Hitomi and Tavernelli, Ivano and Taylor, Charles and Taylour, Pete and Trabing, Kenso and Treinish, Matthew and Turner, Wes and Vogt-Lee, Desiree and Vuillot, Christophe and Wildstrom, Jonathan A. and Wilson, Jessica and Winston, Erick and Wood, Christopher and Wood, Stephen and Wörner, Stefan and Akhalwaya, Ismail Yunus and Zoufal, Christa},
  journal = {},
  title = {Qiskit: An Open-source Framework for Quantum Computing},
  year = {2019}
}

@inproceedings{svore2018q,
  author = {Svore, Krysta and Geller, Alan and Troyer, Matthias and Azariah, John and Granade, Christopher and Heim, Bettina and Kliuchnikov, Vadym and Mykhailova, Mariia and Paz, Andres and Roetteler, Martin},
  booktitle = {Proceedings of the Real World Domain Specific Languages Workshop 2018},
  pages = {1--10},
  title = {Q\# Enabling Scalable Quantum Computing and Development with a High-level dsl},
  year = {2018}
}

@misc{google2018cirq,
  author = {Google AI Quantum Team},
  howpublished = {\url{https://github.com/quantumlib/}},
  title = {Cirq},
  year = {2018}
}

@article{bergholm2018pennylane,
  title={Pennylane: Automatic differentiation of hybrid quantum-classical computations},
  author={Bergholm, Ville and Izaac, Josh and Schuld, Maria and Gogolin, Christian and Ahmed, Shahnawaz and Ajith, Vishnu and Alam, M Sohaib and Alonso-Linaje, Guillermo and AkashNarayanan, B and Asadi, Ali and others},
  journal={arXiv preprint arXiv:1811.04968},
  year={2018}
}

@incollection{feynman2018simulating,
  title={Simulating physics with computers},
  author={Feynman, Richard P},
  booktitle={Feynman and computation},
  pages={133--153},
  year={2018},
  publisher={cRc Press}
}

@article{paltenghi2022bugs,
  title={Bugs in quantum computing platforms: an empirical study},
  author={Paltenghi, Matteo and Pradel, Michael},
  journal={Proceedings of the ACM on Programming Languages},
  volume={6},
  number={OOPSLA1},
  pages={1--27},
  year={2022},
  publisher={ACM New York, NY, USA}
}

@article{bensoussan2025taxonomy,
  title={A Taxonomy of Real Faults for Hybrid Quantum-Classical Software Architectures},
  author={Bensoussan, Avner and Jahangirova, Gunel and Mousavi, Mohammadreza},
  journal={ACM Transactions on Software Engineering and Methodology},
  year={2025},
  publisher={ACM New York, NY}
}

@article{li2026dynamic,
  title={A Dynamic Test Oracle for Quantum Programs with Separable Output States},
  author={Li, Yuechen and Cai, Kai-Yuan and Yin, Beibei},
  journal={IEEE Transactions on Software Engineering},
  year={2026},
  publisher={IEEE}
}

@inproceedings{luo2022comprehensive,
  title={A comprehensive study of bug fixes in quantum programs},
  author={Luo, Junjie and Zhao, Pengzhan and Miao, Zhongtao and Lan, Shuhan and Zhao, Jianjun},
  booktitle={2022 IEEE International conference on software analysis, evolution and reengineering (SANER)},
  pages={1239--1246},
  year={2022},
  organization={IEEE}
}

@article{jia2010analysis,
  title={An analysis and survey of the development of mutation testing},
  author={Jia, Yue and Harman, Mark},
  journal={IEEE transactions on software engineering},
  volume={37},
  number={5},
  pages={649--678},
  year={2010},
  publisher={IEEE}
}

@article{papadakis2015metallaxis,
  title={Metallaxis-FL: mutation-based fault localization},
  author={Papadakis, Mike and Le Traon, Yves},
  journal={Software Testing, Verification and Reliability},
  volume={25},
  number={5-7},
  pages={605--628},
  year={2015},
  publisher={Wiley Online Library}
}

@inproceedings{ma2018deepmutation,
  title={Deepmutation: Mutation testing of deep learning systems},
  author={Ma, Lei and Zhang, Fuyuan and Sun, Jiyuan and Xue, Minhui and Li, Bo and Juefei-Xu, Felix and Xie, Chao and Li, Li and Liu, Yang and Zhao, Jianjun and others},
  booktitle={2018 IEEE 29th international symposium on software reliability engineering (ISSRE)},
  pages={100--111},
  year={2018},
  organization={IEEE}
}

@article{mccabe1976complexity,
  title={A complexity measure},
  author={McCabe, Thomas J},
  journal={IEEE Transactions on software Engineering},
  number={4},
  pages={308--320},
  year={1976},
  publisher={IEEE}
}

@article{zhu1997software,
  title={Software unit test coverage and adequacy},
  author={Zhu, Hong and Hall, Patrick AV and May, John HR},
  journal={Acm computing surveys (csur)},
  volume={29},
  number={4},
  pages={366--427},
  year={1997},
  publisher={Acm New York, NY, USA}
}

@article{rothermel2001prioritizing,
  title={Prioritizing test cases for regression testing},
  author={Rothermel, Gregg and Untch, Roland H. and Chu, Chengyun and Harrold, Mary Jean},
  journal={IEEE Transactions on software engineering},
  volume={27},
  number={10},
  pages={929--948},
  year={2001},
  publisher={IEEE}
}

@inproceedings{mendiluze2022muskit,
  author = {Mendiluze, E\~{n}aut and Ali, Shaukat and Arcaini, Paolo and Yue, Tao},
  booktitle = {Proceedings of the 36th IEEE/ACM International Conference on Automated Software Engineering},
  doi = {10.1109/ASE51524.2021.9678563},
  pages = {1266–1270},
  title = {Muskit: a mutation analysis tool for quantum software testing},
  year = {2022}
}

@inproceedings{wang2022quito,
  author = {Wang, Xinyi and Arcaini, Paolo and Yue, Tao and Ali, Shaukat},
  booktitle = {Proceedings of the 36th IEEE/ACM International Conference on Automated Software Engineering},
  doi = {10.1109/ASE51524.2021.9678798},
  pages = {1237–1241},
  title = {Quito: a coverage-guided test generator for quantum programs},
  year = {2022}
}

@inproceedings{wang2024qucat,
  author = {Wang, Xinyi and Arcaini, Paolo and Yue, Tao and Ali, Shaukat},
  booktitle = {Proceedings of the 38th IEEE/ACM International Conference on Automated Software Engineering},
  doi = {10.1109/ASE56229.2023.00062},
  pages = {2066–2069},
  title = {QuCAT: A Combinatorial Testing Tool for Quantum Software},
  year = {2024}
}

@inproceedings{ye2024quratest,
  author = {Ye, Jiaming and Xia, Shangzhou and Zhang, Fuyuan and Arcaini, Paolo and Ma, Lei and Zhao, Jianjun and Ishikawa, Fuyuki},
  booktitle = {Proceedings of the 38th IEEE/ACM International Conference on Automated Software Engineering},
  doi = {10.1109/ASE56229.2023.00196},
  pages = {1149–1161},
  title = {QuraTest: Integrating Quantum Specific Features in Quantum Program Testing},
  year = {2024}
}

@inproceedings{huang2019statistical,
  author = {Huang, Yipeng and Martonosi, Margaret},
  booktitle = {Proceedings of the 46th International Symposium on Computer Architecture},
  doi = {10.1145/3307650.3322213},
  pages = {541–553},
  title = {Statistical assertions for validating patterns and finding bugs in quantum programs},
  year = {2019}
}

@inproceedings{honarvar2020propertybased,
  author = {Honarvar, Shahin and Mousavi, Mohammad Reza and Nagarajan, Rajagopal},
  booktitle = {Proceedings of the IEEE/ACM 42nd International Conference on Software Engineering Workshops},
  doi = {10.1145/3387940.3391459},
  pages = {430–435},
  title = {Property-based Testing of Quantum Programs in Q\#},
  year = {2020}
}

@article{li2020projectionbased,
  author = {Li, Gushu and Zhou, Li and Yu, Nengkun and Ding, Yufei and Ying, Mingsheng and Xie, Yuan},
  doi = {10.1145/3428218},
  journal = {Proc. ACM Program. Lang.},
  number = {OOPSLA},
  title = {Projection-based runtime assertions for testing and debugging Quantum programs},
  volume = {4},
  year = {2020}
}

@article{acharya2021test,
  author = {Acharya, Nikita and Urbanek, Miroslav and De Jong, Wibe A. and Saeed, Samah Mohamed},
  doi = {10.1145/3477928},
  journal = {J. Emerg. Technol. Comput. Syst.},
  number = {1},
  title = {Test Points for Online Monitoring of Quantum Circuits},
  volume = {18},
  year = {2021}
}

@inproceedings{wang2022qusbt,
  author = {Wang, Xinyi and Arcaini, Paolo and Yue, Tao and Ali, Shaukat},
  booktitle = {Proceedings of the ACM/IEEE 44th International Conference on Software Engineering: Companion Proceedings},
  doi = {10.1145/3510454.3516839},
  pages = {173–177},
  title = {QuSBT: search-based testing of quantum programs},
  year = {2022}
}

@inproceedings{wang2022origin,
  author = {Wang, Xinyi and Yu, Tongxuan and Arcaini, Paolo and Yue, Tao and Ali, Shaukat},
  booktitle = {Proceedings of the Genetic and Evolutionary Computation Conference},
  doi = {10.1145/3512290.3528869},
  pages = {1345–1353},
  title = {Origin program output test generation for quantum programs with multi-objective search},
  year = {2022}
}

@inproceedings{wang2022generating,
  author = {Wang, Xinyi and Arcaini, Paolo and Yue, Tao and Ali, Shaukat},
  booktitle = {International Symposium on Search Based Software Engineering},
  pages = {9--25},
  title = {Generating failing test suites for quantum programs with search},
  year = {2022}
}

@inproceedings{abreu2023metamorphic,
  author = {Abreu, Rui and Fernandes, Jo\~{a}o Paulo and Llana, Luis and Tavares, Guilherme},
  booktitle = {2022 IEEE/ACM 3rd International Workshop on Quantum Software Engineering (Q-SE)},
  doi = {10.1145/3528230.3529189},
  pages = {16–23},
  title = {Metamorphic testing of oracle quantum programs},
  year = {2023}
}

@inproceedings{costa2022asserting,
  author = {Costa, Nuno and Fernandes, Jo\~{a}o Paulo and Abreu, Rui},
  booktitle = {Proceedings of the 1st International Workshop on Quantum Programming for Software Engineering},
  doi = {10.1145/3549036.3562062},
  pages = {32–36},
  title = {Asserting the correctness of Shor implementations using metamorphic testing},
  year = {2022}
}

@inproceedings{sato2024locating,
  author = {Sato, Naoto and Katsube, Ryota},
  booktitle = {Proceedings of the 2024 ACM/IEEE 44th International Conference on Software Engineering: New Ideas and Emerging Results},
  doi = {10.1145/3639476.3639761},
  pages = {26–31},
  title = {Locating Buggy Segments in Quantum Program Debugging},
  year = {2024}
}

@inproceedings{pontolillo2024delta,
  author = {Pontolillo, Gabriel Joseph and Mousavi, Mohammad Reza},
  booktitle = {Proceedings of the 5th ACM/IEEE International Workshop on Quantum Software Engineering},
  doi = {10.1145/3643667.3648219},
  pages = {1–8},
  title = {Delta Debugging for Property-Based Regression Testing of Quantum Programs},
  year = {2024}
}

@inproceedings{fortunato2024gate,
  author = {Fortunato, Daniel and Campos, Jos\'{e} and Abreu, Rui},
  booktitle = {Proceedings of the 1st ACM International Workshop on Quantum Software Engineering:the Next Evolution, Qse-Ne 2024},
  doi = {10.1145/3663531.3664753},
  pages = {15–18},
  title = {Gate Branch Coverage: A Metric for Quantum Software Testing},
  year = {2024}
}

@article{shi2025quantest,
  author = {Shi, Jinjing and Xiao, Zimeng and Shi, Heyuan and Jiang, Yu and Li, Xuelong},
  doi = {10.1145/3688840},
  journal = {ACM Trans. Softw. Eng. Methodol.},
  number = {2},
  title = {QuanTest: Entanglement-Guided Testing of Quantum Neural Network Systems},
  volume = {34},
  year = {2025}
}

@article{chen2024automatic,
  author = {Chen, Kean and Ying, Mingsheng},
  doi = {10.1145/3689333},
  journal = {ACM Trans. Des. Autom. Electron. Syst.},
  number = {6},
  title = {Automatic Test Pattern Generation for Robust Quantum Circuit Testing},
  volume = {29},
  year = {2024}
}

@article{kang2024statistical,
  author = {Kang, Chan Gu and Lee, Joonghoon and Oh, Hakjoo},
  doi = {10.1145/3689716},
  journal = {Proc. ACM Program. Lang.},
  number = {OOPSLA2},
  title = {Statistical Testing of Quantum Programs via Fixed-Point Amplitude Amplification},
  volume = {8},
  year = {2024}
}

@inproceedings{muqeet2024quantum,
  author = {Muqeet, Asmar and Ali, Shaukat and Arcaini, Paolo},
  booktitle = {Proceedings of 2024 39th ACM/IEEE International Conference on Automated Software Engineering, ASE 2024},
  doi = {10.1145/3691620.3695275},
  pages = {2130–2141},
  title = {Quantum Program Testing Through Commuting Pauli Strings on IBM's Quantum Computers},
  year = {2024}
}

@inproceedings{yamaguchi2025practical,
  author = {Yamaguchi, Masaomi and Yoshioka, Nobukazu and Ishikawa, Fuyuki},
  booktitle = {Proceedings of the 33rd ACM International Conference on the Foundations of Software Engineering},
  doi = {10.1145/3696630.3731621},
  pages = {1699–1709},
  title = {Practical Design by Contract Framework for Quantum Applications},
  year = {2025}
}

@article{oldfield2025faster,
  author = {Oldfield, Noah H. and Laaber, Christoph and Yue, Tao and Ali, Shaukat},
  doi = {10.1145/3714468},
  journal = {ACM Trans. Softw. Eng. Methodol.},
  number = {7},
  title = {Faster and Better Quantum Software Testing through Specification Reduction and Projective Measurements},
  volume = {34},
  year = {2025}
}

@article{tan2025hornbro,
  author = {Tan, Siwei and Lu, Liqiang and Xiang, Debin and Chu, Tianyao and Lang, Congliang and Chen, Jintao and Hu, Xing and Yin, Jianwei},
  doi = {10.1145/3715751},
  journal = {Proc. ACM Softw. Eng.},
  number = {FSE},
  title = {HornBro: Homotopy-Like Method for Automated Quantum Program Repair},
  volume = {2},
  year = {2025}
}

@article{xia2025quantum,
  author = {Xia, Shangzhou and Zhao, Jianjun and Zhang, Fuyuan and Guo, Xiaoyu},
  doi = {10.1145/3728926},
  journal = {Proc. ACM Softw. Eng.},
  number = {ISSTA},
  title = {Quantum Concolic Testing},
  volume = {2},
  year = {2025}
}

@article{li2025preparation,
  author = {Li, Yuechen and Cai, Kai-Yuan and Yin, Beibei},
  doi = {10.1145/3736757},
  journal = {ACM Trans. Softw. Eng. Methodol.},
  number = {8},
  title = {Preparation and Utilization of Mixed States for Testing Quantum Programs},
  volume = {34},
  year = {2025}
}

@inproceedings{guo2024on,
  author = {Guo, Xiaoyu and Zhao, Jianjun and Zhao, Pengzhan},
  booktitle = {2024 IEEE/ACM 5th International Workshop on Quantum Software Engineering (Q-SE)},
  doi = {},
  number = {},
  pages = {9-16},
  title = {On Repairing Quantum Programs Using ChatGPT},
  volume = {},
  year = {2024}
}

@inproceedings{huang2024a,
  author = {Huang, Linzhi and Pei, Hanyu and Li, Yuechen and Yin, Beibei and Cai, Kai-Yuan},
  booktitle = {2024 IEEE 24th International Conference on Software Quality, Reliability and Security (QRS)},
  doi = {10.1109/QRS62785.2024.00011},
  number = {},
  pages = {1-12},
  title = {A Strategy of Dynamic Random Testing with Hybrid Distance Metrics for Quantum Programs},
  volume = {},
  year = {2024}
}

@inproceedings{rovara2025a,
  author = {Rovara, Damian and Burgholzer, Lukas and Wille, Robert},
  booktitle = {2025 IEEE International Conference on Quantum Software (QSW)},
  doi = {10.1109/QSW67625.2025.00024},
  number = {},
  pages = {130-136},
  title = {A Framework for Debugging Quantum Programs},
  volume = {},
  year = {2025}
}

@inproceedings{ali2021assessing,
  author = {Ali, Shaukat and Arcaini, Paolo and Wang, Xinyi and Yue, Tao},
  booktitle = {2021 14th IEEE Conference on Software Testing, Verification and Validation (ICST 2021)},
  doi = {10.1109/ICST49551.2021.00014},
  number = {},
  pages = {13-23},
  title = {Assessing the Effectiveness of Input and Output Coverage Criteria for Testing Quantum Programs},
  volume = {},
  year = {2021}
}

@inproceedings{wang2021application,
  author = {Wang, Xinyi and Arcaini, Paolo and Yue, Tao and Ali, Shaukat},
  booktitle = {2021 IEEE 21st International Conference on Software Quality, Reliability and Security (QRS 2021)},
  doi = {10.1109/QRS54544.2021.00029},
  number = {},
  pages = {179-188},
  title = {Application of Combinatorial Testing to Quantum Programs},
  volume = {},
  year = {2021}
}

@inproceedings{pan2023understanding,
  author = {Pan, Zhonghao and Feng, Yang and Li, Zhiyuan and Liu, Yunxin and Li, Yuanchun},
  booktitle = {2023 IEEE International Conference on Software Analysis, Evolution and Reengineering (SANER)},
  doi = {10.1109/SANER56733.2023.00047},
  number = {},
  pages = {426-437},
  title = {Understanding the Impact of Quantum Noise on Quantum Programs},
  volume = {},
  year = {2023}
}

@article{long2024equivalence,
  author = {Long, Peixun and Zhao, Jianjun},
  doi = {10.1016/j.jss.2024.112000},
  journal = {Journal of Systems and Software},
  title = {Equivalence, identity, and unitarity checking in black-box testing of quantum programs},
  volume = {211},
  year = {2024}
}

@article{long2024testing,
  author = {Long, Peixun and Zhao, Jianjun},
  doi = {10.1145/3656339},
  journal = {ACM Transactions on Software Engineering and Methodology},
  number = {6},
  title = {Testing Multi-Subroutine Quantum Programs: From Unit Testing to Integration Testing},
  volume = {33},
  year = {2024}
}

@article{muqeet2024mitigating,
  author = {Muqeet, Asmar and Yue, Tao and Ali, Shaukat and Arcaini, Paolo},
  doi = {10.1109/TSE.2024.3462974},
  journal = {IEEE Transactions on Software Engineering},
  number = {11},
  pages = {2947-2961},
  title = {Mitigating Noise in Quantum Software Testing Using Machine Learning},
  volume = {50},
  year = {2024}
}

@article{fortunato2022mutation,
  author = {Fortunato, Daniel and CAMPOS, JOSÉ and ABREU, RUI},
  doi = {10.1109/TQE.2022.3195061},
  journal = {IEEE Transactions on Quantum Engineering},
  number = {},
  pages = {1-17},
  title = {Mutation Testing of Quantum Programs: A Case Study With Qiskit},
  volume = {3},
  year = {2022}
}

@article{usandizaga2025quantum,
  author = {Usandizaga, Enaut Mendiluze and Ali, Shaukat and Yue, Tao and Arcaini, Paolo},
  doi = {10.1007/s10664-025-10643-z},
  journal = {Empirical Software Engineering},
  number = {3},
  title = {Quantum circuit mutants: Empirical analysis and recommendations},
  volume = {30},
  year = {2025}
}

@article{park2025squad,
  author = {Park, Soohyun and Cho, Jae Hyun and Yook, Hyun Jun and Jhun, Ga San and Lee, Youn Kyu and Kim, Joongheon},
  doi = {10.1007/s11227-025-07556-5},
  journal = {Journal of Supercomputing},
  number = {9},
  title = {SQUAD: software testing for quantum distributed learning software},
  volume = {81},
  year = {2025}
}

@article{li2024automatic,
  author = {Li, Yuechen and Pei, Hanyu and Huang, Linzhi and Yin, Beibei and Cai, Kai-Yuan},
  journal = {ACM Transactions on Software Engineering and Methodology},
  number = {6},
  pages = {1--43},
  title = {Automatic repair of quantum programs via unitary operation},
  volume = {33},
  year = {2024}
}

@article{sato2025buglocating,
  author = {Sato, Naoto and Katsube, Ryota},
  journal = {IEEE Transactions on Software Engineering},
  title = {Bug-locating Method based on Statistical Testing for Quantum Programs},
  year = {2025}
}

@inproceedings{wang2021poster,
  author = {Wang, Jiyuan and Ma, Fucheng and Jiang, Yu},
  booktitle = {2021 14th IEEE Conference on Software Testing, Verification and Validation (ICST)},
  pages = {466--469},
  title = {Poster: Fuzz testing of quantum program},
  year = {2021}
}

@inproceedings{pontolillo2025from,
  author = {Pontolillo, Gabriel and Muqeet, Asmar and Ali, Shaukat and Mousavi, Mohammadreza},
  booktitle = {IEEE International Conference on Quantum Computing and Engineering (QCE)},
  title = {From Ideal to Noisy: Adapting Property-Based Testing for Real-World Noisy Quantum Computers},
  year = {2025}
}

@article{wang2018quanfuzz,
  author = {Wang, Jiyuan and Gao, Ming and Jiang, Yu and Lou, Jianguang and Gao, Yue and Zhang, Dongmei and Sun, Jiaguang},
  journal = {arXiv Preprint arXiv:1810.10310},
  title = {QuanFuzz: Fuzz testing of quantum program},
  year = {2018}
}

@article{jin2025novaq,
  author = {Jin, Tiancheng and Xia, Shangzhou and Zhao, Jianjun},
  journal = {arXiv Preprint arXiv:2509.04763},
  title = {NovaQ: Improving Quantum Program Testing through Diversity-Guided Test Case Generation},
  year = {2025}
}

@article{long2025a,
  author = {Long, Peixun and Zhao, Jianjun},
  journal = {arXiv Preprint arXiv:2505.07243},
  title = {A Black-box Testing Framework for Oracle Quantum Programs},
  year = {2025}
}

@article{shao2024a,
  author = {Shao, Minqi and Zhao, Jianjun},
  journal = {arXiv Preprint arXiv:2411.02450},
  title = {A coverage-guided testing framework for quantum neural networks},
  year = {2024}
}

@article{long2022testing,
  author = {Long, Peixun and Zhao, Jianjun},
  journal = {arXiv Preprint arXiv:2208.09206},
  title = {Testing quantum programs with multiple subroutines},
  year = {2022}
}

@article{ishimoto2025evaluating,
  author = {Ishimoto, Yuta and Kondo, Masanari and Ubayashi, Naoyasu and Kamei, Yasutaka and Katsube, Ryota and Sato, Naoto and Ogawa, Hideto},
  journal = {arXiv Preprint arXiv:2505.09059},
  title = {Evaluating Origin program output Fault Localization for Quantum Programs},
  year = {2025}
}

@article{miranskyy2025on,
  author = {Miranskyy, Andriy and Campos, Jos{\'e} and Mjeda, Anila and Zhang, Lei and de Guzm{\'a}n, Ignacio Garc{\'\i}a Rodr{\'\i}guez},
  journal = {arXiv Preprint arXiv:2507.17235},
  title = {On the Feasibility of Quantum Unit Testing},
  year = {2025}
}

@article{gil2024qcrmut,
  author = {Gil, Sinhu{\'e} Garc{\'\i}a and D{\'\i}az, Luis Llana and Jarabo, Jos{\'e} Ignacio Requeno},
  journal = {arXiv Preprint arXiv:2410.01415},
  title = {QCRMut: Quantum circuit random mutant generator tool},
  year = {2024}
}

@article{klymenko2025qut,
  author = {Klymenko, Mykhailo V and Hoang, Thong and Nguyen, Hoa and Wilkinson, Samuel A and Goldozian, Bahar and Zhenchang, Xing and Lu, Qinghua and Usman, Muhammad and Zhu, Liming},
  journal = {arXiv Preprint arXiv:2509.17538},
  title = {QUT: A Unit Testing Framework for Quantum Subroutines},
  year = {2025}
}

@article{oldfield2025bloch,
  author = {Oldfield, Noah H and Laaber, Christoph and Ali, Shaukat},
  journal = {arXiv Preprint arXiv:2506.18458},
  title = {Bloch Vector Assertions for Debugging Quantum Programs},
  year = {2025}
}

@article{virani2025distinguishing,
  author = {Ahmik Virani and Devraj and Anirudh Suresh and Lei Zhang and M. V. Panduranga Rao},
  journal = {ArXiv},
  title = {Distinguishing Quantum Software Bugs from Hardware Noise: A Statistical Approach},
  volume = {abs/2507.20475},
  year = {2025}
}

@article{klymenko2025contextaware,
  author = {Klymenko, Mykhailo and Hoang, Thong and Wilkinson, Samuel A and Goldozian, Bahar and Ma, Suyu and Xu, Xiwei and Lu, Qinghua and Usman, Muhammad and Zhu, Liming},
  journal = {arXiv Preprint arXiv:2506.10348},
  title = {Context-Aware Unit Testing for Quantum Subroutines},
  year = {2025}
}

@article{pontolillo2025qucheck,
  author = {Pontolillo, Gabriel and Mousavi, Mohammad Reza and Grzesiuk, Marek},
  journal = {arXiv Preprint arXiv:2503.22641},
  title = {QuCheck: A Property-based Testing Framework for Quantum Programs in Qiskit},
  year = {2025}
}

@article{li2019proq,
  author = {Li, Gushu and Zhou, Li and Yu, Nengkun and Ding, Yufei and Ying, Mingsheng and Xie, Yuan},
  journal = {arXiv Preprint arXiv:1911.12855},
  title = {Proq: Projection-based runtime assertions for debugging on a quantum computer},
  year = {2019}
}

@inproceedings{liu2020quantum,
  author = {Liu, Ji and Byrd, Gregory T. and Zhou, Huiyang},
  booktitle = {Proceedings of the Twenty-Fifth International Conference on Architectural Support for Programming Languages and Operating Systems},
  doi = {10.1145/3373376.3378488},
  pages = {1017–1030},
  title = {Quantum Circuits for Dynamic Runtime Assertions in Quantum Computation},
  year = {2020}
}

@article{park2024aqua,
  author = {Park, Soohyun and Baek, Hankyul and Yoon, Jung Won and Lee, Youn Kyu and Kim, Joongheon},
  journal = {Future Generation Computer Systems},
  pages = {545--556},
  title = {AQUA: Analytics-driven quantum neural network (QNN) user assistance for software validation},
  volume = {159},
  year = {2024}
}

@article{chen2022VeriQBench,
  title={VeriQBench: A benchmark for multiple types of quantum circuits},
  author={Chen, Kean and Fang, Wang and Guan, Ji and Hong, Xin and Huang, Mingyu and Liu, Junyi and Wang, Qisheng and Ying, Mingsheng},
  journal={arXiv preprint arXiv:2206.10880},
  year={2022}
}

@article{quetschlich2023mqt,
  title={MQT Bench: Benchmarking software and design automation tools for quantum computing},
  author={Quetschlich, Nils and Burgholzer, Lukas and Wille, Robert},
  journal={Quantum},
  volume={7},
  pages={1062},
  year={2023},
  publisher={Verein zur F{\"o}rderung des Open Access Publizierens in den Quantenwissenschaften}}

@book{rieffel2011quantum,
  title={Quantum computing: A gentle introduction},
  author={Rieffel, Eleanor G and Polak, Wolfgang H},
  year={2011},
  publisher={MIT press}
}

@article{pointing2024optimizing,
  title={Quanto: Optimizing quantum circuits with automatic generation of circuit identities},
  author={Pointing, Jessica and Padon, Oded and Jia, Zhihao and Ma, Henry and Hirth, Auguste and Palsberg, Jens and Aiken, Alex},
  journal={Quantum Science and Technology},
  volume={9},
  number={4},
  pages={045009},
  year={2024},
  publisher={IOP Publishing}
}

@inproceedings{zhao2021bugs4q,
  title={Bugs4Q: A benchmark of real bugs for quantum programs},
  author={Zhao, Pengzhan and Zhao, Jianjun and Miao, Zhongtao and Lan, Shuhan},
  booktitle={2021 36th IEEE/ACM International Conference on Automated Software Engineering (ASE)},
  pages={1373--1376},
  year={2021},
  organization={IEEE}
}

@inproceedings{campos2021qbugs,
  title={Qbugs: A collection of reproducible bugs in quantum algorithms and a supporting infrastructure to enable controlled quantum software testing and debugging experiments},
  author={Campos, Jos{\'e} and Souto, Andr{\'e}},
  booktitle={2021 IEEE/ACM 2nd International Workshop on Quantum Software Engineering (Q-SE)},
  pages={28--32},
  year={2021},
  organization={IEEE}
}

@article{garcia2011equivalent,
  title={Equivalent quantum circuits},
  author={Garcia-Escartin, Juan Carlos and Chamorro-Posada, Pedro},
  journal={arXiv preprint arXiv:1110.2998},
  year={2011}
}

@book{johnston2019programming,
  title={Programming quantum computers: essential algorithms and code samples},
  author={Johnston, Eric R and Harrigan, Nic and Gimeno-Segovia, Mercedes},
  year={2019},
  publisher={O'Reilly Media}
}

@inproceedings{wille2008revlib,
  title={RevLib: An online resource for reversible functions and reversible circuits},
  author={Wille, Robert and Gro{\ss}e, Daniel and Teuber, Lisa and Dueck, Gerhard W and Drechsler, Rolf},
  booktitle={38th International Symposium on Multiple Valued Logic (ismvl 2008)},
  pages={220--225},
  year={2008},
  organization={IEEE}
}

@article{li2023qasmbench,
  title={Qasmbench: A low-level quantum benchmark suite for nisq evaluation and simulation},
  author={Li, Ang and Stein, Samuel and Krishnamoorthy, Sriram and Ang, James},
  journal={ACM Transactions on Quantum Computing},
  volume={4},
  number={2},
  pages={1--26},
  year={2023},
  publisher={ACM New York, NY}
}


\vfill
\end{document}